\documentclass[aps,prb,twocolumn,superscriptaddress,nofootinbib,floatfix]{revtex4-2}
\usepackage{graphicx}
\usepackage{amsmath,amssymb}
\usepackage{bm}
\usepackage[colorlinks=true,linkcolor=blue,citecolor=blue,urlcolor=blue]{hyperref}
\begin{document}

\title{Fractional quantum Hall effect at $\nu=1/2$ in wide GaAs quantum wells: Subband structure, exact diagonalization, and the magnitude of the gap}

\author{Sankar Das Sarma}
\affiliation{Condensed Matter Theory Center and Joint Quantum Institute, Department of Physics, University of Maryland, College Park, Maryland 20742, USA}
\date{\today}

\begin{abstract}
Recent experiments on a wide GaAs quantum well report a fractional quantum Hall state at filling factor $\nu=1/2$ with an unusually large activation gap, a sharp maximum of the gap as a function of density, and a possible transition from the non-Abelian Moore-Read Pfaffian to the Abelian Halperin 331 state. We analyze this system theoretically, with self-consistent subband calculations, exact diagonalization, and a reanalysis of the published transport data. We find that the well has two relevant subbands, that with the second subband frozen out the well most likely has no incompressible state at $\nu=1/2$, that a strict bilayer at the maxima of the charge density does not produce 331, and that an active second subband is essential: the ground state is a paired state of electrons in both subbands, on a continuous path between the Pfaffian and 331, with a neutral gap much larger than that of a one-subband Pfaffian. The position of the state on this path is controlled by electrostatic energies that the accessible system sizes contain only in part, and the standard overlaps do not identify the state at these sizes. Our results are more consistent with a Pfaffian-like state than with 331, but they are not definitive. The transport data do not support the reported magnitude of the gap; the gap we extract is smaller and comparable to the calculated two-subband neutral gap. A benchmark calculation for the Laughlin state at $\nu=1/3$ in the same well provides an experimental test of the electrostatic treatment.

\end{abstract}

\maketitle

\section{Introduction}

The even-denominator fractional quantum Hall effect (FQHE) occupies a special place in condensed matter physics because it is the only experimentally established setting in which non-Abelian topological order is a serious possibility. The $\nu=5/2$ state in the second orbital Landau level (SLL) of high-mobility GaAs two-dimensional (2D) electron systems~\cite{Willett1987} is widely believed to be in the universality class of the Moore-Read (MR) Pfaffian~\cite{MooreRead1991,GreiterWenWilczek1991,ReadGreen2000}, a spin-polarized one-component paired state of composite fermions whose quasiparticles are Ising anyons, and it is the original platform proposed for topological quantum computation~\cite{DasSarmaFreedmanNayak2005,NayakRMP2008}.

The half-filled lowest Landau level (LLL) is different. In a thin 2D layer the $\nu=1/2$ state is the compressible composite-fermion Fermi liquid of Halperin, Lee, and Read~\cite{HLR1993}, and no FQHE exists. An incompressible $\nu=1/2$ state does appear in two related systems: double quantum wells~\cite{Eisenstein1992} and wide single quantum wells~\cite{Suen1992,Suen1992b,Suen1994,Shabani2013}, where the electrostatic repulsion among the electrons produces a bilayer-like charge distribution. In both cases the natural incompressible state is the two-component Abelian Halperin $(3,3,1)$ state~\cite{Halperin1983}, denoted 331 below. In a wide well, however, the two ``layers'' are connected by tunneling that is not small, and sufficiently strong tunneling converts any two-component state into a one-component one. A one-component incompressible state at $\nu=1/2$ in the LLL would be a paired state, the leading candidate being the MR Pfaffian~\cite{GreiterWenWilczek1992,Ho1995}. Whether the wide-well $\nu=1/2$ FQHE is the Abelian 331 state or the non-Abelian Pfaffian has been an open question for more than thirty years.

The question has been sharpened by recent experiments from the Princeton group on an ultra-high-quality 72.5-nm-wide GaAs quantum well whose density $n$ is tuned by gates~\cite{Singh2024,Singh2025}. Singh \textit{et al.}~\cite{Singh2025} report (i) a $\nu=1/2$ FQHE with a transport activation gap of up to about 6~K at $n=1.37\times10^{11}$~cm$^{-2}$; (ii) a sharp maximum of the gap as a function of density, the gap falling by half for a change of less than 7\% in density on either side; and (iii) evidence, from the states flanking $\nu=1/2$, that the electron system changes from one-component to two-component character close to the density of the gap maximum. They suggest that the $\nu=1/2$ state itself undergoes a transition from the Pfaffian on the low-density side to 331 on the high-density side, while noting that it could remain one-component throughout.

These are important claims. At the field of the experiment ($B=11.3$~T at $n=1.37$) the Coulomb energy is $E_C=e^2/\epsilon\ell=170$~K, where $\ell=(\hbar/eB)^{1/2}$ is the magnetic length and $\epsilon$ the dielectric constant, so that 6~K is $0.035\,E_C$. The calculated gap of the $\nu=5/2$ state in the ideal zero-width, disorder-free limit is $0.025\,E_C$~\cite{Morf1998,MorfdAmbrumenilDasSarma2002}, and the most complete exact-diagonalization study of the one-component $\nu=1/2$ problem as a function of layer thickness found gaps several times smaller than that, decreasing with system size~\cite{StorniMorfDasSarma2010}. Measured FQHE gaps are smaller than the corresponding ideal theoretical gaps, because disorder, finite thickness, and Landau-level mixing all reduce the gap; at $\nu=5/2$ the reduction is by a factor of 2 to more than 10. To our knowledge a measured even-denominator gap exceeding the ideal $5/2$ gap in Coulomb units has no precedent, and if the state is non-Abelian a gap of this size would change the experimental outlook of the field.

In this paper we provide a theoretical analysis of the experiment. We address three questions in turn. What is the single-particle (subband) structure of the wide well, and how sensitive is it to the electrostatic model used? What does exact diagonalization, which makes no assumption about the nature of the ground state, give for the gap and for the character of the $\nu=1/2$ state as a function of density, and how far can its results be trusted at the system sizes that are accessible? And what do the published transport data establish about the size of the gap and its density dependence?

Our main results are the following.

(1) The subband structure depends strongly on how the neutralizing positive charge is modeled, and the well has two relevant subbands. A calculation in which the electrons alone determine the electrostatic potential in the well, as appropriate for a symmetric modulation-doped structure, gives a symmetric-antisymmetric splitting $\Delta_{\rm SAS}=0.94$~meV at $n=1.37$, i.e., $\alpha\equiv\Delta_{\rm SAS}/E_C=0.064$, consistent with the values of $\alpha$ quoted in Ref.~\onlinecite{Singh2025}. The second subband is thus only $0.064\,E_C$ above the first, and a description that omits it is not justified a priori. A uniform jellium background in the well gives 2.26~meV, a factor of 2.4 larger, and should not be used for this sample.

(2) Exact diagonalization of the Hamiltonian of the actual well, with both subbands retained, for 8 electrons on the sphere, gives a uniform $L=0$ ground state at all densities, with 7--18\% of the electrons in the antisymmetric subband over the experimental range. It is a paired state that is closer to the Pfaffian than to 331 (overlaps of 0.85--0.66 and 0.44--0.65, respectively, between $n=1.0$ and 1.8), and its gap is 2--4~K, rising with density, with no maximum.

(3) That result is not converged in a specific and identifiable way. The population of the antisymmetric subband is controlled by long-wavelength electrostatic energies, and a sphere of radius 2--2.5 magnetic lengths contains only a quarter to a half of them. When these terms are replaced by their exactly known values for the infinite system, the antisymmetric population at $n=1.37$ rises from 12\% to 37--43\%, and the gap acquires a maximum as a function of density. Exact diagonalization at these sizes therefore does not determine whether the state at a given density is closer to the Pfaffian or to 331.

(4) It does determine the structure. As the antisymmetric subband is populated the ground state evolves continuously along the one-parameter family of triplet-paired states introduced by Ho~\cite{Ho1995}, from the Pfaffian toward 331. For 8 electrons the neutral gap has a maximum of $0.025\,E_C$, three times its value when the second subband is frozen out, when a quarter of the electrons are in the antisymmetric subband; the maximum is a crossing of two excitation branches of different reflection parity. As functions of the antisymmetric population these results are the same in all treatments of the electrostatic terms. We also show that the overlaps with the Pfaffian and with 331, the standard diagnostics, do not identify the state at these sizes.

(5) If the antisymmetric subband is frozen out altogether, which is the one-component description used in part of the literature, the ground state is Pfaffian-like for all densities, for widths $w\ge5\,\ell$, and for any model of the transverse wave function, and its gap decreases with system size to 0.6--1.0~K for 16 electrons, is nearly independent of density, and has no maximum.

(6) The published Arrhenius plots are curved: a single activation gap does not span the measured temperature range. At the density of the maximum the slope runs from 1.6--1.7~K at the high-temperature end of the trace to 5--7~K at the low-temperature end, and the reported gaps are the low-temperature slopes. Neither is the gap. The high-temperature slope is suppressed by the saturation of the resistance at its value in the compressible state. The low-temperature slope of 6~K requires a prefactor 25 times larger than activated transport allows, and a state with a 6~K gap would not lose its resistance minimum at 0.4--0.5~K, as this one does. The gap that follows from the magnitude of the resistance, with the theoretical prefactor of activated transport, which fits the data at the three densities away from the maximum without adjustment, is $3.9\pm0.5$~K at $n=1.37$, 3.5~K at $n=1.31$, and 2.3--3.0~K at $n=1.21$, 1.49, and 1.59, in agreement with the 3.7~K reported earlier by the same group~\cite{Singh2024}. The maximum is real, and it is lower and broader than reported.

(7) A gap of about 4~K, $0.023\,E_C$, is four to six times the gap of the frozen-subband limit for 16 electrons. A one-component Pfaffian does not have a gap of this size in this well. It is equal, within the uncertainties, to the maximum of the neutral gap of the two-subband model for 8 electrons, but that agreement is for one system size and for the ideal system, and measured even-denominator gaps are otherwise well below the ideal ones. The magnitude of the gap is not yet explained.

(8) We conclude that the $\nu=1/2$ state of the wide well is neither a one-component Pfaffian nor a demonstrated large-gap non-Abelian state. With the second subband frozen out the well most likely has no Pfaffian FQHE at all; a strict bilayer at the maxima of the charge density does not give 331; and an active second subband is essential. The state is a paired state of electrons in two subbands on a continuous path between the Pfaffian and 331. Our finite-size results taken together are more consistent with the Pfaffian side of that path than with 331, but they are not definitive, because the two-subband calculation is limited to eight electrons.

(9) All gaps that we calculate are neutral gaps: the lowest excitation energy at fixed flux in a finite system. The charge (transport) gap is the energy of a well-separated quasiparticle-quasihole pair in the infinite system, which is typically larger, since a pair in a finite system is not well separated and is lowered by the electrostatic attraction; moreover, some low-lying neutral excitations, here the intersubband ones, are not quasiparticle-quasihole pairs at all, and the two gaps need not vanish together at a transition. The activation gap measured in transport is usually identified with the charge gap, but this identification is problematic, because the transport process in the fractional quantum Hall regime is not understood and disorder certainly reduces the charge gap.

The rest of the paper is organized as follows. Section~\ref{sec:background} summarizes the background and the earlier theory. Section~\ref{sec:exp} summarizes the experiment. Section~\ref{sec:hartree} presents the self-consistent Hartree calculations. Section~\ref{sec:method} describes the exact-diagonalization models, including the treatment of the uniform interaction terms on which the two-subband results depend. Section~\ref{sec:twocomp} gives the results for the bilayer and for the two-subband model of the well, and Sec.~\ref{sec:onecomp} those of the frozen-subband limit. Section~\ref{sec:transport} analyzes the transport data. Section~\ref{sec:discussion} discusses the implications, and Sec.~\ref{sec:conclusion} lists our conclusions, the approximations on which they rest, and the theoretical and experimental work needed for further progress. The Appendices contain the jellium results, the Hartree subband structure of the earlier Princeton wide wells, numerical details and validation, the width dependence of the one-component gap in three models of the transverse wave function, a strict zero-width bilayer placed at the maxima of the Hartree density, a benchmark calculation for the Laughlin state at $\nu=1/3$ in the same well, and the entanglement spectrum of the two-subband ground state at the density of the gap maximum.

\section{Background}
\label{sec:background}

\subsection{$\nu=5/2$ and the Moore-Read Pfaffian}

The Moore-Read state~\cite{MooreRead1991} is the exact zero-energy ground state of a three-body interaction that penalizes three electrons in the closest allowed configuration~\cite{GreiterWenWilczek1991}; on the sphere it occurs at $2Q=2N-3$, where $2Q$ is the number of flux quanta through the surface~\cite{Haldane1983}. It is a chiral $p$-wave paired state of composite fermions~\cite{ReadGreen2000}, with quasiparticles of charge $e/4$. Numerical work on the Coulomb interaction in the SLL has established that the ground state at this shift is uniform, has a large overlap with the Pfaffian or its particle-hole conjugate, and is adiabatically connected to it~\cite{Morf1998,RezayiHaldane2000,MorfdAmbrumenilDasSarma2002,Feiguin2008,PetersonJolicoeurDasSarma2008,PetersonJolicoeurDasSarma2008b,StorniMorfDasSarma2010,WojsQuinn2006,WojsTokeJain2010,Pakrouski2015,Rezayi2017}. The excitation gap of the ideal system (zero width, no disorder, no Landau-level mixing) extrapolates to $\Delta_{5/2}\approx0.025\,e^2/\epsilon\ell$~\cite{Morf1998,MorfdAmbrumenilDasSarma2002}; a density-matrix renormalization-group study obtained 0.030 and described it as an upper bound~\cite{Feiguin2008}.

Two experimental facts about $5/2$ are benchmarks for what follows. First, the measured activation gaps, in the convention $R_{xx}\propto\exp(-\Delta/2k_BT)$, are 0.11--0.45~K~\cite{Pan2008}, 0.27 and 0.54~K~\cite{Choi2008}, and 0.26~K~\cite{Dean2008} in samples with $5/2$ at 2.6--5.3~T, and 0.82~K in the highest-quality sample to date, with $5/2$ at about 1.7~T~\cite{Chung2021}. With $0.025\,e^2/\epsilon\ell=1.26\sqrt{B[{\rm T}]}$~K these are between a factor of 2 and a factor of more than 10 below the ideal gap. No measured gap of an even-denominator state in GaAs has equaled the ideal one. Part of the reduction is intrinsic to activated transport in a smooth disorder potential, where the measured activation energy is that of the saddle points and is further lowered by tunneling~\cite{dAmbrumenilHalperinMorf2011}. Second, transport cannot determine the topological order. The evidence that the $5/2$ state is non-Abelian comes from numerics and from thermal Hall and interference experiments~\cite{Banerjee2018,Willett2009}.

\subsection{$\nu=1/2$ in bilayers and wide wells}

In a thin single layer the half-filled LLL is compressible~\cite{HLR1993}. A FQHE at total filling $1/2$ was discovered in 1992 in double quantum wells~\cite{Eisenstein1992} and in wide single wells~\cite{Suen1992,Suen1992b}, both of which have a second, layer or subband, degree of freedom. The natural two-component state is Halperin's~\cite{Halperin1983}
\begin{equation}
\Psi_{331}=\prod_{i<j}(z_i-z_j)^3\prod_{i<j}(w_i-w_j)^3\prod_{i,j}(z_i-w_j),
\label{eq:331}
\end{equation}
where $z$ and $w$ are the coordinates of electrons in the two layers. It is Abelian and occurs on the sphere at the same shift $2Q=2N-3$ as the Pfaffian. The possibility of an incompressible state at total filling $1/2$ in double layers had been raised by numerical work before the experiments~\cite{YoshiokaMacDonaldGirvin1989,HeXieDasSarmaZhang1991}, and the observed states were identified with 331 immediately afterwards~\cite{HeDasSarmaXie1993}.

A wide well differs from a double well in that the tunneling between the two halves is not small, and the interpretation of its $\nu=1/2$ state has gone back and forth. Suen \textit{et al.}~\cite{Suen1994} studied the dependence of the gap on the subband separation and on charge asymmetry and concluded that the state has a two-component origin. Shabani \textit{et al.}~\cite{Shabani2013} mapped the stability of the state in symmetric wells of width 41--96~nm, found it to be bounded by a compressible state at low density and an insulating phase at high density, and considered it very likely to be the 331 state. On the other hand, a $\nu=1/2$ state that appears in asymmetric wide wells was attributed to a one-component, possibly Moore-Read, state~\cite{Shabani2009b}, and the observation of FQHE states at $\nu=8/17$ and $7/13$ next to $\nu=1/2$ in a 72.5-nm well was interpreted as evidence for a Pfaffian parent state~\cite{Singh2024}, these being the fillings of the simplest daughter states of the Pfaffian~\cite{LevinHalperin2009}. Of the two, $7/13$ is also a daughter of the 331 state~\cite{LevinHalperin2009,Singh2024}; $8/17$ is the one that distinguishes the Pfaffian.

On the theory side there are three lines of work. (i) \textit{One-component.} A paired one-component state was proposed for the double-layer and wide-well systems at the outset~\cite{GreiterWenWilczek1992}. Storni, Morf, and Das Sarma~\cite{StorniMorfDasSarma2010} showed by exact diagonalization for up to 18 electrons that a finite layer width adiabatically connects the LLL Coulomb ground state at $\nu=1/2$ to the Pfaffian, but that the gaps are small, decrease with system size in a nonmonotonic way, and open at a width that increases with system size; they concluded that a gap is unlikely to survive in the thermodynamic limit for any width supporting a single-layer system. (ii) \textit{Bilayer with tunneling.} Tunneling polarizes the layer pseudospin and must eventually convert a two-component state into a one-component one. Ho~\cite{Ho1995} constructed a one-parameter family of triplet-paired states that interpolates between 331 and the Pfaffian, Read and Rezayi~\cite{ReadRezayi1996} showed that Ho's model Hamiltonian has the degeneracies of the 331 state except at its Pfaffian point, where its gap collapses, and Read and Green~\cite{ReadGreen2000} showed that the two phases are separated by a transition at which the gap of the antisymmetric fermions closes. Exact-diagonalization studies of the bilayer with tunneling reached differing conclusions about the one-component side. Papi\'c \textit{et al.}~\cite{Papic2010} found that tunneling of order $0.1\,e^2/\epsilon\ell$ at $d=\ell$ drives 331 into a state with a large Pfaffian overlap on the sphere but a very small gap, and did not find evidence that it is a stable phase. Peterson and Das Sarma~\cite{PetersonDasSarma2010}, for eight electrons, found the Pfaffian only at strong tunneling and small layer separation and with a very small gap, found the gap to be largest close to the transition line on the 331 side, and concluded that the states observed up to then were most likely 331. Peterson, Papi\'c, and Das Sarma~\cite{PetersonPapicDasSarma2010} added a charge imbalance and found the 331 state to be remarkably robust to it. Zhu \textit{et al.}~\cite{Zhu2016}, using the density-matrix renormalization group for up to 24 electrons in a bilayer with layer separation $3\ell$ and layer width $1.5\ell$, found a continuous transition at a tunneling amplitude of 0.04--0.07\,$e^2/\epsilon\ell$ to a phase that they identified as the Moore-Read state from its entanglement properties. (iii) \textit{Variational.} Zhao \textit{et al.}~\cite{Zhao2021}, by three-dimensional fixed-phase diffusion Monte Carlo at high field with Landau-level mixing, concluded that the wide-well state is likely to be the one-component Pfaffian, with the caveats that the result depends on the choice of phase and that the two-component states were evaluated with $\Delta_{\rm SAS}=0$; their own exact diagonalization of the one-component problem for 14--18 electrons gave very small or negative gaps. Sharma \textit{et al.}~\cite{Sharma2024} found a $p$-wave pairing instability of the one-component composite-fermion Fermi sea for wide-well interactions, with the 331 state higher in energy, again without the subband splitting in the two-component energy.

The one-component calculations are of two kinds, which should be distinguished: a strict single-layer model~\cite{StorniMorfDasSarma2010}, which is well defined, and calculations for a well with two occupied subbands that keep a single component~\cite{Zhao2021,Sharma2024}, which assume what is to be shown. The wide well itself has been treated as a two-subband system in two studies, both with the subbands of an infinite square well and with the splitting as a free parameter: Papi\'c \textit{et al.}~\cite{Papic2009}, for eight electrons, found a crossover from 331 at small splitting to a state of large Pfaffian overlap at large splitting, and Thiebaut \textit{et al.}~\cite{Thiebaut2013}, for a biased well with up to ten electrons, found a 331 state rotated in the subband pseudospin. None of these studies treats the actual wide well, with its self-consistent subband wave functions, its measured subband splitting, and both subbands retained. That is the calculation attempted here.

\section{The experiment}
\label{sec:exp}

We summarize the aspects of Ref.~\onlinecite{Singh2025} that enter our analysis. The sample is a 72.5-nm-wide GaAs quantum well with Al$_{0.3}$Ga$_{0.7}$As barriers, of mobility about $10\times10^6$~cm$^2$/Vs. The density is varied with front and back gates between 0.98 and 1.82 while the charge distribution is kept symmetric. (Densities are in units of $10^{11}$~cm$^{-2}$ throughout.) The subband splitting $\Delta_{\rm SAS}$ is obtained from the two frequencies of the low-field Shubnikov-de Haas oscillations, i.e., from the difference of the subband densities at zero field, and the system is characterized by $\alpha=\Delta_{\rm SAS}/(e^2/\epsilon\ell)$. Reference~\onlinecite{Singh2025} presents $\Delta_{\rm SAS}$ and $\alpha$ graphically; the values of $\alpha$ quoted in its text are those at which the behavior changes.

The observations are as follows. (i) For $\alpha>0.11$, at the lowest density, there is no FQHE at $\nu=1/2$ and the state is a composite-fermion Fermi sea. (ii) With increasing density a $\nu=1/2$ FQHE develops. Its gap, obtained from fits of the longitudinal resistance at the minimum to $R_{xx}\propto\exp(-\Delta/2k_BT)$, is reported to reach about 6~K at $n=1.37$ and to fall by about 50\%, to about 3~K, at $n=1.28$ and $n=1.45$, a change of less than 7\% in density. The Arrhenius plots for five densities are shown in the Supplemental Material of that paper together with the gap at 17 densities, the highest being $n\approx1.59$, where the gap is about 2.1~K; we analyze them in Sec.~\ref{sec:transport}. (iii) For $0.064\lesssim\alpha\lesssim0.073$ FQHE states are observed at $\nu=8/17$ and $7/13$; they are fully developed at $n=1.35$. (iv) At $\alpha\simeq0.064$ the electron system as a whole changes from one-component to two-component, as indicated by the appearance of insulating phases attributed to a bilayer Wigner crystal on the flanks of $\nu=1/2$ and by the disappearance of several odd-numerator FQHE states. This value of $\alpha$ lies just on the high-density side of the gap maximum, which is thus on the one-component side of the transition. (v) The $\nu=1/2$ state is destroyed by a deliberately imposed charge asymmetry of 18\% at $n=1.17$, 27\% at $n=1.28$, and 13\% at $n=1.45$. (vi) The $\nu=1/2$ FQHE is still present at the highest density at which its gap was measured; its expected disappearance near $n\approx1.78$ could not be reached with the available magnetic field.

The authors conclude that the $\nu=1/2$ state has a one-component origin for $\alpha>0.064$, suggest that at higher density it is a two-component state, so that a transition between the Pfaffian and 331 occurs as $\alpha$ is reduced, and state that they cannot rule out that it remains one-component over the whole range. They note that the measured gaps near the maximum exceed those calculated for the bilayer with tunneling~\cite{PetersonDasSarma2010,Zhu2016}, and that the maximum is much sharper than calculated. They estimate a disorder broadening of about 0.5~K.

Two further facts are relevant. First, the same group earlier reported, for a 72.5-nm well and with the same definition of the gap, a $\nu=1/2$ gap that increases from 1.48~K at $n=1.17$ to 3.73~K at $n=1.35$~\cite{Singh2024}. At $n=1.17$ this agrees with the value in Ref.~\onlinecite{Singh2025}; at $n=1.35$ Ref.~\onlinecite{Singh2025} shows about 5.7~K. Second, Ref.~\onlinecite{Singh2025} compares its results with those of Ref.~\onlinecite{Suen1994} for a 77-nm well of lower mobility, in which the gap also has a maximum as a function of $\alpha$, at nearly the same $\alpha$, but is about eight times smaller.

\section{Self-consistent Hartree theory of the wide well}
\label{sec:hartree}

\subsection{Model}

We treat the motion along the growth direction $z$ in the effective-mass approximation with $m^*=0.067\,m_e$ and $\epsilon=12.9$. The confining potential is a square well of width $w$ with barrier height $V_0=224$~meV, appropriate for Al$_{0.3}$Ga$_{0.7}$As. The subband wave functions $\phi_i(z)$ and energies $E_i$ satisfy
\begin{equation}
\left[-\frac{\hbar^2}{2m^*}\frac{d^2}{dz^2}+V_{\rm conf}(z)+U_H(z)\right]\phi_i(z)=E_i\phi_i(z),
\label{eq:schr}
\end{equation}
with the Hartree potential determined by Poisson's equation,
\begin{equation}
\frac{d^2U_H}{dz^2}=-\frac{4\pi e^2}{\epsilon}\left[n(z)-n_b(z)\right],
\label{eq:poisson}
\end{equation}
where $n(z)=\sum_in_i|\phi_i(z)|^2$ is the electron density and $n_b(z)$ the neutralizing positive charge. At zero magnetic field the subband densities follow from the 2D density of states, $n_i=(m^*/\pi\hbar^2)(E_F-E_i)\theta(E_F-E_i)$, with $\sum_in_i=n$. In the density range of interest two subbands are occupied, the symmetric ($S$) and antisymmetric ($A$) ones, so that
\begin{equation}
\Delta_{\rm SAS}=E_A-E_S=\frac{\pi\hbar^2}{m^*}\,(n_S-n_A).
\label{eq:sas}
\end{equation}
Equation~(\ref{eq:sas}) is the relation used in Ref.~\onlinecite{Singh2025} to extract $\Delta_{\rm SAS}$ from the Shubnikov-de Haas frequencies. The measured quantity is therefore directly the zero-field subband population difference, which the Hartree calculation predicts.

We consider two models for $n_b$. In the \textit{remote-doping} model the positive charge resides in two symmetric sheets outside the well, as in the actual modulation-doped and gated structure. The sheets produce no electric field between them, so the potential in the region occupied by the electrons is determined by the electrons alone, and the result is independent of the setback distance. This is the physically appropriate model for a symmetric well. In the \textit{jellium} model the positive charge is smeared uniformly over the well, $n_b=n/w$ for $|z|<w/2$. This model, which is often used for convenience, cancels most of the Hartree repulsion that pushes the electrons toward the walls. We present both, because the comparison shows which conclusions depend on the electrostatic model and which do not. Exchange-correlation corrections to the subband structure are not included in either model.

Equations~(\ref{eq:schr}) and (\ref{eq:poisson}) are solved by finite differences (2001 points on a 200~nm domain) with linear mixing to a self-consistency of $10^{-4}$~meV in $U_H$.

\subsection{Results}

\begin{figure*}[t]
\includegraphics[width=0.92\textwidth]{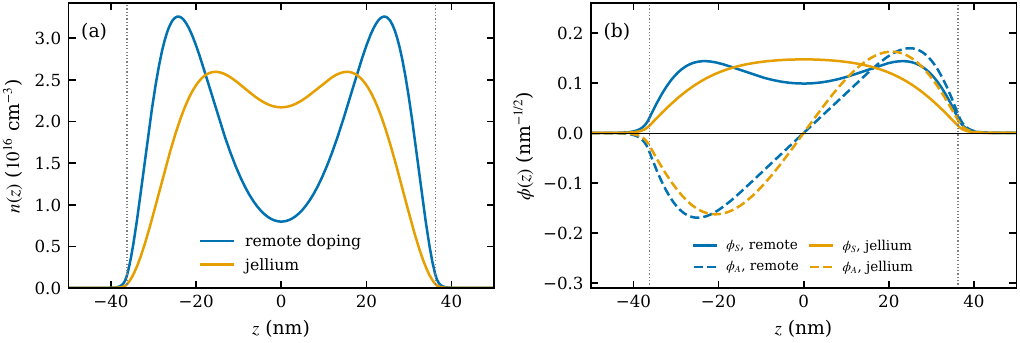}
\caption{Self-consistent Hartree results for $w=72.5$~nm and $n=1.37\times10^{11}$~cm$^{-2}$ in the remote-doping model (blue) and the jellium model (orange). (a) Electron density $n(z)$. (b) Symmetric ($\phi_S$, solid) and antisymmetric ($\phi_A$, dashed) subband wave functions. Dotted vertical lines mark the well edges. With remote doping the density at the well center is 24\% of its maximum; with jellium it is 84\%.}
\label{fig:profiles}
\end{figure*}

\begin{table}[t]
\caption{Hartree results at $w=72.5$~nm, $n=1.37\times10^{11}$~cm$^{-2}$ ($B=11.33$~T at $\nu=1/2$, $\ell=7.62$~nm, $E_C=14.65$~meV $=170$~K). $d$ is the separation of the two density maxima.}
\label{tab:hartree}
\begin{ruledtabular}
\begin{tabular}{lcc}
\hline\hline
 & Remote doping & Jellium\\
\hline
$\Delta_{\rm SAS}$ (meV) & 0.94 & 2.26\\
$\alpha=\Delta_{\rm SAS}/E_C$ & 0.064 & 0.154\\
$n_S$ ($10^{11}$ cm$^{-2}$) & 0.816 & 1.001\\
$n_A$ ($10^{11}$ cm$^{-2}$) & 0.554 & 0.369\\
$n_A/n$ & 40.4\% & 27.0\%\\
$d$ (nm) & 48.4 & 30.9\\
$d/\ell$ & 6.35 & 4.05\\
$n(0)/n_{\rm max}$ & 0.24 & 0.84\\
\hline\hline
\end{tabular}
\end{ruledtabular}
\end{table}

Figure~\ref{fig:profiles} and Table~\ref{tab:hartree} show the results at the density of the reported gap maximum. The two models differ substantially. With remote doping the charge distribution is strongly bilayer-like: two layers of full width at half maximum $\lambda\approx2.6\,\ell$ centered at $z=\pm24$~nm, a peak separation $d=6.35\,\ell$, and a density at the well center that is 24\% of the maximum. The subband splitting is $\Delta_{\rm SAS}=0.94$~meV and 40\% of the electrons are in the antisymmetric subband at zero field. With jellium the distribution has only a shallow minimum at the center, $\Delta_{\rm SAS}=2.26$~meV, and 27\% of the electrons are in the antisymmetric subband.

\begin{figure*}[t]
\includegraphics[width=0.92\textwidth]{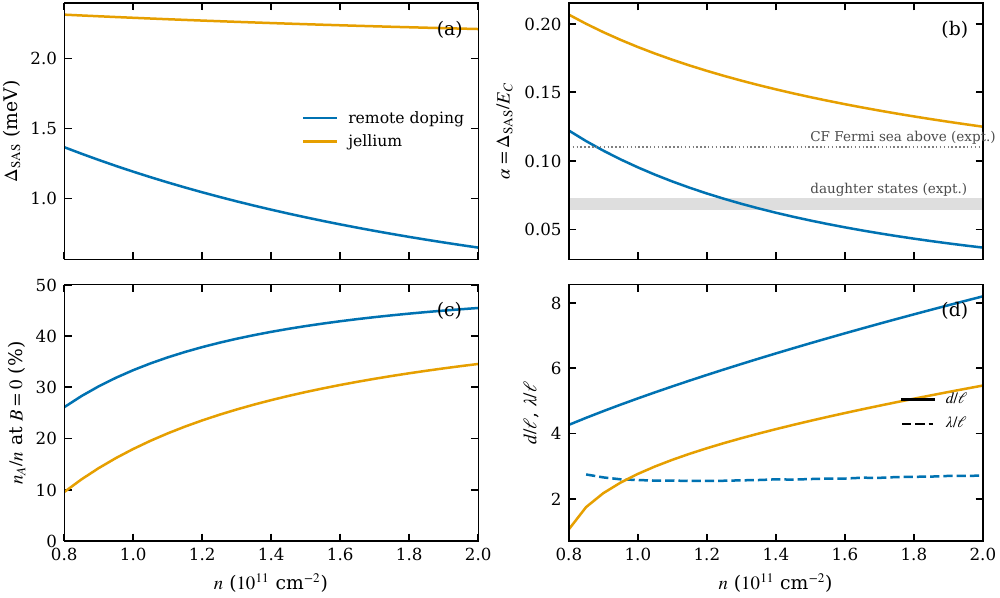}
\caption{Density dependence of the Hartree results for $w=72.5$~nm in the remote-doping (blue) and jellium (orange) models. (a) Subband splitting $\Delta_{\rm SAS}$. (b) $\alpha=\Delta_{\rm SAS}/E_C$ with $E_C$ evaluated at $\nu=1/2$; the shaded band is the range $0.064<\alpha<0.073$ in which Ref.~\onlinecite{Singh2025} observes the daughter states (fully developed at $n=1.35$), and the dotted line is the value above which that experiment finds a composite-fermion Fermi sea. (c) Fraction of electrons in the antisymmetric subband at zero field. (d) Separation $d$ of the density maxima (solid) and full width at half maximum $\lambda$ of each layer (dashed), in units of the magnetic length at $\nu=1/2$; $\lambda$ is defined only when the central density is below half the maximum.}
\label{fig:hartreedens}
\end{figure*}

Figure~\ref{fig:hartreedens} shows the density dependence. In the remote-doping model $\Delta_{\rm SAS}$ falls from 1.19~meV at $n=1.0$ to 0.73~meV at $n=1.8$, and $\alpha$ falls from 0.095 to 0.043, as the increasing Hartree repulsion raises the barrier at the well center; the antisymmetric-subband fraction rises from 33\% to 44\% and $d/\ell$ from 5.1 to 7.6. In the jellium model $\Delta_{\rm SAS}$ is nearly independent of density (2.29 to 2.22~meV) and $\alpha$ stays between 0.18 and 0.13.

Reference~\onlinecite{Singh2025} gives $\Delta_{\rm SAS}$ and $\alpha$ only graphically, so the comparison with the measurement is limited to the values of $\alpha$ quoted in its text. The experiment finds that $\Delta_{\rm SAS}$ decreases with increasing density, as in the remote-doping model. It observes the daughter states for $0.064\lesssim\alpha\lesssim0.073$ and reports them fully developed at $n=1.35$, where the remote-doping model gives $\alpha=0.065$ and the jellium model 0.155. It finds a compressible state for $\alpha>0.11$ at the lowest density, $n=0.98$, where the remote-doping model gives 0.097 and the jellium model 0.185. The remote-doping values are thus consistent with the experimental ones at the higher density and at least 12\% low at the lowest density. The jellium values are too large by a factor of 1.7 to 2.4 and nearly independent of density. We therefore use the remote-doping model as our primary model and present the jellium results for comparison. The remote-doping model has no adjustable parameter, but its agreement with the measured $\alpha$ is established only at the level just stated; a direct comparison with the measured $\Delta_{\rm SAS}(n)$ should be made by those who have the data.

\begin{figure*}[t]
\includegraphics[width=0.92\textwidth]{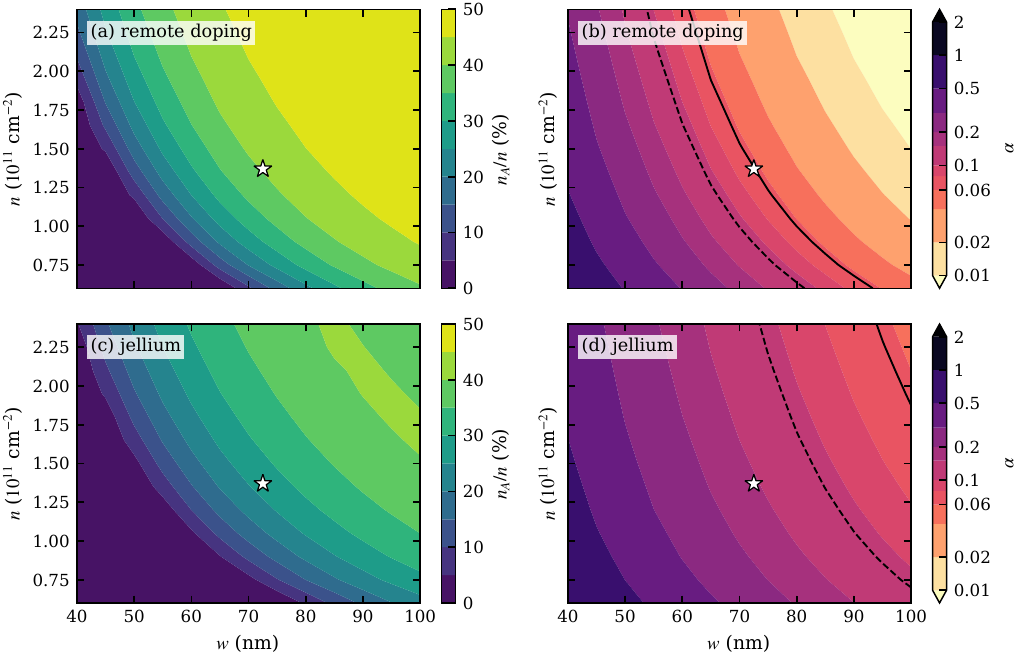}
\caption{Hartree results over the plane of well width $w$ and density $n$. Left: fraction of electrons in the antisymmetric subband at zero field. Right: $\alpha=\Delta_{\rm SAS}/E_C$ at $\nu=1/2$ on a logarithmic color scale; the black solid and dashed contours are $\alpha=0.064$ and $0.11$. (a),(b) Remote-doping model. (c),(d) Jellium model. The star marks $w=72.5$~nm, $n=1.37$. In the remote-doping model a third subband is not occupied anywhere in the range shown. In the jellium model it is occupied for $w\ge85$~nm and $n\ge1.8$, by up to 10\% of the electrons, and Eq.~(\ref{eq:sas}) does not apply there.}
\label{fig:hartreemaps}
\end{figure*}

Figure~\ref{fig:hartreemaps} gives the antisymmetric-subband fraction and $\alpha$ over a wide range of width and density. In the remote-doping model the contours of constant $\alpha$ run diagonally across the plane, wider wells reaching a given $\alpha$ at lower density. The range of $\alpha$ over which the $\nu=1/2$ FQHE is observed in the 72.5-nm well, from 0.11 down to at least 0.05 (the value at $n=1.59$), corresponds to a diagonal strip in the $(w,n)$ plane, as found empirically in the earlier wide-well experiments~\cite{Shabani2013}. In Appendix~\ref{app:earlier} the same calculation is applied to the 68-nm well of Ref.~\onlinecite{Suen1992}, in which the wide-well $\nu=1/2$ FQHE was first observed, and to the 77-nm well of Ref.~\onlinecite{Suen1994}; the state was observed in these wells in the same window of $\alpha$ and of antisymmetric-subband occupancy as in the 72.5-nm well, and the calculated $\Delta_{\rm SAS}$ and layer separation agree reasonably with the values measured in those experiments. In all three samples the density range in which the $\nu=1/2$ FQHE is seen is a range in which the second subband is substantially occupied, which is the main conclusion of this paper: the wide-well $\nu=1/2$ FQHE manifests only when the second subband plays an active role.

Two points about the meaning of the antisymmetric-subband population deserve emphasis. First, a 60:40 division of the electrons between the symmetric and antisymmetric subbands is not a 60:40 division between the two layers. The well is symmetric, so the two layers hold equal charge regardless of the subband populations. In the language of the layer pseudospin, with the two layers as the eigenstates of $\tau_z$, tunneling couples to $\tau_x$, the symmetric and antisymmetric states are the eigenstates of $\tau_x$, and the subband population difference is the pseudospin polarization along $x$, $\langle\tau_x\rangle=(n_S-n_A)/n$. The 331 state has $\langle\tau_x\rangle=0$, i.e., equal populations of the two subbands, and the Pfaffian in the symmetric subband has $\langle\tau_x\rangle=1$. Unequal subband populations are therefore a partial pseudospin polarization along the tunneling direction, which the 331 state acquires continuously when tunneling is switched on (Sec.~\ref{sec:bilayer}); they are not a charge imbalance between the layers, which is a polarization along $z$ and is absent in the experiment except when deliberately imposed.

Second, the zero-field populations are not the populations at $\nu=1/2$. At zero field the antisymmetric subband is occupied because of the Pauli principle: the Fermi energy $\pi\hbar^2n_S/m^*$ exceeds $\Delta_{\rm SAS}$. At $\nu=1/2$ the lowest Landau level of the symmetric subband alone can accommodate all the electrons, and noninteracting electrons would all occupy it. Whether the antisymmetric subband is populated in the high-field state is decided entirely by the competition between $\Delta_{\rm SAS}$ and the interaction energy, and this is a many-body question. The Hartree calculation provides the single-particle input: the wave functions $\phi_S$, $\phi_A$ and the splitting $\Delta_{\rm SAS}$.

\section{Exact diagonalization: Models and methods}
\label{sec:method}

All many-body calculations are done in the spherical geometry~\cite{Haldane1983,FanoOrtolaniColombo1986} for $N$ electrons in the lowest Landau level with $2Q$ flux quanta through the sphere, at the shift of the Pfaffian and of the 331 state, which coincide: $2Q=2N-3$. (The benchmark calculation at $\nu=1/3$ in Appendix~\ref{app:third} is at $2Q=3N-3$.) The electrons are fully spin polarized and Landau-level mixing is neglected. Energies are in units of $E_C=e^2/\epsilon\ell$ unless given in kelvin; at $\nu=1/2$, $E_C=50.5\,\sqrt{B[{\rm T}]}$~K, which is 145~K at $n=1.0$, 170~K at $n=1.37$, and 195~K at $n=1.8$. The density of $N$ electrons on a sphere with $2Q=2N-3$ exceeds that of the infinite system at $\nu=1/2$ by the factor $2N/2Q$. We do not apply the customary rescaling of the magnetic length that compensates for this; it would multiply all gaps quoted below by $(2Q/2N)^{1/2}$, which is 0.90 for $N=8$ and 0.95 for $N=16$.

\subsection{One-component models}
\label{sec:onecompmethod}

Two different one-component calculations appear in the literature on wide wells, and they must be kept apart. The first is a strict single-layer model: a layer of finite width with one transverse state and nothing else, as in Ref.~\onlinecite{StorniMorfDasSarma2010}. It is a well-defined model, appropriate when the second subband is far away. The second takes a self-consistent charge profile of a well in which a second subband lies close by, and is occupied at zero field, and nevertheless places all the electrons in one transverse state, as in Ref.~\onlinecite{Sharma2024} and in the exact-diagonalization part of Ref.~\onlinecite{Zhao2021}. This is not a model of the well but a hypothesis about it, namely that at $\nu=1/2$ the second subband is empty; we call it the frozen-subband limit. It must be tested against a calculation that retains both subbands, which is the two-subband model of Sec.~\ref{sec:twosubmethod}. We report both one-component calculations, the first in Appendix~\ref{app:width} and the second in Sec.~\ref{sec:onecomp}, for comparison with the earlier work and because they reach larger sizes; the calculation for the actual well is the two-subband one. The principal calculation of Ref.~\onlinecite{Zhao2021}, a three-dimensional fixed-phase diffusion Monte Carlo at high field, does not fix the transverse wave function, but its phase is that of a one-component trial state.

In either one-component calculation all electrons are in a single transverse state with density $\rho(z)$, and the interaction between two electrons is
\begin{equation}
V(r)=\frac{e^2}{\epsilon}\int\!dz\,dz'\,\frac{\rho(z)\rho(z')}{\sqrt{r^2+(z-z')^2}} .
\label{eq:vr}
\end{equation}
On the sphere $r$ is the chord distance, $V$ is a function of the angle $\theta$ between the two electrons, $V=\sum_kv_kP_k(\cos\theta)$, and the pseudopotential for a pair of total angular momentum $L$ is~\cite{FanoOrtolaniColombo1986}
\begin{align}
V_L={}&(-1)^{2Q+L}(2Q+1)^2\nonumber\\
&\times\sum_{k=0}^{2Q}v_k\begin{Bmatrix}Q&Q&L\\Q&Q&k\end{Bmatrix}
\begin{pmatrix}Q&k&Q\\-Q&0&Q\end{pmatrix}^{2}.
\label{eq:sphericalpp}
\end{align}
For the pure Coulomb interaction Eq.~(\ref{eq:sphericalpp}) reproduces the closed form of Ref.~\onlinecite{FanoOrtolaniColombo1986} to machine precision. We refer to this as the \textit{spherical} convention and use it for all results unless stated otherwise. For $\rho$ we use the density of the lowest (symmetric) subband, $\rho_S=|\phi_S|^2$, which is the consistent choice when all the electrons are in that subband; the total zero-field density $\rho_{\rm tot}=n(z)/n$ gives gaps that differ by 0--9\% (Sec.~\ref{sec:onecomp}).

Much of the literature on finite-thickness effects~\cite{PetersonJolicoeurDasSarma2008,PetersonJolicoeurDasSarma2008b} uses instead the pseudopotentials of the infinite plane,
\begin{align}
V_m&=\int_0^\infty\!dq\,\ell\,F(q)L_m(q^2\ell^2)e^{-q^2\ell^2},\nonumber\\
F(q)&=\int\!dz\,dz'\,\rho(z)\rho(z')e^{-q|z-z'|},
\label{eq:vm}
\end{align}
for pairs of relative angular momentum $m=2Q-L$ on the sphere. We call this the \textit{planar} convention. The two conventions define different finite-size Hamiltonians with the same thermodynamic limit. For the wide well they give gaps that differ by a factor of up to two at the accessible sizes, and we report both, because the difference is a direct measure of the finite-size uncertainty.

The Pfaffian is obtained as the unique zero-energy $L=0$ eigenstate at $2Q=2N-3$ of the three-body Hamiltonian $\sum_{i<j<k}P_{ijk}(3Q-3)$, which projects on the closest approach of three electrons~\cite{GreiterWenWilczek1991}. We report the overlap $O_{\rm MR}=|\langle\Psi_{\rm MR}|\Psi_0\rangle|$ with the ground state $\Psi_0$.

The gap is defined as the difference between the two lowest energies at fixed flux when the ground state has $L=0$, as in Refs.~\onlinecite{Morf1998,StorniMorfDasSarma2010}, and is not defined otherwise. Two remarks are needed on its meaning. First, it is a neutral gap. The activation energy measured in transport is usually identified with that of a separated quasiparticle-quasihole pair, which we have not computed. Second, at this flux composite fermions carrying two flux quanta experience an effective flux $2Q^*=2Q-2(N-1)=-1$ and fill shells of degeneracy 2, 4, 6, 8. The sizes $N=6$ and 12 are closed-shell configurations; they are the sizes known to be aliased, with the $\nu=2/3$ and $\nu=3/5$ states respectively~\cite{MorfdAmbrumenilDasSarma2002}. For the other sizes the partially filled shell contains two particles or two holes in the shell of degeneracy 6 ($N=8$, 10; allowed $L=0$, 2, 4), two particles in the shell of degeneracy 8 ($N=14$; $L=0$, 2, 4, 6), or four ($N=16$; $L=0$, 2, 2, 4, 4, 5, 6, 8). The low-lying states that we find are members of these multiplets in every case (Sec.~\ref{sec:onecomp}). The gap of the one-component problem at these sizes is therefore the splitting between the $L=0$ member of the composite-fermion multiplet and the next one, i.e., a measure of the pairing energy, and a ground state with $L\neq0$ means that the system at that size is not paired. We exclude $N=6$ and 12 from all statements about the size dependence. The one-component calculation is carried to $N=16$, where the $L_z=0$ Hilbert space has dimension $2\,374\,753$.

\subsection{Bilayer model}
\label{sec:bilayermethod}

The second model is the zero-thickness bilayer with tunneling,
\begin{equation}
H_2=\sum_{i<j}\left[\frac{P^{\parallel}_{ij}}{r_{ij}}+\frac{P^{\perp}_{ij}}{\sqrt{r_{ij}^2+d^2}}\right]-\frac{\Delta_{\rm SAS}}{2}\sum_i\tau^x_i,
\label{eq:bilayer}
\end{equation}
in units of $E_C$ and $\ell$, where $P^{\parallel}$ ($P^{\perp}$) projects on pairs in the same (different) layers and $r_{ij}$ is the chord distance; both interactions are treated in the spherical convention. With tunneling the number of electrons in each layer is not conserved. We work in the basis of symmetric ($S$) and antisymmetric ($A$) states, in which the conserved quantity is the parity of the number $N_A$ of electrons in the antisymmetric state, which is the parity of the many-body state under reflection in the midplane, and we diagonalize separately in the even and odd sectors (total dimension 2496 for $N=6$ and $123\,711$ for $N=8$ at $L_z=0$). The 331 state is the unique zero-energy state of the Hamiltonian that penalizes relative angular momentum 1 for pairs in the same layer and 0 for pairs in different layers.

We characterize the ground state $\Psi_0$, which is in the even sector in all cases of interest, by the following quantities. (i) The antisymmetric fraction $f_A=\langle N_A\rangle/N$ and the probability $P_0$ that $N_A=0$. (ii) The overlap $O_{331}=|\langle\Psi_{331}|\Psi_0\rangle|$. (iii) The overlap $O_{\rm MR}=|\langle\Psi_{\rm MR}|\Psi_0\rangle|$ with the Pfaffian formed entirely in the symmetric state. Since $\Psi_{\rm MR}$ has $N_A=0$, $O_{\rm MR}\le\sqrt{P_0}$, and we also give the overlap of the normalized $N_A=0$ component of the ground state with the Pfaffian, $\tilde O_{\rm MR}=O_{\rm MR}/\sqrt{P_0}$, which measures the correlations of the electrons when they are all in the symmetric state, separately from how often that happens. (iv) The overlap with the one-parameter family of triplet-paired states of Ho~\cite{Ho1995}. In the layer basis the 331 state is a Pfaffian of pairs with the pseudospin wave function $|{\uparrow\downarrow}\rangle+|{\downarrow\uparrow}\rangle$, which in the $S$, $A$ basis is $|SS\rangle-|AA\rangle$. Ho's family replaces it by $|SS\rangle-t\,|AA\rangle$:
\begin{equation}
\Psi_{\rm Ho}(t)={\rm Pf}\left[\frac{e_ie_j-t\,o_io_j}{z_i-z_j}\right]\prod_{i<j}(z_i-z_j)^2 ,
\label{eq:hoexplicit}
\end{equation}
where $e_i$ and $o_i$ denote the symmetric and antisymmetric subband (pseudospin) states of electron $i$. This is the pair function of Eq.~(40) of Ref.~\onlinecite{ReadGreen2000} with $t=\tan(\pi/4-\theta)$; $t=1$ ($\theta=0$) is the 331 state. The spin state of the pairs, $\cos(\theta-\pi/4)\,e_ie_j+\sin(\theta-\pi/4)\,o_io_j$, was suggested by Halperin~\cite{Halperin1994}, and Read and Rezayi~\cite{ReadRezayi1996} analyzed Ho's model Hamiltonian, of which $\Psi_{\rm Ho}$ is the exact zero-energy ground state: away from the Pfaffian point its zero-energy quasihole states and torus ground states have the degeneracies of the 331 state, and at the Pfaffian point the Hamiltonian is pathological, with a collapsing gap. Expanding the Pfaffian, the component with $N_A=2m$ carries the factor $(-t)^m$; since the same component of the 331 state carries $(-1)^m$,
\begin{equation}
|\Psi_{\rm Ho}(t)\rangle\propto\sum_{m=0}^{N/2}t^m\,\mathcal{P}_{2m}|\Psi_{331}\rangle ,
\label{eq:ho}
\end{equation}
where $\mathcal{P}_{2m}$ projects on $N_A=2m$. For $t=1$ this is 331 and for $t\to0$ it is the Pfaffian in the symmetric state; we have verified that $\mathcal{P}_0|\Psi_{331}\rangle$ coincides with $|\Psi_{\rm MR}\rangle$ to machine precision. We report $O_{\rm Ho}=\max_t|\langle\Psi_{\rm Ho}(t)|\Psi_0\rangle|$ and the maximizing $t^*$. A large $O_{\rm Ho}$ identifies the state as a triplet-paired state on the path between the two, and $t^*$ locates it on the path. In the analysis of Read and Green~\cite{ReadGreen2000}, consistent with these degeneracies, the states of this family with $t>0$ are all in the Abelian 331 phase; the transition to the Moore-Read phase occurs when the antisymmetric component enters strong pairing, and $o$-$o$ pairs remain in the Moore-Read ground state as tightly bound pairs. The parameter $t^*$ therefore measures the proximity of a finite-size ground state to the two end states, not its topological phase. Read and Rezayi also pointed out that, with tunneling present, Ho's state with its parameter used variationally is a better trial state than 331, and that a large overlap of a ground state with a trial state does not establish its universality class~\cite{ReadRezayi1996}; both remarks apply to the use we make of the family here. The two end states are not orthogonal: $\langle\Psi_{331}|\Psi_{\rm MR}\rangle=0.28$ for $N=6$ and 0.17 for $N=8$.

We give two gaps: $\Delta_e$, the lowest excitation energy in the even sector, and $\Delta_o$, the energy of the lowest state of the odd sector relative to the ground state. Excitations in the odd sector change the reflection parity, i.e., they are intersubband (pseudospin) excitations. The neutral gap is the smaller of the two.

\subsection{Two-subband model of the wide well}
\label{sec:twosubmethod}

This is the model of the actual well. The single-particle basis consists of the LLL orbitals of the symmetric and antisymmetric Hartree subbands with the actual wave functions $\phi_S(z)$ and $\phi_A(z)$. The interaction has four independent channels allowed by the reflection symmetry of the well: three direct ones, in which each electron stays in its subband ($SS$, $AA$, $SA$), and one, denoted $X$, in which both electrons change subband. The corresponding potentials are
\begin{equation}
U_c(r)=\frac{e^2}{\epsilon}\int\!dz\,dz'\,\frac{f_c(z)\,g_c(z')}{\sqrt{r^2+(z-z')^2}},
\label{eq:channels}
\end{equation}
with $(f,g)=(\phi_S^2,\phi_S^2)$, $(\phi_A^2,\phi_A^2)$, $(\phi_S^2,\phi_A^2)$, and $(\phi_S\phi_A,\phi_S\phi_A)$ for $c=SS$, $AA$, $SA$, and $X$, and their pseudopotentials follow from Eq.~(\ref{eq:sphericalpp}). For two thin layers the model reduces to Eq.~(\ref{eq:bilayer}); the calculation reproduces the bilayer results in that limit to four digits [Appendix~\ref{app:numerics}, check (iii)].

\textit{The uniform terms.} The component $k=0$ of each channel in Eq.~(\ref{eq:sphericalpp}), $v_0^c$, is the average of $U_c$ over the sphere. Its contribution to the Hamiltonian is, exactly,
\begin{align}
H_{k=0}=&\ \tfrac12v_0^{SS}N_S(N_S-1)+\tfrac12v_0^{AA}N_A(N_A-1)\nonumber\\
&+v_0^{SA}N_SN_A+\tfrac12v_0^{X}\Big[\Big(\sum_i\tau^z_i\Big)^2-N\Big],
\label{eq:k0}
\end{align}
where $\sum_i\tau^z_i=N_L-N_R$ is the difference of the numbers of electrons in the two halves of the well. The first three terms are the electrostatic (Hartree) energy of a uniform transfer of charge between the subbands and the last is the charging energy of a uniform transfer between the two halves. These are long-wavelength energies, and a sphere whose radius, $\sqrt Q\,\ell=2.1\,\ell$ for $N=6$ and $2.5\,\ell$ for $N=8$, is less than half the separation of the two charge layers ($6.35\,\ell$ at $n=1.37$) does not represent them. For an infinite plane the corresponding quantities are known exactly: per unit area $A$, $v_0^c\to-(2\pi e^2/\epsilon A)\langle|z-z'|\rangle_c$, with $\langle|z-z'|\rangle_c=\int dz\,dz'f_c(z)g_c(z')|z-z'|$, up to a constant common to the three direct channels. At $n=1.37$ the differences that matter on the sphere are the following fractions of those of a plane of the same area: 24\% ($N=6$) and 31\% ($N=8$) for $v_0^{SS}-v_0^{SA}$, less than 10\% in magnitude for $v_0^{SA}-v_0^{AA}$, and 42\% and 48\% for $v_0^X$. The uniform terms of the finite sphere thus contain a fraction of the electrostatic energy that drives charge from the symmetric into the antisymmetric subband.

The straightforward calculation is to diagonalize the Hamiltonian on the sphere as it stands, with the uniform terms of the sphere itself. We call this the plain diagonalization (S) and present its results first. Because its uniform terms are deficient in the way just described, we also give two corrected versions in which they are replaced by those of the infinite system. The three are:

(S) the uniform terms of the sphere itself, $v_0^c$ (plain diagonalization);

(T) those of a plane of the area of the sphere, $v_0^c=-(E_C/2Q)\langle|z-z'|\rangle_c/\ell$;

(T$'$) those of a plane at the physical density of $\nu=1/2$, $v_0^c=-(E_C/2N)\langle|z-z'|\rangle_c/\ell$.

In all three the $k\ge1$ components are those of the sphere. All three have the same thermodynamic limit. Treatments T and T$'$ differ by the factor $2N/2Q$, which is 1.33 for $N=6$ and 1.23 for $N=8$ and is the same density mismatch discussed at the beginning of this section.

\textit{The single-particle term.} The Hartree splitting $\Delta_{\rm SAS}$ contains the mean-field interaction among the electrons, which the many-body Hamiltonian already includes. The single-particle splitting to be used is the bare one,
\begin{align}
\varepsilon_0&=\Delta_{\rm SAS}-\delta_H ,\nonumber\\
\delta_H&=-\frac{E_C}{2}\sum_{b=S,A}\frac{n_b}{n}\,\frac{\langle|z-z'|\rangle_{Ab}-\langle|z-z'|\rangle_{Sb}}{\ell},
\label{eq:eps0}
\end{align}
where $\delta_H$ is the electron-electron Hartree contribution to the splitting at the zero-field populations $n_S$, $n_A$. In the remote-doping model $\varepsilon_0$ is the difference of the kinetic plus confinement energies of the two subband wave functions; in the jellium model it also contains the potential of the positive background. In the remote-doping model at $n=1.37$, $\varepsilon_0=2.45$~meV ($0.168\,E_C$) and $\delta_H=-1.52$~meV: the Hartree energy lowers the antisymmetric subband, whose charge sits farther from the center, by more than the measured splitting itself. The Hamiltonian is
\begin{equation}
H=\varepsilon_0N_A+H_{k=0}+H_{k\ge1}.
\label{eq:twosub}
\end{equation}
In the thermodynamic limit, and with the zero-field populations, $\varepsilon_0$ plus the derivative of the first three terms of Eq.~(\ref{eq:k0}) with respect to $N_A$ reproduces $\Delta_{\rm SAS}$. The subband wave functions are those of the zero-field Hartree potential and are not allowed to relax in the high-field state, and higher subbands are neglected.

The ground state is characterized by the quantities defined in Sec.~\ref{sec:bilayermethod}. The two-subband calculation is limited to $N=6$ and $N=8$; the next size, $N=10$, has an $L_z=0$ space of several million states with a much denser Hamiltonian than in the one-component problem and was beyond our resources. Of the two sizes, $N=6$ is a closed-shell (aliased) size in its one-component limit, where its gap of about 5~K has no significance; we show it because it is the only check on the size dependence, and we mark it wherever it appears.

\section{Results: the bilayer and the two-subband model of the well}
\label{sec:twocomp}

\subsection{Idealized bilayer}
\label{sec:bilayer}

\begin{figure*}[t]
\includegraphics[width=0.92\textwidth]{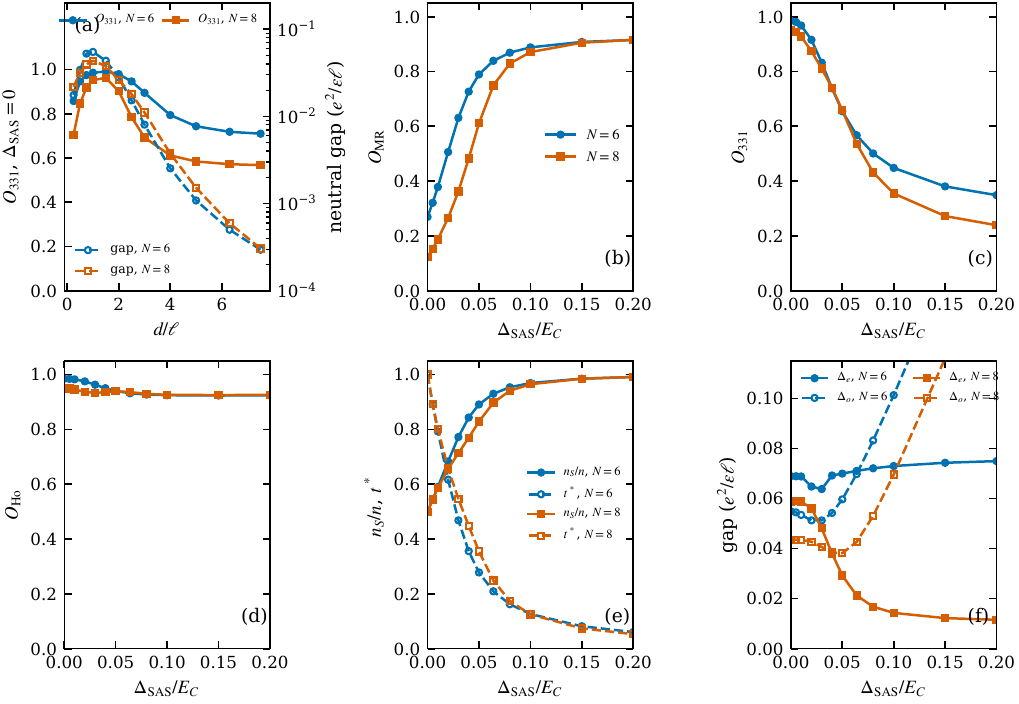}
\caption{Zero-thickness bilayer at total filling $1/2$ on the sphere, $N=6$ (blue; aliased in the one-component limit) and $N=8$ (red). (a) Without tunneling, versus layer separation: overlap with 331 (solid, left axis) and neutral gap (dashed, right axis). (b)--(f) At $d=\ell$, versus tunneling: (b) overlap $O_{\rm MR}$ of the ground state with the Pfaffian in the symmetric state; (c) overlap $O_{331}$ with the 331 state; (d) overlap $O_{\rm Ho}$ with the best Ho state, Eq.~(\ref{eq:ho}); (e) fraction of electrons in the symmetric state (solid) and the parameter $t^*$ of the best Ho state (dashed); (f) lowest excitation energy in the even sector, $\Delta_e$ (solid), and in the odd sector, $\Delta_o$ (dashed).}
\label{fig:bilayer}
\end{figure*}

We begin with the zero-thickness bilayer, Eq.~(\ref{eq:bilayer}), where the questions can be posed cleanly: how does tunneling take the 331 state to the Pfaffian, and what do the diagnostics of Sec.~\ref{sec:bilayermethod} show along the way. The results are in Fig.~\ref{fig:bilayer} and Table~\ref{tab:bilayer}. The size $N=6$ is aliased in its one-component limit and is included for comparison only.

\begin{table*}[t]
\caption{Zero-thickness bilayer at $d=\ell$ with tunneling. $n_S/n$: fraction of electrons in the symmetric state. $O_{331}$, $O_{\rm MR}$: overlaps with 331 and with the Pfaffian in the symmetric state; $\tilde O_{\rm MR}=O_{\rm MR}/\sqrt{P_0}$. $O_{\rm Ho}$, $t^*$: overlap with the best state of the Ho family, Eq.~(\ref{eq:ho}), and its parameter. $\Delta_e$, $\Delta_o$: lowest excitation energies in the even and odd sectors. Energies in units of $E_C$. $N=6$ is aliased in the one-component limit.}
\label{tab:bilayer}
\begin{ruledtabular}
\begin{tabular}{c|cccccc|cccccccc}
\hline\hline
 & \multicolumn{6}{c|}{$N=6$} & \multicolumn{8}{c}{$N=8$}\\
$\Delta_{\rm SAS}$ & $n_S/n$ & $O_{331}$ & $O_{\rm Ho}$ & $t^*$ & $\Delta_e$ & $\Delta_o$ & $n_S/n$ & $O_{331}$ & $O_{\rm MR}$ & $\tilde O_{\rm MR}$ & $O_{\rm Ho}$ & $t^*$ & $\Delta_e$ & $\Delta_o$\\
\hline
0.000 & 0.500 & 0.985 & 0.985 & 1.00 & 0.069 & 0.055 & 0.500 & 0.951 & 0.123 & 0.980 & 0.951 & 1.00 & 0.059 & 0.044\\
0.005 & 0.545 & 0.981 & 0.985 & 0.89 & 0.069 & 0.054 & 0.544 & 0.945 & 0.154 & 0.976 & 0.950 & 0.89 & 0.059 & 0.044\\
0.010 & 0.590 & 0.968 & 0.983 & 0.79 & 0.069 & 0.053 & 0.586 & 0.928 & 0.187 & 0.972 & 0.946 & 0.80 & 0.059 & 0.044\\
0.020 & 0.683 & 0.916 & 0.975 & 0.62 & 0.065 & 0.051 & 0.656 & 0.874 & 0.266 & 0.965 & 0.937 & 0.66 & 0.056 & 0.043\\
0.030 & 0.772 & 0.833 & 0.963 & 0.47 & 0.064 & 0.051 & 0.714 & 0.811 & 0.363 & 0.959 & 0.933 & 0.55 & 0.048 & 0.041\\
0.040 & 0.843 & 0.739 & 0.950 & 0.36 & 0.069 & 0.054 & 0.770 & 0.741 & 0.482 & 0.953 & 0.936 & 0.45 & 0.038 & 0.038\\
0.050 & 0.891 & 0.655 & 0.940 & 0.28 & 0.070 & 0.060 & 0.829 & 0.659 & 0.611 & 0.948 & 0.940 & 0.36 & 0.029 & 0.038\\
0.064 & 0.930 & 0.567 & 0.932 & 0.21 & 0.071 & 0.070 & 0.898 & 0.536 & 0.750 & 0.942 & 0.936 & 0.25 & 0.021 & 0.043\\
0.080 & 0.954 & 0.501 & 0.927 & 0.16 & 0.072 & 0.083 & 0.941 & 0.432 & 0.829 & 0.938 & 0.930 & 0.17 & 0.017 & 0.053\\
0.100 & 0.969 & 0.448 & 0.925 & 0.13 & 0.073 & 0.101 & 0.965 & 0.355 & 0.871 & 0.936 & 0.926 & 0.13 & 0.014 & 0.070\\
0.150 & 0.985 & 0.381 & 0.924 & 0.08 & 0.074 & 0.149 & 0.985 & 0.273 & 0.905 & 0.933 & 0.925 & 0.07 & 0.012 & 0.116\\
0.200 & 0.991 & 0.349 & 0.924 & 0.06 & 0.075 & 0.198 & 0.992 & 0.239 & 0.916 & 0.931 & 0.926 & 0.05 & 0.012 & 0.165\\
\hline\hline
\end{tabular}
\end{ruledtabular}
\end{table*}

\textit{Without tunneling} [Fig.~\ref{fig:bilayer}(a)] the ground state is the 331 state for layer separations of order the magnetic length: the overlap is 0.985 ($N=6$) and 0.951 ($N=8$) at $d=\ell$ and exceeds 0.9 for $0.5\le d/\ell\le2.5$ ($N=6$) and $0.75\le d/\ell\le2$ ($N=8$). The gap is largest near $d=\ell$, 0.055 and 0.044. At large separation the layers decouple into two compressible quarter-filled systems; the gap falls below $10^{-3}$ for $d\ge6\,\ell$, and the overlap with 331 tends to a constant that is not small, 0.72 for $N=6$ and 0.57 for $N=8$ at $d=6.3\,\ell$. An overlap with 331 of this size therefore carries no information for $N\le8$; it is the baseline against which the wide-well values must be judged. The bilayer defined by the two maxima of the Hartree density of the well, for widths of 60--100~nm, is treated in Appendix~\ref{app:modelD}: its layers are 3.5--11 magnetic lengths apart and it is in this decoupled regime throughout.

\textit{With tunneling at $d=\ell$} [Fig.~\ref{fig:bilayer}(b)--(f)] the electrons are transferred continuously into the symmetric state, and four facts stand out.

(i) The ground state follows Ho's path. Its overlap with the best state of the family of Eq.~(\ref{eq:ho}) is at least 0.92 for both sizes at every value of the tunneling, from the 331 state ($t^*=1$) to the Pfaffian in the symmetric state ($t^*\to0$), and $t^*$ decreases smoothly and monotonically. The overlaps with the two end states cross at $\Delta_{\rm SAS}=0.041\,E_C$ for $N=6$ and $0.053\,E_C$ for $N=8$, where about 85\% of the electrons are in the symmetric state and $t^*\approx0.35$. Nothing singular happens there in a system of this size.

(ii) Unequal populations of the symmetric and antisymmetric states do not by themselves indicate a departure from 331. At $\Delta_{\rm SAS}=0.01\,E_C$ the populations are 59:41 and the overlap with 331 is still 0.97 ($N=6$) and 0.93 ($N=8$).

(iii) The overlap with the Pfaffian is not a useful diagnostic once the antisymmetric state is populated. The overlap $O_{\rm MR}$ is bounded by $\sqrt{P_0}$, the amplitude for finding no electron in the antisymmetric state, and equals it to within a few percent: the normalized $N_A=0$ component of the ground state has an overlap $\tilde O_{\rm MR}$ of 0.93--0.98 with the Pfaffian at every value of the tunneling, including zero, where the ground state is 331. This is a property of the 331 state itself, whose $N_A=0$ component is exactly the Pfaffian. A small $O_{\rm MR}$ thus says that the antisymmetric state is populated and nothing more.

(iv) There are two branches of excitations, and they exchange roles. For weak tunneling the lowest excitation is in the odd sector, i.e., it is an intersubband (pseudospin) excitation; its energy is nearly constant, $0.044$--$0.038\,E_C$ for $N=8$ up to $\Delta_{\rm SAS}=0.05\,E_C$, and then rises linearly with the tunneling. The lowest even-sector excitation lies higher for weak tunneling ($0.059\,E_C$) and falls steadily for $\Delta_{\rm SAS}>0.02\,E_C$, to $0.012\,E_C$ at strong tunneling, which is the small gap of the one-component system that remains when all the electrons are in the symmetric state. The two cross at $\Delta_{\rm SAS}\approx0.04\,E_C$ for $N=8$. The neutral gap, the lower of the two, therefore decreases monotonically from the 331 side to the Pfaffian side with a kink at the crossing, and has no maximum for $N=8$ at this layer separation.

These results are consistent with the earlier bilayer studies~\cite{Papic2010,PetersonDasSarma2010}: tunneling converts 331 into a state with a large Pfaffian overlap, and the gap on the Pfaffian side is small.

\subsection{The wide well}
\label{sec:twosub}

We now turn to the two-subband Hamiltonian of the 72.5-nm well, Eq.~(\ref{eq:twosub}), with the Hartree wave functions and bare splitting of the remote-doping model. The results for $N=8$ are in Fig.~\ref{fig:twosub} and Tables~\ref{tab:twosub} and \ref{tab:twosubov}; those for $N=6$ are in Appendix~\ref{app:numerics}, and those for the jellium model in Appendix~\ref{app:jellium}.

\begin{figure*}[t]
\includegraphics[width=0.92\textwidth]{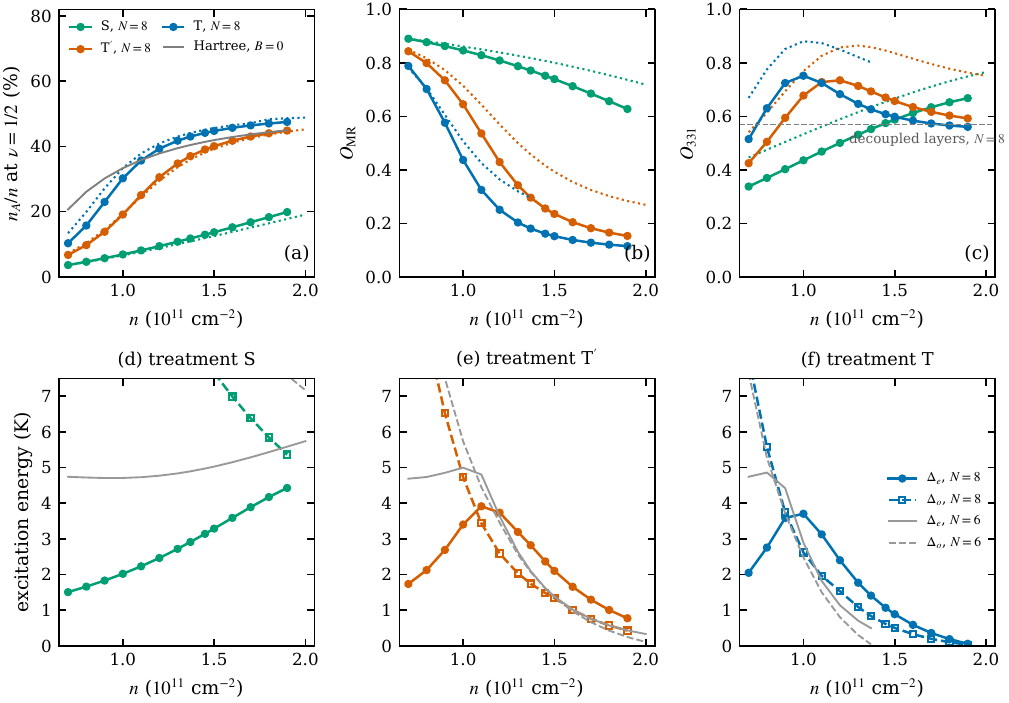}
\caption{Two-subband exact diagonalization for the 72.5-nm well (remote-doping model) versus density, for the three treatments of the uniform interaction terms defined in Sec.~\ref{sec:twosubmethod}: plain diagonalization S (green), T$'$ (red), and T (blue); symbols are $N=8$ and dotted lines $N=6$. (a) Fraction of electrons in the antisymmetric subband in the $\nu=1/2$ ground state; the gray line is the zero-field Hartree value. (b) Overlap $O_{\rm MR}$ of the ground state with the Pfaffian in the symmetric subband. (c) Overlap $O_{331}$ with the 331 state; the dashed line is the value for two decoupled layers at $N=8$. (d)--(f) Lowest excitation energy in the even sector ($\Delta_e$, solid) and in the odd sector ($\Delta_o$, dashed) for $N=8$ (colored) and for $N=6$ (gray; aliased in the one-component limit), for each treatment. Points are shown only where the ground state has $L=0$ and is in the even sector. The overlap with the best Ho state and its parameter $t^*$ are given in Table~\ref{tab:twosub} and Fig.~\ref{fig:twosubfA}.}
\label{fig:twosub}
\end{figure*}

\begin{table*}[t]
\caption{Two-subband exact diagonalization for $w=72.5$~nm, remote-doping model, $N=8$, for the three treatments of the uniform terms (Sec.~\ref{sec:twosubmethod}). $\alpha=\Delta_{\rm SAS}/E_C$ and $n_A/n$ at $B=0$ are the Hartree values. For each treatment: $n_A/n$ in the $\nu=1/2$ ground state; overlap $O_{\rm Ho}$ with the best Ho state and its parameter $t^*$; lowest excitation energies $\Delta_e$ and $\Delta_o$ in the even and odd sectors, in kelvin. The ground state has $L=0$ and is in the even sector in all cases.}
\label{tab:twosub}
\begin{ruledtabular}
\begin{tabular}{ccc|ccccc|ccccc|ccccc}
\hline\hline
 & & $B=0$ & \multicolumn{5}{c|}{Treatment T$'$} & \multicolumn{5}{c|}{Treatment T} & \multicolumn{5}{c}{Treatment S}\\
$n$ & $\alpha$ & $n_A/n$ & $n_A/n$ & $O_{\rm Ho}$ & $t^*$ & $\Delta_e$ & $\Delta_o$ & $n_A/n$ & $O_{\rm Ho}$ & $t^*$ & $\Delta_e$ & $\Delta_o$ & $n_A/n$ & $O_{\rm Ho}$ & $t^*$ & $\Delta_e$ & $\Delta_o$\\
\hline
0.70 & 0.140 & 21\% & 7\% & 0.93 & 0.16 & 1.74 & 10.98 & 10\% & 0.92 & 0.21 & 2.05 & 8.08 & 4\% & 0.93 & 0.11 & 1.51 & 15.95\\
0.80 & 0.122 & 26\% & 10\% & 0.93 & 0.21 & 2.13 & 8.78 & 16\% & 0.93 & 0.30 & 2.76 & 5.58 & 5\% & 0.93 & 0.13 & 1.66 & 14.50\\
0.90 & 0.107 & 30\% & 14\% & 0.93 & 0.27 & 2.69 & 6.52 & 23\% & 0.91 & 0.41 & 3.59 & 3.74 & 6\% & 0.93 & 0.15 & 1.83 & 13.29\\
1.00 & 0.095 & 33\% & 19\% & 0.92 & 0.35 & 3.40 & 4.73 & 30\% & 0.86 & 0.52 & 3.70 & 2.62 & 7\% & 0.94 & 0.16 & 2.02 & 12.29\\
1.10 & 0.085 & 36\% & 25\% & 0.89 & 0.44 & 3.91 & 3.44 & 36\% & 0.79 & 0.61 & 3.13 & 1.97 & 8\% & 0.94 & 0.18 & 2.23 & 11.22\\
1.20 & 0.076 & 38\% & 31\% & 0.84 & 0.52 & 3.74 & 2.59 & 39\% & 0.72 & 0.68 & 2.40 & 1.54 & 9\% & 0.94 & 0.20 & 2.47 & 10.18\\
1.30 & 0.069 & 39\% & 35\% & 0.78 & 0.59 & 3.20 & 2.03 & 42\% & 0.67 & 0.74 & 1.77 & 1.09 & 11\% & 0.94 & 0.22 & 2.72 & 9.25\\
1.37 & 0.064 & 40\% & 37\% & 0.74 & 0.63 & 2.82 & 1.75 & 43\% & 0.64 & 0.77 & 1.41 & 0.84 & 12\% & 0.94 & 0.24 & 2.91 & 8.65\\
1.45 & 0.059 & 41\% & 39\% & 0.71 & 0.67 & 2.37 & 1.48 & 44\% & 0.62 & 0.80 & 1.07 & 0.62 & 13\% & 0.94 & 0.26 & 3.14 & 8.03\\
1.50 & 0.056 & 42\% & 40\% & 0.69 & 0.70 & 2.10 & 1.34 & 45\% & 0.61 & 0.82 & 0.89 & 0.51 & 14\% & 0.93 & 0.27 & 3.29 & 7.66\\
1.60 & 0.051 & 43\% & 42\% & 0.66 & 0.74 & 1.66 & 1.01 & 46\% & 0.59 & 0.85 & 0.59 & 0.33 & 15\% & 0.93 & 0.29 & 3.59 & 6.99\\
1.70 & 0.047 & 44\% & 43\% & 0.63 & 0.77 & 1.30 & 0.76 & 46\% & 0.58 & 0.87 & 0.36 & 0.20 & 17\% & 0.92 & 0.31 & 3.89 & 6.38\\
1.80 & 0.043 & 44\% & 44\% & 0.61 & 0.80 & 1.01 & 0.57 & 47\% & 0.57 & 0.89 & 0.19 & 0.10 & 18\% & 0.92 & 0.33 & 4.17 & 5.84\\
1.90 & 0.040 & 45\% & 45\% & 0.60 & 0.82 & 0.78 & 0.43 & 47\% & 0.56 & 0.91 & 0.06 & 0.03 & 20\% & 0.91 & 0.35 & 4.43 & 5.36\\
\hline\hline
\end{tabular}
\end{ruledtabular}
\end{table*}

\begin{table*}[t]
\caption{Same calculations as Table~\ref{tab:twosub}: probability $P_0$ that no electron is in the antisymmetric subband, overlap $O_{\rm MR}$ with the Pfaffian in the symmetric subband, $\tilde O_{\rm MR}=O_{\rm MR}/\sqrt{P_0}$, and overlap $O_{331}$ with 331.}
\label{tab:twosubov}
\begin{ruledtabular}
\begin{tabular}{c|cccc|cccc|cccc}
\hline\hline
 & \multicolumn{4}{c|}{T$'$} & \multicolumn{4}{c|}{T} & \multicolumn{4}{c}{S}\\
$n$ & $P_0$ & $O_{\rm MR}$ & $\tilde O_{\rm MR}$ & $O_{331}$ & $P_0$ & $O_{\rm MR}$ & $\tilde O_{\rm MR}$ & $O_{331}$ & $P_0$ & $O_{\rm MR}$ & $\tilde O_{\rm MR}$ & $O_{331}$\\
\hline
0.70 & 0.76 & 0.84 & 0.967 & 0.43 & 0.65 & 0.79 & 0.975 & 0.52 & 0.86 & 0.89 & 0.959 & 0.34\\
0.80 & 0.67 & 0.80 & 0.978 & 0.50 & 0.51 & 0.70 & 0.988 & 0.63 & 0.83 & 0.88 & 0.964 & 0.37\\
0.90 & 0.55 & 0.73 & 0.988 & 0.59 & 0.33 & 0.58 & 0.996 & 0.72 & 0.79 & 0.86 & 0.970 & 0.40\\
1.00 & 0.42 & 0.65 & 0.995 & 0.68 & 0.19 & 0.44 & 0.998 & 0.75 & 0.75 & 0.85 & 0.976 & 0.44\\
1.10 & 0.29 & 0.54 & 0.998 & 0.73 & 0.11 & 0.33 & 0.996 & 0.72 & 0.71 & 0.83 & 0.981 & 0.47\\
1.20 & 0.19 & 0.43 & 0.997 & 0.73 & 0.06 & 0.25 & 0.993 & 0.68 & 0.67 & 0.81 & 0.986 & 0.50\\
1.30 & 0.12 & 0.34 & 0.995 & 0.71 & 0.04 & 0.20 & 0.990 & 0.65 & 0.63 & 0.79 & 0.990 & 0.53\\
1.37 & 0.09 & 0.30 & 0.993 & 0.69 & 0.03 & 0.18 & 0.988 & 0.63 & 0.60 & 0.77 & 0.992 & 0.55\\
1.45 & 0.07 & 0.26 & 0.990 & 0.67 & 0.03 & 0.16 & 0.986 & 0.61 & 0.57 & 0.75 & 0.995 & 0.57\\
1.50 & 0.06 & 0.24 & 0.989 & 0.66 & 0.02 & 0.15 & 0.985 & 0.60 & 0.55 & 0.74 & 0.996 & 0.59\\
1.60 & 0.04 & 0.20 & 0.986 & 0.63 & 0.02 & 0.14 & 0.983 & 0.58 & 0.51 & 0.71 & 0.997 & 0.61\\
1.70 & 0.03 & 0.18 & 0.984 & 0.62 & 0.02 & 0.13 & 0.982 & 0.57 & 0.47 & 0.69 & 0.998 & 0.63\\
1.80 & 0.03 & 0.17 & 0.983 & 0.60 & 0.02 & 0.12 & 0.981 & 0.57 & 0.43 & 0.66 & 0.998 & 0.65\\
1.90 & 0.02 & 0.15 & 0.981 & 0.59 & 0.01 & 0.12 & 0.980 & 0.56 & 0.40 & 0.63 & 0.998 & 0.67\\
\hline\hline
\end{tabular}
\end{ruledtabular}
\end{table*}

\textit{Plain diagonalization.} We begin with the Hamiltonian on the sphere as it stands (treatment S; green in Fig.~\ref{fig:twosub}, last blocks of Tables~\ref{tab:twosub} and \ref{tab:twosubov}). For $N=8$ the ground state has $L=0$ at every density. The antisymmetric subband holds 7\% of the electrons at $n=1.0$, 12\% at $n=1.37$, and 18\% at $n=1.8$, against 33--44\% at zero field. The overlap with the Pfaffian in the symmetric subband is 0.85, 0.77, and 0.66 at these three densities and the overlap with 331 is 0.44, 0.55, and 0.65; the two become equal at $n=1.81$. The overlap with the best Ho state is 0.92--0.94 throughout, with $t^*$ rising from 0.16 to 0.33. The lowest excitation is in the even sector, and its energy increases monotonically with density: 2.0~K at $n=1.0$, 2.9~K at $n=1.37$, and 4.2~K at $n=1.8$. The intersubband excitations are at 5.8~K or more. The plain diagonalization for eight electrons thus gives, over the experimental range, a predominantly one-component paired state that is closer to the Pfaffian than to 331, with a growing admixture of the antisymmetric subband, and a gap of 2--4~K that rises with density and has no maximum. Compared with the frozen-subband limit at the same size (Sec.~\ref{sec:onecomp}), the admixture of the second subband raises the gap from 1.4~K to 2.9~K at $n=1.37$.

\textit{The electrostatic deficiency of the sphere.} This result cannot be taken at face value, for the reason given in Sec.~\ref{sec:twosubmethod}: the transfer of charge between the subbands is governed by long-wavelength electrostatic energies, and a sphere of radius $2$--$2.5\,\ell$ contains between a quarter and a half of the electrostatic energy of a charge distribution $6\,\ell$ wide. The deficiency is in the direction of keeping the electrons in the symmetric subband. When the uniform terms of the sphere are replaced by those of the infinite system, which are known exactly, the result changes substantially [Fig.~\ref{fig:twosub}(a)]. With the terms of a plane at the physical density (T$'$) the antisymmetric subband holds 19\%, 37\%, and 44\% of the electrons at $n=1.0$, 1.37, and 1.8, and with those of a plane of the area of the sphere (T) it holds 30\%, 43\%, and 47\%: the population rises steeply between $n=0.8$ and $n=1.3$ and then follows the zero-field Hartree value. The two sizes agree with each other for each treatment (to within 2\% of the total density for S and T$'$ and to within 4\% for T), so the difference between the treatments is not removed by going from $N=6$ to $N=8$.

The corrected treatments are closer to the thermodynamic limit than the plain one in the one respect in which the plain one is known to be deficient. They are not unambiguous: they combine the long-wavelength terms of a plane with the remaining terms of a small sphere, and T and T$'$ differ from each other at the lower densities by as much as T$'$ differs from S. We therefore report the plain result as the result of the diagonalization, the corrected ones as the measure of its finite-size uncertainty, and state our conclusions in a form that does not depend on the choice.

\begin{figure*}[t]
\includegraphics[width=0.92\textwidth]{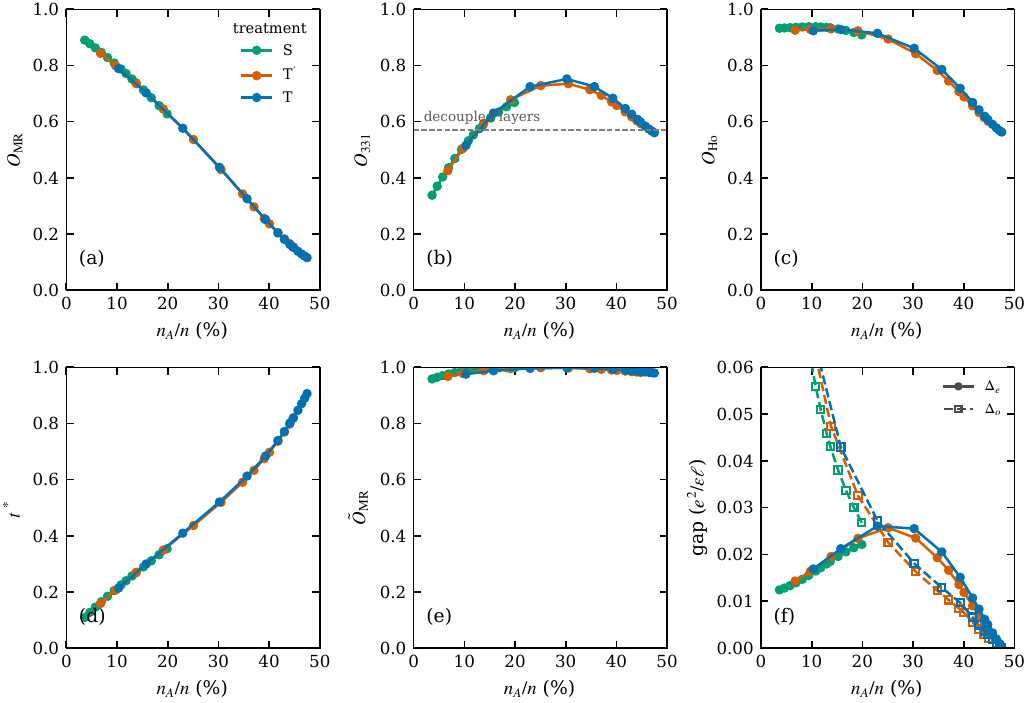}
\caption{The $N=8$ results of Fig.~\ref{fig:twosub} for all densities and all three treatments, plotted against the fraction of electrons in the antisymmetric subband. (a) Overlap $O_{\rm MR}$ with the Pfaffian in the symmetric subband. (b) Overlap $O_{331}$ with the 331 state; the dashed line is the value for two decoupled layers. (c) Overlap $O_{\rm Ho}$ with the best Ho state. (d) Its parameter $t^*$. (e) Overlap $\tilde O_{\rm MR}$ of the normalized $N_A=0$ component of the ground state with the Pfaffian. (f) Lowest excitation energies in the even sector ($\Delta_e$, solid) and in the odd sector ($\Delta_o$, dashed), in units of $E_C$.}
\label{fig:twosubfA}
\end{figure*}

\textit{As functions of the antisymmetric population the three treatments agree.} Figure~\ref{fig:twosubfA} shows the same data plotted against $f_A=n_A/n$ instead of the density. The points of the three treatments, which at a given $f_A$ correspond to densities that differ by as much as a factor of two, fall on common curves. The treatment of the uniform terms, and the density, enter almost entirely through the single number $f_A$. This allows the following statements, which hold for all three treatments.

(i) The ground state has $L=0$ and even parity for $f_A$ up to at least 45\%.

(ii) It is a triplet-paired state on Ho's path between the Pfaffian and 331 as long as the antisymmetric population is not too large. The overlap with the best Ho state is 0.91--0.94 for $f_A\le20\%$ and 0.85 at $f_A=30\%$, and the parameter $t^*$ is a universal increasing function of $f_A$: 0.21 at 10\%, 0.36 at 20\%, 0.51 at 30\%, and 0.70 at 40\%. In the bilayer at $d=\ell$ the overlaps with the two end states cross at $t^*\approx0.35$; in the wide well they cross at $f_A=18\%$ in all three treatments, where $t^*=0.33$. For $f_A>30\%$ the overlap with the Ho family falls, to 0.69 at 40\% and 0.6 at 45\%, and becomes equal to the overlap with 331 itself, which is then within 0.1 of the value for two decoupled layers (0.57). At large antisymmetric population the state is thus no longer well described by either end state or by anything between them at $N=8$; it is approaching two weakly coupled layers.

(iii) The normalized $N_A=0$ component of the ground state has an overlap of 0.96 or more with the Pfaffian at all densities in all treatments (Table~\ref{tab:twosubov}), larger than the overlap of the one-component ground state itself (0.94). As in the bilayer, the decrease of $O_{\rm MR}$ with density measures the population of the antisymmetric subband and nothing else.

(iv) The two branches of excitations behave as in the bilayer. The even-sector gap $\Delta_e$ increases with the antisymmetric population, from the one-component value to a maximum of $0.026\,E_C$ at $f_A\approx25\%$, three times the gap of the frozen-subband limit at the same size ($0.008\,E_C$), and then decreases toward zero as $f_A$ approaches 50\%. The odd-sector gap $\Delta_o$ decreases monotonically. They cross at $f_A=24\%$, where both are $0.025\,E_C$; for larger $f_A$ the lowest excitation is the intersubband one. The neutral gap therefore has a maximum of $0.025\,E_C$ at $f_A=24\%$, with a kink, since the maximum is a level crossing. In kelvin this is 3.6--3.8~K at the densities at which the crossing occurs in treatments T and T$'$, and it would be 4.3~K at $n=1.37$.

The density at which $f_A$ reaches 24\% is what the treatments disagree about: $n=0.91$ ($\alpha=0.106$) for T, $n=1.07$ ($\alpha=0.088$) for T$'$, and beyond the highest density studied, $n>1.9$ ($\alpha<0.04$), for S. At the density of the reported gap maximum, $n=1.37$, the three treatments give $f_A=43\%$, 37\%, and 12\%, neutral gaps of 0.8, 1.8, and 2.9~K, and states that are, respectively, past the crossing and approaching decoupled layers, past the crossing with $t^*=0.63$, and well before it with $t^*=0.24$.

\textit{Size dependence.} The $N=6$ results (Appendix~\ref{app:numerics}) agree with $N=8$ for the populations and for $t^*(f_A)$, and place the crossing of the two excitation branches at the same $f_A$ (23\%). The gaps differ: for $N=6$ the even-sector gap is $0.03$--$0.04\,E_C$ on the one-component side, where this size is aliased, and decreases monotonically with $f_A$ instead of rising to a maximum. The enhancement of the even-sector gap by the antisymmetric admixture, and with it the gap maximum, is therefore established for a single system size. We could not reach $N=10$.

\textit{Benchmark at $\nu=1/3$.} The same calculation at $\nu=1/3$, where the one-subband ground state is the Laughlin state and is not in doubt, is given in Appendix~\ref{app:third}. The three treatments differ there as they do here. In the plain diagonalization the Laughlin state survives over the whole density range, with 4--11\% of the electrons in the antisymmetric subband and a gap that the second subband reduces, by a quarter at $n=1.0$ and by 60--70\% at $n=1.8$. In the corrected treatments the Laughlin state is lost at $n=1.1$--1.4, depending on the treatment and the size, close to the density at which the experiment finds insulating phases below $\nu=1/2$.

\textit{Jellium.} With the jellium subband structure the bare splitting is more than twice as large ($\varepsilon_0=0.39\,E_C$ at $n=1.37$), and the system is one-component at all densities in all treatments: for $N=8$ the antisymmetric fraction is 2--7\% in treatments S and T$'$ and 5--16\% in treatment T, $t^*\le0.3$, and the even-sector gap is 1.6--3.3~K, increasing with density, with no crossing of the excitation branches (Appendix~\ref{app:jellium}).

\subsection{Summary of the two-component results}

The two-subband calculation gives a consistent picture of how the state evolves when the antisymmetric subband is populated: continuously, along Ho's path from the Pfaffian toward 331, with an even-sector gap that at $N=8$ first triples and then collapses, an intersubband gap that decreases throughout, and a maximum of the neutral gap of $0.025\,E_C$ where a quarter of the electrons are in the antisymmetric subband. It does not determine where on this path the experimental sample is at a given density. The plain diagonalization puts it at 12\% antisymmetric population at $n=1.37$, on the Pfaffian side, with a gap of 2.9~K; but that number is decided by electrostatic energies that a system of eight electrons on a sphere contains only in part, and when they are supplied from the infinite system the population becomes 37--43\%, on the 331 side of the gap maximum, with a gap of 0.8--1.8~K.

\section{The frozen-subband limit}
\label{sec:onecomp}

We now ask what the calculation gives if the antisymmetric subband is removed from the Hilbert space, so that all the electrons are in the symmetric subband. For this well that is not the physical problem; Sec.~\ref{sec:twosub} shows that the antisymmetric subband is occupied at $\nu=1/2$. It is the hypothesis that the state is a one-component one, it is the form in which the wide-well problem has been treated in part of the literature~\cite{Zhao2021,Sharma2024}, and it can be carried to twice the system size. Its value here is as a reference: it shows what a one-component Pfaffian in this well can and cannot do.

\begin{figure*}[t]
\includegraphics[width=0.92\textwidth]{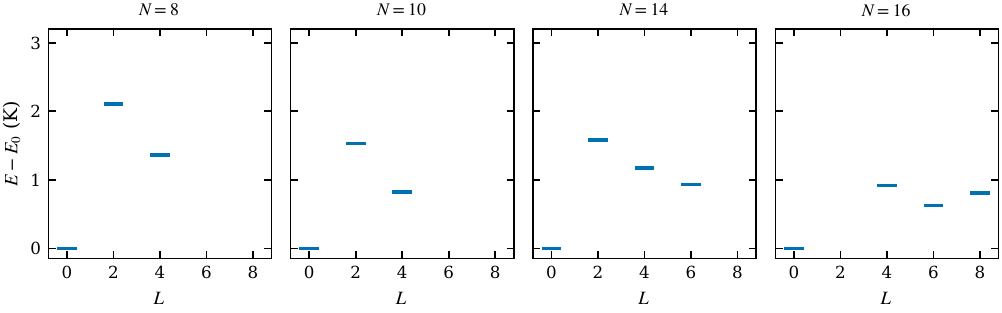}
\caption{Low-lying spectra in the frozen-subband limit at $w=72.5$~nm, $n=1.37$ (remote-doping profile, $\rho_S$, spherical convention) versus total angular momentum $L$, for the four unaliased sizes. The states shown are the lowest five ($N=8$, 10, 14) or four ($N=16$) found; they are members of the composite-fermion multiplets listed in Sec.~\ref{sec:onecompmethod}.}
\label{fig:onecompspectra}
\end{figure*}

\begin{table*}[t]
\caption{Frozen-subband limit at $w=72.5$~nm, $n=1.37$ ($E_C=170$~K). $D$: dimension of the $L_z=0$ space. $L_1$: angular momentum of the lowest excited state (remote doping, spherical convention). Spherical convention, remote-doping profile with $\rho_S$: gap $\Delta$ in units of $E_C$ and in kelvin and overlap $O_{\rm MR}$ with the Pfaffian; gap with $\rho_{\rm tot}$; gap for the jellium profile with $\rho_S$. Planar convention, remote-doping profile with $\rho_{\rm tot}$: gap and overlap. The ground state has $L=0$ in every case. Sizes marked with an asterisk are aliased. The Pfaffian overlap was not computed for $N=16$.}
\label{tab:onecomp}
\begin{ruledtabular}
\begin{tabular}{ccc|ccc|c|c|cc}
\hline\hline
 & & & \multicolumn{3}{c|}{Spherical, remote, $\rho_S$} & $\rho_{\rm tot}$ & Jellium & \multicolumn{2}{c}{Planar, remote, $\rho_{\rm tot}$}\\
$N$ & $D$ & $L_1$ & $\Delta$ ($E_C$) & $\Delta$ (K) & $O_{\rm MR}$ & $\Delta$ (K) & $\Delta$ (K) & $\Delta$ (K) & $O_{\rm MR}$\\
\hline
6$^*$ & 18 & 3 & 0.0319 & 5.43 & 0.929 & 5.43 & 6.51 & 6.12 & 0.867\\
8 & 151 & 4 & 0.0080 & 1.36 & 0.942 & 1.30 & 1.59 & 2.57 & 0.878\\
10 & 1\,514 & 4 & 0.0048 & 0.82 & 0.921 & 0.75 & 0.95 & 1.57 & 0.879\\
12$^*$ & 16\,660 & 5 & 0.0218 & 3.71 & 0.693 & 3.70 & 4.42 & 3.86 & 0.646\\
14 & 194\,668 & 6 & 0.0055 & 0.93 & 0.794 & 0.87 & 1.06 & 1.48 & 0.727\\
16 & 2\,374\,753 & 6 & 0.0037 & 0.63 & -- & -- & 0.70 & 1.04 & --\\
\hline\hline
\end{tabular}
\end{ruledtabular}
\end{table*}

\begin{figure*}[t]
\includegraphics[width=0.92\textwidth]{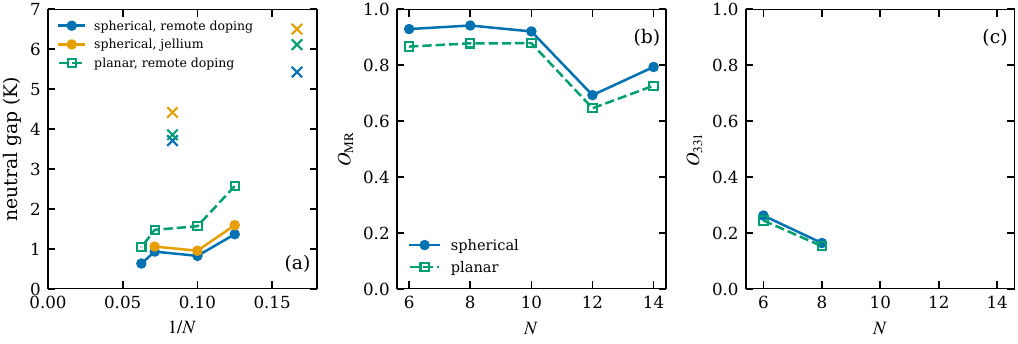}
\caption{System-size dependence in the frozen-subband limit at $w=72.5$~nm, $n=1.37$. (a) Neutral gap versus $1/N$ in the spherical convention for the remote-doping (blue) and jellium (orange) profiles and in the planar convention for the remote-doping profile (green). Crosses are the aliased sizes $N=6$ and 12. (b) Overlap $O_{\rm MR}$ with the Pfaffian. (c) Overlap $O_{331}$ with the 331 state, which in this limit equals $|\langle\Psi_{331}|\Psi_{\rm MR}\rangle|\,O_{\rm MR}$ and is shown for the sizes for which the 331 state was constructed ($N=6$ and 8).}
\label{fig:onecompN}
\end{figure*}

Figure~\ref{fig:onecompspectra}, Table~\ref{tab:onecomp}, and Fig.~\ref{fig:onecompN} show the results at the density of the reported gap maximum.

(i) The ground state has $L=0$ for every system size, in both pseudopotential conventions and for both electrostatic models.

(ii) Its overlap with the Pfaffian is large: 0.94, 0.92, and 0.79 for $N=8$, 10, and 14 in the spherical convention, and 0.88, 0.88, and 0.73 in the planar convention. The Hilbert-space dimension for $N=14$ is $1.9\times10^5$. The overlap decreases with size, and we have no value for $N=16$. These overlaps do not discriminate between the Pfaffian and 331, because in this limit there is nothing to discriminate: a state with all electrons in the symmetric subband overlaps with 331 only through the component of 331 that has no electron in the antisymmetric subband, which is the Pfaffian, so that $O_{331}=|\langle\Psi_{331}|\Psi_{\rm MR}\rangle|\,O_{\rm MR}$ exactly. The factor is 0.28 for $N=6$ and 0.17 for $N=8$, giving $O_{331}=0.26$ and 0.16 [Fig.~\ref{fig:onecompN}(c)]. The frozen-subband state is Pfaffian-like by construction; whether the actual state is closer to the Pfaffian or to 331 can be asked only of the two-subband calculation (Figs.~\ref{fig:twosub} and \ref{fig:twosubfA}).

(iii) The gap is small and decreases with system size. In the spherical convention it is $0.0080\,E_C$ (1.36~K) for $N=8$, 0.0048 (0.82~K) for $N=10$, 0.0055 (0.93~K) for $N=14$, and 0.0037 (0.63~K) for $N=16$. In the planar convention it is 2.57, 1.57, 1.48, and 1.04~K. The two conventions differ by a factor of 1.9--2.0 at $N=8$ and 10 and of 1.6--1.8 at $N=14$ and 16 (comparing like with like, the planar values with $\rho_S$ are 2.66, 1.65, 1.54, and 1.11~K), and both have fallen by a factor of 2.2--2.5 between $N=8$ and $N=16$. The dependence on $N$ is not monotonic, and the angular momentum of the lowest excitation changes with $N$ ($L_1=4$, 4, 6, 6), so that an extrapolation is not meaningful. The gap at the largest size we can reach is 0.6--1.0~K and has not converged.

These gaps are the splittings of the composite-fermion multiplets described in Sec.~\ref{sec:onecompmethod}: the lowest states have $L=0$, 4, 2 for $N=8$ and 10, $L=0$, 6, 4, 2 for $N=14$, and $L=0$, 6, 8, 4 for $N=16$ (Fig.~\ref{fig:onecompspectra}). The $L=0$ member is lowest, which is the signature of pairing, but by an energy that is a small fraction of the scale $0.02$--$0.03\,E_C$ set by the aliased closed-shell sizes $N=6$ and 12. The gap of the aliased size $N=6$ is 5.4--6.1~K, close to the gap reported in Ref.~\onlinecite{Singh2025}; this is a coincidence without significance.

(iv) The result is insensitive to everything except the system size. The remote-doping and jellium profiles differ by a factor of 2.4 in $\Delta_{\rm SAS}$, yet the gaps differ by 11--17\% and the overlaps by less than 0.001. Using the total density $\rho_{\rm tot}$ in place of $\rho_S$ lowers the gap by 5--9\% for the unaliased sizes. Over the experimental range $n=1.0$--1.8 the gap in kelvin increases by 7--10\% for $N=8$, 10, and 14, with no maximum, and over $w=40$--100~nm and $n=0.7$--2.3 the $N=10$ gap stays between 0.5 and 1.25~K (Appendix~\ref{app:width}). For $N=16$ the gap is 0.58, 0.63, and 0.65~K at $n=1.0$, 1.37, and 1.8. The same is true of a strict single-layer model with no self-consistency at all, which is the calculation of Ref.~\onlinecite{StorniMorfDasSarma2010} and is also given in Appendix~\ref{app:width}. That reference found small gaps that decrease nonmonotonically with size and concluded that the gap is unlikely to survive in the thermodynamic limit; our results agree.

The frozen-subband limit therefore gives a definite answer: a Pfaffian-like paired state, with a gap that is below 1~K at the largest size studied and still decreasing, nearly independent of density, and with no maximum. A one-component Pfaffian in this well has neither a gap of several kelvin nor a gap maximum as a function of density. Both require the second subband.

\section{The activation gap: Analysis of the published transport data}
\label{sec:transport}

\begin{figure*}[t]
\includegraphics[width=0.92\textwidth]{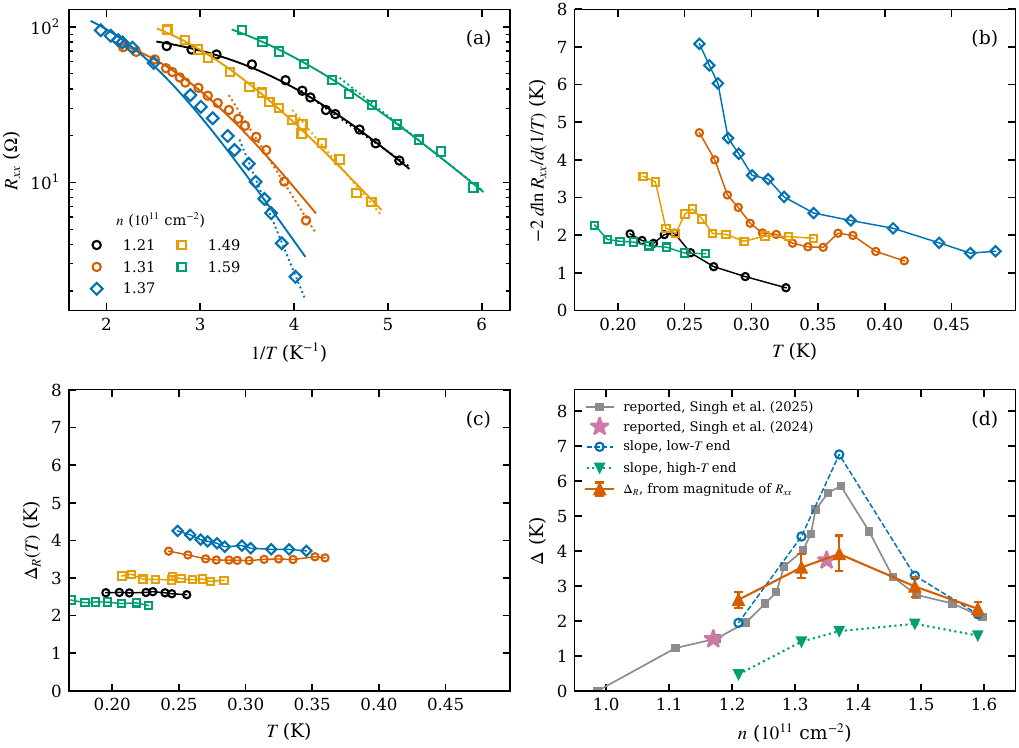}
\caption{Analysis of the Arrhenius data of Ref.~\onlinecite{Singh2025} (their Fig.~S4(b)). (a) Digitized $R_{xx}$ at the $\nu=1/2$ minimum versus $1/T$ for five densities. Dotted lines: straight lines through the five lowest-temperature points. Solid lines: Eq.~(\ref{eq:sat}) with the prefactor fixed at $R_0=12.9$~k$\Omega$ and $\Delta$ and $R_N$ fitted. (b) Local slope $-2\,d\ln R_{xx}/d(1/T)$ over four consecutive points; for activation across a single gap this would be a constant. (c) Gap $\Delta_R(T)$ obtained from the magnitude of $R_{xx}$ at each temperature, Eq.~(\ref{eq:deltaR}), with $R_0=12.9$~k$\Omega$ and $R_N=95~\Omega$. (d) Gap versus density: values reported in Ref.~\onlinecite{Singh2025} (gray squares) and in Ref.~\onlinecite{Singh2024} (stars); slope of the low-temperature end of each trace (open circles) and of its high-temperature end (green); $\Delta_R$ (red triangles; the bars cover the variation with temperature and with $R_0$ and $R_N$ over the ranges given in the text).}
\label{fig:transport}
\end{figure*}

The gaps of Ref.~\onlinecite{Singh2025} are obtained from Arrhenius fits to $R_{xx}$ at the $\nu=1/2$ minimum, shown for five densities in Fig.~S4(b) of that paper. We have digitized those data by automated matching of the plot symbols and analyzed them in the same convention, $R_{xx}\propto\exp(-\Delta/2k_BT)$. The digitized traces contain 11 to 17 points each. The accuracy is about $\pm0.02$~K$^{-1}$ in $1/T$ and $\pm3\%$ in $R_{xx}$ for isolated symbols. It is worse where symbols overlap: four spurious matches were removed by hand from the $n=1.37$ trace, some points of that trace between $1/T=2.5$ and 2.9~K$^{-1}$ are hidden behind other symbols and are missing, and the $n=1.49$ trace contains two points at the same temperature that differ by 15\%. The results are in Fig.~\ref{fig:transport} and Table~\ref{tab:transport}. They are an analysis of a published figure and not a substitute for an analysis of the original data.

\begin{table*}[t]
\caption{Analysis of the digitized Arrhenius data of Ref.~\onlinecite{Singh2025}. $N_p$: number of digitized points. $T$, $R_{xx}$: ranges covered. $\Delta_{\rm rep}$: gap reported in Ref.~\onlinecite{Singh2025}, read from its Fig.~S4(c). $\Delta_{\rm HT}$: slope of the high-temperature end of the trace (the first 0.6~K$^{-1}$ in $1/T$). $\Delta_{\rm LT}$, $R_0^{\rm LT}$: slope and prefactor of the low-temperature end (the five coldest points). Free fit: Eq.~(\ref{eq:sat}) with $\Delta$, $R_0$, and $R_N$ all free. Fixed prefactor: Eq.~(\ref{eq:sat}) with $R_0=12.9$~k$\Omega$, with the rms deviation in $\ln R_{xx}$. $\Delta_R$: gap from the magnitude of $R_{xx}$, Eq.~(\ref{eq:deltaR}), mean over the trace and, in parentheses, full range (see text). Gaps in K, densities in $10^{11}$~cm$^{-2}$.}
\label{tab:transport}
\begin{ruledtabular}
\begin{tabular}{cccc|c|c|cc|cc|ccc|c}
\hline\hline
 & & & & & High $T$ & \multicolumn{2}{c|}{Low $T$} & \multicolumn{2}{c|}{Free fit} & \multicolumn{3}{c|}{$R_0=12.9$~k$\Omega$} & \\
$n$ & $N_p$ & $T$ (K) & $R_{xx}$ ($\Omega$) & $\Delta_{\rm rep}$ & $\Delta_{\rm HT}$ & $\Delta_{\rm LT}$ & $R_0^{\rm LT}$ (k$\Omega$) & $\Delta$ & $R_0$ (k$\Omega$) & $\Delta$ & $R_N$ ($\Omega$) & rms & $\Delta_R$\\
\hline
1.21 & 12 & 0.20--0.38 & 13.8--75 & 2.0 & 0.5 & 1.9 & 2.0 & $2.8\pm0.2$ & 19 & 2.6 & $97\pm3$ & 0.03 & 2.6 (2.4--2.8)\\
1.31 & 17 & 0.24--0.46 & 5.7--74 & 4.0 & 1.4 & 4.4 & 53 & $4.6\pm0.7$ & 86 & 3.6 & $108\pm8$ & 0.09 & 3.5 (3.2--3.9)\\
1.37 & 17 & 0.25--0.52 & 2.5--95 & 5.9 & 1.7 & 6.8 & 1900 & $5.7\pm0.5$ & 261 & 4.0 & $163\pm14$ & 0.18 & 3.9 (3.4--4.4)\\
1.49 & 16 & 0.21--0.38 & 7.5--96 & 2.7 & 1.9 & 3.3 & 21 & $3.1\pm0.2$ & 14 & 3.1 & $154\pm10$ & 0.06 & 3.0 (2.7--3.3)\\
1.59 & 11 & 0.17--0.29 & 9.3--95 & 2.1 & 1.6 & 2.2 & 6.4 & $2.3\pm0.3$ & 9.7 & 2.4 & $165\pm11$ & 0.04 & 2.3 (2.1--2.5)\\
\hline\hline
\end{tabular}
\end{ruledtabular}
\end{table*}

\subsection{A single activation gap does not span the measured range}

Figure~\ref{fig:transport}(b) shows the local slope of the Arrhenius plot, $-2\,d\ln R_{xx}/d(1/T)$, which for activation across a single gap would be a constant equal to the gap. At $n=1.37$ it increases on cooling from 1.6~K at 0.48~K through 2.4~K at 0.37~K and 4.2~K at 0.29~K to between 5 and 7~K at 0.25--0.26~K, and there is no temperature interval over which it is constant. A straight line through the whole trace gives 3.2~K with an rms deviation of 0.26 in $\ln R_{xx}$, nine times the digitization error. The other traces are curved in the same sense, less strongly.

The two ends of each trace therefore give very different apparent gaps, with different density dependences. The slope of the high-temperature end is $\Delta_{\rm HT}=0.5$, 1.4, 1.7, 1.9, and 1.6~K for $n=1.21$, 1.31, 1.37, 1.49, and 1.59: between 0.5 and 2~K, with no maximum at $n=1.37$. The slope of the low-temperature end is $\Delta_{\rm LT}=1.9$, 4.4, 6.8, 3.3, and 2.2~K, with a pronounced maximum at $n=1.37$. The reported gaps, 2.0, 4.0, 5.9, 2.7, and 2.1~K, are close to the low-temperature slopes; at $n=1.37$ a line through the eight coldest points ($T\le0.30$~K) gives 5.7~K. The reported gaps, and the reported sharp maximum, are the slopes of the low-temperature ends of these traces.

Neither slope is the gap, for the following reasons.

\subsection{The high-temperature slope: saturation}

Each trace starts at a resistance of 75--95~$\Omega$, which is the resistance on either side of the minimum in Fig.~S4(a) of Ref.~\onlinecite{Singh2025} at 0.4~K, where the minimum has all but disappeared. The resistance at the minimum cannot rise above that of the surrounding compressible state, $R_N$. The simplest form that contains this is
\begin{equation}
\frac{1}{R_{xx}(T)}=\frac{1}{R_N}+\frac{1}{R_0\exp(-\Delta/2k_BT)} ,
\label{eq:sat}
\end{equation}
whose local slope, $\Delta\,R_N/(R_N+R_0e^{-\Delta/2k_BT})$, goes to zero at high temperature whatever the gap. Over the high-temperature end of the traces $R_{xx}$ is 50--95~$\Omega$, within a factor of two of $R_N$, and the small values of $\Delta_{\rm HT}$ reflect how close each trace is to saturation where it starts. They are lower bounds on the gap. A gap of 1.3~K, for example, would make $R_{xx}$ fall by a factor of 3.7 between 0.5 and 0.25~K; at $n=1.37$ it falls by a factor of 38.

\subsection{The low-temperature slope: the prefactor, and the disappearance of the minimum}

The slope of an Arrhenius plot determines the gap only together with a prefactor, and the prefactor is not free. For activated transport of quasiparticles of charge $e^*$ across a mobility gap in a smooth disorder potential, $\sigma_{xx}=\sigma_0\exp(-\Delta/2k_BT)$ with $\sigma_0=2(e^*)^2/h$~\cite{PolyakovShklovskii1995}, reduced to 0.8--1.25~$(e^*)^2/h$ when tunneling through the saddle points is included~\cite{dAmbrumenilHalperinMorf2011}; the prefactors measured for other FQHE states are of this size or smaller~\cite{dAmbrumenilHalperinMorf2011}. With $\rho_{xx}=\sigma_{xx}\rho_{xy}^2$, $\rho_{xy}=2h/e^2$, and $e^*=e/4$, the resistivity prefactor is $\rho_0=5$--13~k$\Omega$, the upper value, $h/2e^2=12.9$~k$\Omega$, corresponding to $\sigma_0=2(e^*)^2/h$. The measured $R_{xx}$ is related to $\rho_{xx}$ by a geometrical factor that depends on the sample and contacts, which Ref.~\onlinecite{Singh2025} does not specify and which cannot exceed a number of order unity (it is $\ln2/\pi=0.22$ for a van der Pauw geometry).

The straight lines through the low-temperature ends of the traces have prefactors $R_0^{\rm LT}=2.0$, 53, 1900, 21, and 6.4~k$\Omega$ (Table~\ref{tab:transport}). At $n=1.37$ the prefactor is 150 times the largest value that activated transport allows; a line with the reported slope of 5.85~K through the coldest point needs 320~k$\Omega$, 25 times that value. Fits of Eq.~(\ref{eq:sat}) with all three parameters free, which remove the effect of the saturation, tell the same story: at $n=1.21$, 1.49, and 1.59 they give $R_0=19$, 14, and 10~k$\Omega$, which is the expected value, and gaps of $2.8\pm0.2$, $3.1\pm0.2$, and $2.3\pm0.3$~K; at $n=1.31$ and 1.37 they give 86 and 260~k$\Omega$, with gaps of $4.6\pm0.7$ and $5.7\pm0.5$~K (jackknife uncertainties). The large gaps reported at the two densities nearest the maximum exist only in combination with prefactors that are one to two orders of magnitude too large.

The physical content of this statement is simple, and can be checked without any fit. A gap of 5.85~K with the largest allowed prefactor gives $R_{xx}=12.9~{\rm k}\Omega\times e^{-5.85/0.8}=9~\Omega$ at 0.40~K. The measured value at $n=1.37$ and 0.40~K is 59~$\Omega$, and Fig.~S4(a) of Ref.~\onlinecite{Singh2025} shows that by 0.4~K the minimum has nearly disappeared. A state with a gap of 6~K does not lose its resistance minimum at a temperature fifteen times smaller than the gap; the $\nu=1/2$ minimum in this sample disappears at 0.4--0.5~K.

\subsection{The gap from the magnitude of the resistance}

With the prefactor known, the gap follows from the magnitude of $R_{xx}$ at each temperature, with no reference to the slope:
\begin{equation}
\Delta_R(T)=2k_BT\,\ln\frac{R_0}{R_{\rm act}(T)},\qquad \frac{1}{R_{\rm act}}=\frac{1}{R_{xx}}-\frac{1}{R_N}.
\label{eq:deltaR}
\end{equation}
If the transport is activated across a temperature-independent gap, $\Delta_R(T)$ is a constant; if the gap depends on temperature, $\Delta_R(T)$ is the gap at temperature $T$, whereas the Arrhenius slope is $\Delta-T\,d\Delta/dT$ and overstates a gap that decreases with temperature.

Figure~\ref{fig:transport}(c) shows $\Delta_R(T)$ for $R_0=12.9$~k$\Omega$ and $R_N=95~\Omega$, for the points with $R_{xx}\le R_N/2$. Two things are apparent. First, at $n=1.21$, 1.49, and 1.59 it is constant to within 0.1--0.2~K over the whole trace: 2.6, 3.0, and 2.3~K. At these densities the data are activated transport across a fixed gap with the theoretical prefactor, and the fit of Eq.~(\ref{eq:sat}) with $R_0$ fixed at 12.9~k$\Omega$ is as good as the fit with $R_0$ free (Table~\ref{tab:transport}). Second, with the same prefactor the gap is 3.5--3.7~K at $n=1.31$ and 3.7--4.3~K at $n=1.37$, increasing on cooling at $n=1.37$ from 3.7~K at 0.35~K to 4.3~K at 0.25~K. At these two densities the fit with fixed prefactor is poorer (rms 0.09 and 0.18), which is the same statement as the anomalous prefactor of the free fit: the resistance falls on cooling faster than a fixed gap allows, i.e., the gap grows on cooling.

The values of $\Delta_R$ depend logarithmically on $R_0$ and weakly on $R_N$. Varying $R_0$ between 9.7 and 18.7~k$\Omega$, the range of the free fits at the three densities where the transport is simply activated (the upper end is 1.4 times the theoretical maximum, within what the unknown sample geometry permits), and $R_N$ between 80 and 110~$\Omega$, gives the ranges in the last column of Table~\ref{tab:transport}: 2.4--2.8, 3.2--3.9, 3.4--4.4, 2.7--3.3, and 2.1--2.5~K. The fits with $R_0$ fixed return a larger $R_N$, 154--165~$\Omega$, at $n=1.37$, 1.49, and 1.59 (Table~\ref{tab:transport}); using it in place of 95~$\Omega$ raises $\Delta_R$ by less than 0.2~K at any temperature and by 0.07~K on average, within these ranges. A smaller prefactor lowers all the gaps: for $R_0=5.2$~k$\Omega$, the lower end of the theoretical range, the fits of Eq.~(\ref{eq:sat}) give 2.2, 3.1, 3.5, 2.7, and 2.1~K, with poorer fits at every density.

\subsection{Conclusions from the transport data}

(a) The gap of the $\nu=1/2$ state has a maximum as a function of density near $n=1.37$, as reported.

(b) The gap at the maximum is $3.9\pm0.5$~K at the temperatures of the measurement, not 6~K. The reported value is the low-temperature slope of a curved Arrhenius plot; it requires a prefactor 25 times larger than activated transport allows and is contradicted by the disappearance of the resistance minimum at 0.4--0.5~K.

(c) Away from the maximum the gap is 2.3--3.0~K, somewhat larger than the reported 2.0--2.7~K. The maximum is therefore lower and broader than reported: the ratio of the gap at $n=1.37$ to the mean of those at $n=1.21$ and 1.59 is 1.6, against 2.9 for the reported values.

(d) The gap is not about 1~K either. The slopes of the high-temperature ends of the traces, 0.5--1.9~K, are suppressed by the saturation of $R_{xx}$.

(e) The value at the maximum agrees with the earlier measurement of the same group for a 72.5-nm well, 3.73~K at $n=1.35$~\cite{Singh2024} [Fig.~\ref{fig:transport}(d)].

(f) At the two densities nearest the maximum, and only there, the gap increases on cooling in the measured range, by about 0.5~K between 0.35 and 0.25~K at $n=1.37$. Its zero-temperature value is not determined by these data and is larger than 4.3~K. A gap that is still developing at a temperature of a tenth of its own magnitude is unusual, and it distinguishes the state at the gap maximum from the states on either side.

(g) The analysis rests on the theoretical prefactor of activated transport, which is confirmed in this sample by the three densities at which the transport is simply activated. It uses a digitized figure. An analysis of the original data along these lines is needed.

\section{Discussion}
\label{sec:discussion}

\subsection{What is established and what is not}

\begin{figure*}[t]
\includegraphics[width=0.92\textwidth]{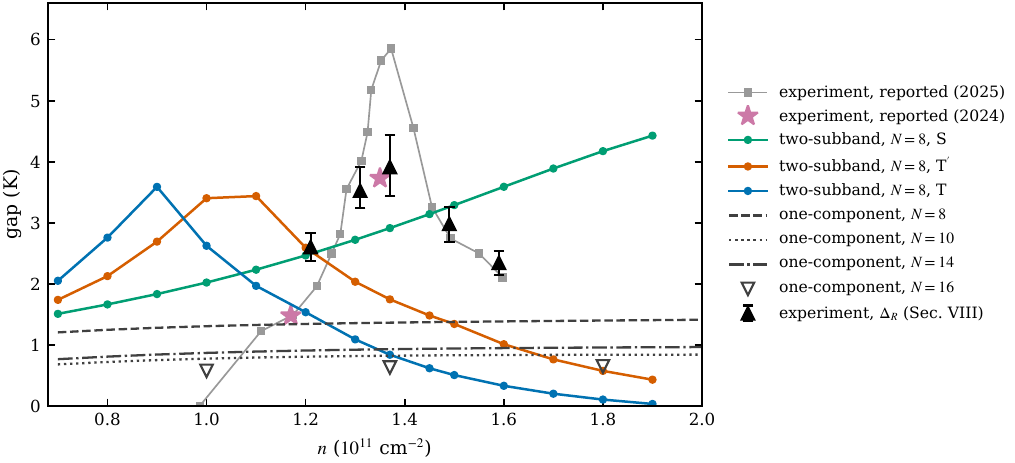}
\caption{Gap of the $\nu=1/2$ state in the 72.5-nm well versus density: experiment and theory. Experiment: activation gaps reported in Ref.~\onlinecite{Singh2025} (gray squares) and in Ref.~\onlinecite{Singh2024} (stars); the gap $\Delta_R$ obtained in Sec.~\ref{sec:transport} from the magnitude of $R_{xx}$ in the data of Ref.~\onlinecite{Singh2025} (black triangles, Table~\ref{tab:transport}). Theory, neutral gaps of the ideal system in the spherical convention: two-subband model for $N=8$, plain diagonalization (S) and with the uniform terms of the infinite system (T$'$ and T), the smaller of $\Delta_e$ and $\Delta_o$ of Table~\ref{tab:twosub}; frozen-subband limit for $N=8$, 10, 14 (black lines) and $N=16$ (open triangles).}
\label{fig:summary}
\end{figure*}

The calculations of this paper establish the following about the $\nu=1/2$ state of the 72.5-nm well, summarized in Fig.~\ref{fig:summary}.

\textit{Established.} (1) The single-particle structure: with the electrons alone determining the potential in the well, the calculated $\alpha$ is consistent with the values quoted in the experiment, the charge distribution is strongly bilayer-like, and the second subband is $0.064\,E_C$ above the first at $n=1.37$. (2) The two-subband Hamiltonian for eight electrons on the sphere has a uniform paired ground state at all densities, whose character changes continuously with the population of the antisymmetric subband along Ho's path from the Pfaffian toward 331, with the gap structure described in Sec.~\ref{sec:twosub}; as functions of that population these properties do not depend on how the uniform interaction terms are treated. (3) If the antisymmetric subband is frozen out, the state is Pfaffian-like for every density, for widths $w\ge5\,\ell$, and for either electrostatic model, with a gap that is below 1~K for $N=16$ and still decreasing. (4) The transport data show a gap maximum near $n=1.37$, of $3.9\pm0.5$~K and not 6~K, with 3.5~K at $n=1.31$ and 2.3--3.0~K at $n=1.21$, 1.49, and 1.59.

\textit{Not established.} (1) The antisymmetric population at $\nu=1/2$ as a function of density, and hence where on the path a given density lies. (2) The size dependence of the two-subband results. (3) The topological order of the state at any density.

\subsection{The magnitude of the gap}

The gap at the maximum is $3.9\pm0.5$~K (Sec.~\ref{sec:transport}), i.e., $0.023\pm0.003\,E_C$. For comparison, in the same units: the gap of the ideal $\nu=5/2$ state is 0.025~\cite{Morf1998,MorfdAmbrumenilDasSarma2002}; the gap of the frozen-subband limit in this well is 0.004--0.006 for $N=16$; the largest neutral gap of the two-subband model for $N=8$ is 0.025; the gap of the ideal bilayer 331 state at $d=\ell$ is 0.044 for $N=8$.

Three conclusions follow. First, a one-component Pfaffian state in this well does not have a gap of the observed size. At equal system size the gap of the frozen-subband limit is a factor of three below the two-subband maximum, and for $N=16$ it is a factor of four to six below the measured gap, before disorder and Landau-level mixing are considered, and it is still decreasing with size. This is a firm statement, and it does not depend on any of the uncertain elements of the two-subband calculation. The one-component model also has no maximum of the gap as a function of density.

Second, the two-subband model contains what the one-component model lacks. Its neutral gap for $N=8$ has a maximum as a function of the antisymmetric population, where the even-sector and intersubband excitations cross, and the value at the maximum, $0.025\,E_C$, or 4.3~K at $n=1.37$, equals the measured maximum within the uncertainties. If the antisymmetric population in the sample is about a quarter at $n=1.37$, the calculation accounts for the existence, the position, and the height of the observed maximum. Away from the maximum the calculated neutral gap falls to half its peak value when the antisymmetric population increases by ten percentage points, and to three quarters of it when the population decreases by ten points; the measured gap falls by a factor of 1.6 between the maximum and the densities $n=1.21$ and 1.59.

Third, this agreement must not be overinterpreted, for four reasons. (i) Whether the antisymmetric population is a quarter at $n=1.37$ is exactly what the calculation cannot determine; the three treatments of the uniform terms give 12\%, 37\%, and 43\%. (ii) The calculated value is for a single system size. The one-component gap falls by a factor of 2.2--2.5 between $N=8$ and 16, and if the two-subband gap does the same the calculated maximum is 1.7--2~K. (iii) The calculated gap is that of an ideal system, and measured activation gaps of even-denominator states are otherwise a factor of 2 to more than 10 below the ideal ones (Sec.~\ref{sec:background}); a measured gap equal to the ideal one would be unprecedented, although this sample is of unprecedented quality, with an estimated disorder broadening of 0.5~K~\cite{Singh2025}. (iv) The measured quantity is an activation energy for charged excitations, whereas the calculated maximum is a crossing of two neutral branches, one of which is an intersubband excitation that need not limit dc transport. The magnitude of the gap is therefore not yet explained. A gap of about 4~K at a maximum is, however, within reach of a two-subband description and out of reach of a one-component one.

The temperature dependence of the gap at the two densities nearest the maximum (Sec.~\ref{sec:transport}) is a further indication that the state there is special. At a crossing of two excitation branches of different character the low-energy excitations change with density, and near such a point a strong temperature dependence of the transport gap would not be surprising; we have no theory of it.

\subsection{One component or two; Pfaffian or 331}

The experiment infers, from the insulating phases and the odd-numerator states, that the electron system changes from one-component to two-component at $\alpha\simeq0.064$, just on the high-density side of the gap maximum~\cite{Singh2025}, and takes the $\nu=8/17$ state, which is a daughter of the Pfaffian and not of 331, as evidence for a Pfaffian at the densities where it is seen, around $n=1.35$.

Our results bear on this as follows. In the calculation there is no sharp boundary between one and two components at $\nu=1/2$: the antisymmetric population grows continuously with density in every treatment, and the state moves continuously along Ho's path. A state at the gap maximum of the calculation has a quarter of its electrons in the antisymmetric subband and $t^*\approx0.4$; it is neither the Pfaffian in the symmetric subband nor 331; its overlaps with the two are 0.55 and 0.71 for $N=8$, and its overlap with the best Ho state is 0.90. On the high-density side of that point the antisymmetric population approaches the zero-field value of 40--45\%, the state moves toward 331 ($t^*=0.6$--0.9), and for the largest populations it approaches two weakly coupled layers. On the low-density side it approaches the Pfaffian.

The plain diagonalization places the sample on the Pfaffian side of this point over the whole experimental range, reaching it only beyond $n=1.9$; the calculations with the electrostatic terms of the infinite system place the point at $n=0.9$--1.1 and the sample at $n=1.37$ on the 331 side. If the observed gap maximum is identified with the calculated one, then the sample at $n=1.37$ is at the midpoint of the path, the plain diagonalization underestimates the antisymmetric population at that density (12\%), and the corrected ones overestimate it (37\% and 43\%). This identification is an inference from the experiment, not a result of the calculation.

On the question of topological order the calculation is silent. The Pfaffian and 331 phases are separated by a transition at which a neutral fermion gap closes~\cite{ReadGreen2000}, and which may not be visible in the charge gap. In the systems we can diagonalize the evolution is smooth, the overlap with the Pfaffian does not discriminate (Sec.~\ref{sec:bilayer}), and the entanglement spectrum for eight electrons (Appendix~\ref{app:es}) locates the state on Ho's path consistently with the overlaps but does not identify the phase. We therefore make no statement about where along the path the transition occurs. We can say the following. The $\nu=1/2$ state in this well has electrons in both subbands, 12\% in the antisymmetric one at $n=1.37$ in the plain diagonalization and three times as many when the electrostatic energy of the infinite system is used. The Abelian 331 phase is therefore a real possibility at $n=1.37$ and above, and so is the Pfaffian phase; the experimental observation of the $8/17$ state favors the latter near $n=1.35$, and nothing presently decides the matter at higher density.

\subsection{Unequal subband populations and the 331 state}

It is sometimes argued that 331 is excluded in a wide well because the subband populations are unequal, 60:40 at zero field in the present sample, whereas 331 is a balanced state. The argument confuses two different imbalances. The balance required by 331 is between the two layers, which hold equal charge by symmetry. Unequal subband populations are a pseudospin polarization along the tunneling direction, which the 331 state acquires continuously under tunneling: in the ideal bilayer at $d=\ell$ the overlap with 331 is still 0.97 ($N=6$) and 0.93 ($N=8$) when the populations are 59:41, and the state remains on Ho's path with an overlap above 0.92 for all polarizations (Table~\ref{tab:bilayer}). A charge imbalance between the layers is a different perturbation; the experiment imposes it deliberately and finds that 13--27\% destroys the state~\cite{Singh2025}, and an earlier calculation for the bilayer with tunneling found the 331 state to be remarkably robust to it and the Pfaffian side less so~\cite{PetersonPapicDasSarma2010}. We have not studied it.

\subsection{Relation to earlier theory}

Our strict single-layer results agree with Ref.~\onlinecite{StorniMorfDasSarma2010}, quantitatively when the widths are compared as rms widths of the transverse density (Appendix~\ref{app:width}), and extend them to the self-consistent charge profile, which makes little difference.

Our bilayer results agree with Refs.~\onlinecite{Papic2010,PetersonDasSarma2010} on the conversion of 331 into a state of large Pfaffian overlap by tunneling and on the smallness of the gap on the Pfaffian side. Reference~\onlinecite{PetersonDasSarma2010} found the gap in the plane of tunneling and layer separation to be largest close to the transition on the 331 side; at $d=\ell$ we find it monotonic. The density-matrix renormalization-group study of Ref.~\onlinecite{Zhu2016}, for much larger systems in a bilayer with layer separation $3\ell$, identified a Moore-Read phase at intermediate tunneling from its entanglement properties, and a gap maximum near the transition. Our wide-well results, with the maximum of the neutral gap at the crossing of two excitation branches, are in line with a maximum near the crossover; they add nothing to the identification of the phases.

The variational studies of Refs.~\onlinecite{Zhao2021,Sharma2024} find a one-component paired state to be lower in energy than 331 for wide-well parameters. They compare the two end points of the path, the fully polarized state and the unpolarized 331 state, and evaluate the latter without the subband splitting. The exact ground states we find have intermediate polarization and an overlap of 0.85--0.94 with the states of Ho's family for antisymmetric populations up to 30\%. A variational calculation with Ho's states and the splitting included would connect the two approaches, and would reach sizes at which the electrostatic problem that limits the present calculation does not arise.

\subsection{Earlier wide-well experiments}

The earlier experiments~\cite{Suen1992,Suen1992b,Suen1994,Shabani2013} found the $\nu=1/2$ state in a window of density for each well width, bounded by a compressible state at low density and an insulating phase at high density, and, in the 77-nm well of Ref.~\onlinecite{Suen1994}, a gap with a maximum at nearly the same $\alpha$ as in Ref.~\onlinecite{Singh2025} but about eight times smaller~\cite{Singh2025}. A maximum at the same $\alpha$ in a sample of much lower mobility supports an intrinsic origin of the maximum, tied to the subband structure (Appendix~\ref{app:earlier}). The factor of eight in magnitude between two samples of nearly the same width at the same $\alpha$ is then entirely an effect of sample quality, and its size, compared with the disorder broadening of about 0.5~K estimated for the newer sample, is another indication that the magnitudes of these gaps are not understood.

\section{Conclusions}
\label{sec:conclusion}

We have presented self-consistent Hartree calculations, exact diagonalization in one-component, bilayer, and two-subband models, and an analysis of the published transport data for the $\nu=1/2$ fractional quantum Hall state in the 72.5-nm-wide GaAs quantum well of Ref.~\onlinecite{Singh2025}. Our conclusions are these.

(1) \textit{Subband structure.} A Hartree calculation in which the electrons alone determine the potential in the well gives $\alpha=\Delta_{\rm SAS}/E_C=0.064$ at $n=1.37$, consistent with the values quoted in the experiment, and a strongly bilayer-like charge distribution. The well has two relevant subbands. A jellium background gives a splitting 2.4 times larger and should not be used for this sample.

(2) \textit{Exact diagonalization of the two-subband Hamiltonian.} For 8 electrons on the sphere the ground state is a uniform paired state at all densities, with 7--18\% of the electrons in the antisymmetric subband for $n=1.0$--1.8. It is closer to the Pfaffian than to 331, and its neutral gap is 2--4~K, rising with density, with no maximum.

(3) \textit{That result is not converged, for an identifiable reason.} The antisymmetric population is controlled by electrostatic energies that systems of six and eight electrons on a sphere contain only in part. Replacing them by their values for the infinite system raises the population at $n=1.37$ from 12\% to 37\% or 43\%, depending on how it is done, and moves the state from the Pfaffian side to the 331 side. The location of the sample between the two is not determined.

(4) \textit{Structure.} When the antisymmetric subband is populated the ground state evolves continuously along Ho's path of triplet-paired states from the Pfaffian toward 331. For eight electrons the gap to even-parity excitations triples as the antisymmetric population grows to a quarter and then collapses; the gap to intersubband excitations decreases throughout; the two cross at an antisymmetric population of 24\%, where the neutral gap has its maximum, $0.025\,E_C$. As functions of the antisymmetric population these results do not depend on how the electrostatic terms are treated.

(5) \textit{Frozen-subband limit.} If the antisymmetric subband is removed from the calculation, the ground state for $N\le16$ is Pfaffian-like at all densities and for widths $w\ge5\,\ell$, for any model of the transverse wave function, including a strict single layer with no self-consistency. Its neutral gap is nearly independent of density and is 0.6--1.0~K for 16 electrons, depending on the pseudopotential convention. A gap in a finite system is not a fractional quantum Hall effect, and the size dependence argues against one. Between $N=8$ and $N=16$ the gap at the width of the well falls by a factor of 2.2--2.5, though not monotonically; the largest system, $N=16$, has no gap for widths below $4\,\ell$, where the smaller ones do ($N=10$ from $1.1$--$1.6\,\ell$, $N=8$ and 14 from zero width); and the overlap with the Pfaffian falls from 0.94 for $N=8$ to 0.79 for $N=14$. Nothing in these results indicates that the gap survives in the thermodynamic limit, which is the conclusion reached for the strict single-layer model in Ref.~\onlinecite{StorniMorfDasSarma2010}. With the second subband frozen out the wide well most likely has no Pfaffian FQHE at $\nu=1/2$ and is a compressible composite-fermion Fermi sea, which is what the experiment finds when the second subband is far away, $\alpha>0.11$~\cite{Singh2025}. A one-component Pfaffian accounts neither for a gap of several kelvin nor for a maximum of the gap as a function of density.

(6) \textit{Diagnostics.} The overlap with the Pfaffian in the symmetric subband measures the probability that the antisymmetric subband is empty and nothing else; the component of the ground state with all electrons in the symmetric subband is the Pfaffian to 0.96 or better at all densities, and so is that of the 331 state itself. The overlap with 331 is 0.57 for two decoupled layers of four electrons each. Neither overlap, at these sizes, identifies the state. The overlap with Ho's family does better, and the parameter $t^*$ gives a continuous measure of the position between the two. The orbital entanglement spectrum (Appendix~\ref{app:es}) places the ground state on Ho's path where the overlap with the Ho family does.

(7) \textit{Subband populations and 331.} Unequal populations of the symmetric and antisymmetric subbands are a pseudospin polarization, not a charge imbalance between the layers, and do not exclude 331-like correlations.

(8) \textit{Transport.} A single activation gap does not span the measured temperature range. At $n=1.37$ the slope of the published Arrhenius plot runs from 1.6--1.7~K at its high-temperature end to 5--7~K at its low-temperature end. The high-temperature slope is suppressed by the saturation of $R_{xx}$ at the resistance of the compressible state. The low-temperature slope, which is the reported 6~K, requires a prefactor 25 times larger than activated transport allows and is contradicted by the disappearance of the resistance minimum at 0.4--0.5~K. The gap obtained from the magnitude of $R_{xx}$ with the theoretical prefactor, which describes the data at the three densities away from the maximum with no adjustment, is $3.9\pm0.5$~K at $n=1.37$ and 2.3--3.0~K at $n=1.21$, 1.49, and 1.59, in agreement with the 3.7~K reported earlier by the same group. The gap has a maximum near $n=1.37$ that is lower and broader than reported, and at the two densities nearest the maximum it increases on cooling in the measured range.

(9) \textit{The magnitude of the gap.} The experimental activation gap, about 4~K (and not 6~K as claimed in the experiment), is comparable to the neutral gap of the plain two-subband diagonalization for eight electrons, about 3~K, and much larger than that of the frozen-subband limit, about 1~K for 16 electrons. The measured maximum, $0.023\pm0.003\,E_C$, is four to six times the neutral gap of the frozen-subband limit for 16 electrons. It equals, within the uncertainties, the maximum of the neutral gap of the two-subband model for 8 electrons, which occurs where a quarter of the electrons are in the antisymmetric subband. This agreement is for one system size and the ideal system, and it presupposes an antisymmetric population that the calculation cannot confirm. The magnitude of the gap is not yet explained, but it is within reach of a two-subband description and out of reach of a one-component one. The claim of a Moore-Read state with a gap of 6~K does not stand. It is not supported by the data from which it was obtained, for the reasons given in conclusion (8), and it is not supported by theory: 6~K is $0.035\,E_C$, which is 40\% above the gap of the ideal $\nu=5/2$ state in the same units, 40\% above the largest neutral gap of the two-subband model of the well for eight electrons, and six to nine times the gap of the one-component Pfaffian in this well for 16 electrons.

(10) \textit{Pfaffian or 331.} Taken together, our results are more consistent with the observed state being a Pfaffian-like paired state than with its being 331. First, the diagonalization of the two-subband Hamiltonian as it stands, with nothing added, gives an overlap of 0.85--0.71 with the Pfaffian against 0.44--0.61 with 331 for $n=1.0$--1.6, and a position $t^*=0.16$--0.29 on Ho's path, on which the Pfaffian is at 0 and 331 at 1 (as noted in Sec.~\ref{sec:bilayermethod}, $t$ measures proximity to the end states, not the phase). Second, no calculation of ours for the well produces a developed 331 state: for eight electrons the overlap with 331 is 0.75 at most at any density in any treatment of the electrostatic terms, against 0.57 for two decoupled layers and 0.95 for the bilayer at $d=\ell$, and the strict bilayer of conclusion (11) does not give 331 at all. Third, the experiment observes the $8/17$ state, a daughter of the Pfaffian and not of 331, near the density of the gap maximum~\cite{Singh2024,Singh2025}. The state in question is not the one-component Pfaffian of conclusion (5). It has a minority of its electrons in the antisymmetric subband and owes its gap to them (conclusion 4).

This conclusion is not definitive, for three reasons. It rests on eight electrons, and for the reason given in conclusion (3) the antisymmetric population is underestimated at that size; with the electrostatic terms of the infinite system the same calculation places the sample on the 331 side of the path above $n\approx1.0$. Overlaps do not determine topological order; the entanglement spectrum [conclusion (13)] confirms where each ground state lies on the path but does not choose between the treatments. And the two-subband calculations for 10 and 12 electrons, which would decide whether the plain diagonalization or the corrected one is closer to the truth, were beyond the computing resources available for this work. The benchmark calculation at $\nu=1/3$ (Appendix~\ref{app:third}) bears on the first of these reasons. There the plain diagonalization shows no sign of the loss of the Laughlin state, which the corrected treatments place near the density at which the experiment finds insulating phases below $\nu=1/2$. This suggests that at $\nu=1/2$ as well the true antisymmetric population lies between the plain and the corrected values.

(11) \textit{Strict bilayer at the Hartree maxima.} Two layers of zero width at the maxima of the Hartree density, without tunneling and with equal populations, is the model most favorable to 331. It does not give 331. The layers are 5.1--7.1 magnetic lengths apart in the 72.5-nm well for $n=1.0$--1.6, and 3.5--11 for wells of 60--100~nm, against the 1--2 at which the bilayer ground state is 331. The overlap with 331 is that of two decoupled layers, 0.57 for eight electrons, and the gap is 0.07--0.2~K (Appendix~\ref{app:modelD}). The 331-like correlations of the two-subband calculation depend on the width of the layers and on the tunneling.

(12) \textit{The second subband is essential.} The $\nu=1/2$ FQHE of the Princeton sample cannot be understood without an active second subband. With the antisymmetric subband frozen out there is most likely no FQHE [conclusion (5)]; with the two halves of the well decoupled there is none [conclusion (11)]. The state exists between these limits, where the antisymmetric subband lies within about $0.1\,E_C$ of the symmetric one and is partly occupied in the correlated state, and the gap is largest when about a quarter of the electrons are in it. This is why the effect requires a wide well, and why width alone is not sufficient: the well must be wide enough, at the given density, for the two subbands to be nearly degenerate on the scale of the Coulomb energy, and not so wide that the two charge layers cease to communicate. The window of $\alpha$ in which the state is observed, bounded by a composite-fermion Fermi sea at large $\alpha$ and by insulating phases at small $\alpha$~\cite{Suen1994,Shabani2013,Singh2025}, is this window. The self-consistent Hartree calculations of Appendix~\ref{app:earlier} for the earlier Princeton wells, the 68-nm well of 1992~\cite{Suen1992} and the 77-nm well of 1994~\cite{Suen1994}, show that the $\nu=1/2$ FQHE was observed in them in the same window of $\alpha$ and of second-subband occupancy as in the 72.5-nm well, with the calculated subband splittings and layer separations in reasonable agreement with the measured ones. In all three Princeton wells the density range of the state is a range in which the second subband is substantially occupied; the second subband is essential for the existence of the $\nu=1/2$ FQHE, and a wide well without an active second subband does not show it. This is the main new conclusion of this paper. Its empirical part, that the state is observed only where the second subband is active, rests on the Hartree calculation and the experimental data alone, independently of the exact diagonalization.

(13) \textit{Entanglement spectrum.} The orbital entanglement spectrum of the two-subband ground state at the density of the gap maximum (Appendix~\ref{app:es}) is a diagnostic independent of any trial state, and it confirms the picture obtained from the overlaps. For each treatment of the electrostatic terms it places the ground state on Ho's path where the overlaps do: for the plain diagonalization at the same point, $t=0.24$, with the counting of the Pfaffian in the even sector, the 331-type levels at higher entanglement energy, and a clear entanglement gap; for the corrected treatments closer to 331, without an entanglement gap. The two diagnostics agree because the location of the sample on the path is not a question about the character of a given ground state, which both determine consistently, but about which ground state is the right one, and that is fixed by the long-wavelength electrostatic energy of charge transfer between the subbands. The entanglement spectrum thus reduces the open problem to a single quantity, the population of the antisymmetric subband at $\nu=1/2$, which a larger calculation or a measurement can supply.

\textit{Relation to earlier work, and what is new.} The wide well has been treated as a two-subband system before. Papi\'c \textit{et al.}~\cite{Papic2009} retained the two lowest subbands of an infinite square well and diagonalized the resulting Hamiltonian for eight electrons at $\nu=1/2$, with the width and $\Delta_{\rm SAS}$ as independent parameters; they found a state with an overlap of up to 0.95 with 331 where the subband polarization vanishes, a state with an overlap of up to 0.92 with the Pfaffian at large splitting in narrower wells, and a continuous crossover between them. Thiebaut \textit{et al.}~\cite{Thiebaut2013} did the same for a biased square well with up to ten electrons and found the incompressible state to be a 331 state rotated in the subband pseudospin, with the Pfaffian having the largest overlap only where the well is effectively one-component and, in their judgment, compressible. The bilayer with tunneling~\cite{Papic2010,PetersonDasSarma2010,PetersonPapicDasSarma2010,Zhu2016} is the same problem for thin layers with the separation and the tunneling as free parameters.

We agree with these works on three points: the evolution between 331 and the Pfaffian with increasing subband splitting is continuous; the gap is small at the Pfaffian end; and the one-component limit, where the Pfaffian overlap is largest, is the limit in which an incompressible state is least likely (Refs.~\onlinecite{StorniMorfDasSarma2010,Thiebaut2013} and conclusion 5). We disagree with them, or go beyond them, on the following. (i) The subband wave functions and the splitting used here are those of the sample, from a self-consistent calculation, and are not free parameters. With them the region of well-developed 331 found in Refs.~\onlinecite{Papic2009,PetersonDasSarma2010} is not reached at any density: the charge layers are five to seven magnetic lengths apart. The conclusion of the bilayer studies that the observed wide-well states are 331 does not carry over to this sample. (ii) The subband populations of the correlated state are controlled by electrostatic energies that arise from the different charge profiles of the two subbands, and a sphere of six or eight electrons contains only a quarter to a half of them [conclusion (3)]. This, and not the correlation energy, is what limits the calculation, and it applies equally to the earlier two-subband calculations at these sizes. (iii) The overlaps with the Pfaffian and with 331 on which the earlier identifications rest do not identify the state at these sizes [conclusion (6)]; the overlap with Ho's family does. (iv) The maximum of the neutral gap is a crossing of two excitation branches of opposite reflection parity at an antisymmetric population of a quarter. (v) The one-component calculations for wells with two occupied subbands~\cite{Zhao2021,Sharma2024}, which have been taken to support a one-component Pfaffian in wide wells, do not contain the second subband as an independent component, which is what produces the gap. Within a one-component space a Pfaffian-like state is the only paired candidate, and our calculation in the same limit reproduces it, with a gap of 1~K or less that decreases with size. Because the second subband is not treated as an independent component, these calculations are unreliable for this problem, in their quantitative predictions and in their qualitative picture of the state. The bilayer calculations with which Ref.~\onlinecite{Singh2025} compares its gaps~\cite{PetersonDasSarma2010,Zhu2016} do contain two components, but for layers at most a few magnetic lengths apart with the separation and the tunneling as free parameters, and a quantitative comparison with this well is not justified. (vi) The reported gaps had not been reexamined.

\textit{In-plane magnetic field.} A magnetic field parallel to the layer suppresses the tunneling between the two halves of the well and makes the charge distribution more bilayer-like~\cite{Hasdemir2015}. For two thin layers a distance $d$ apart the tunneling amplitude in the lowest Landau level is reduced by the factor $\exp[-(d/\ell)^2\tan^2\theta/4]$ at tilt angle $\theta$; with $d=6.35\,\ell$ this is 0.26 at $20^\circ$ and 0.03 at $30^\circ$. To that extent tilting the sample at fixed density and filling is equivalent to lowering $\alpha$, and it moves the state along the same path as increasing the density does: toward a larger antisymmetric population, away from the Pfaffian, toward 331, and finally to two decoupled layers. We therefore expect a compressible state at low density to become an FQHE state under tilt; the gap of a state on the low-density side of the maximum to rise and then fall; the gap of a state beyond the maximum to fall monotonically; and an insulating phase at large tilt. The sequence compressible state, FQHE, insulator with increasing tilt has been observed in 60- and 65-nm wells and was interpreted there as evidence for 331~\cite{Hasdemir2015}. In our picture the sequence does not by itself distinguish the Pfaffian from 331, because both lie on the path that the tilt traverses. The tilt provides a second, independent way of traversing the path, and with it a test. If the picture is right, the gap maximum should follow a single curve of constant effective tunneling in the plane of density and tilt: the tilt at which the gap is largest should decrease as the density approaches $n=1.37$ from below and vanish there, and beyond $n=1.37$ the gap should only fall. The $8/17$ state, a daughter of the Pfaffian and not of 331, marks the Pfaffian side of the path; a $\nu=1/2$ state that remains strong at tilts at which $8/17$ has disappeared is on the other side. A 331 state of two layers that does not depend on tunneling would behave differently: a small tilt would leave its gap unchanged or increase it at every density. We do not expect that in this well, because without tunneling the layers are too far apart to support any $\nu=1/2$ state [conclusion (11)], and the insulating phase observed at large tilt is consistent with this. Two caveats apply. The thin-layer factor overstates the suppression in a wide well, where the in-plane field also deforms the subband wave functions, mixes the subband and cyclotron motions, makes the interaction anisotropic, and leaves $\Delta_{\rm SAS}$ without a direct definition~\cite{Hasdemir2015}. And the tilt raises the Zeeman energy at fixed filling, which tests the assumption of full spin polarization at the same time. We have done no calculation in a tilted field; the statements above are expectations, and a two-subband calculation with an in-plane field is needed to make them quantitative.

These conclusions rest on approximations and limitations, which we list.

(a) \textit{System size.} The two-subband calculations are for $N=6$ and 8. The first is aliased in its one-component limit; the gap maximum of conclusion (4) is found for $N=8$ only. The radius of the sphere, 2.1--2.6 magnetic lengths, is less than half the separation of the two charge layers, and this is the origin of the indeterminacy stated in conclusion (3). The frozen-subband calculations extend to $N=16$ and have not converged; the two pseudopotential conventions differ by a factor of 1.6--2.0 in the gap.

(b) \textit{Hartree input.} The subband wave functions and the bare splitting are those of a Hartree calculation at zero field. Exchange and correlation corrections to the subband structure, the relaxation of the wave functions in the high-field state, and the higher subbands are neglected. The agreement of the calculated $\alpha$ with experiment is established only at the few values quoted in Ref.~\onlinecite{Singh2025}, and at the lowest density the calculated value is at least 12\% low.

(c) \textit{Landau-level mixing.} At 11~T the ratio of the Coulomb energy to the cyclotron energy is 0.75. Landau-level mixing is neglected. It also decides between the Pfaffian and its particle-hole conjugate, which we have not considered.

(d) \textit{Disorder} is neglected.

(e) \textit{Spin.} Full spin polarization is assumed. The Zeeman energy at 11~T is 3.3~K, comparable to the gaps.

(f) \textit{Gaps.} All calculated gaps are neutral gaps at fixed flux. The energy of a separated quasiparticle-quasihole pair, with which the transport gap is usually identified, has not been computed, and the intersubband branch need not be relevant to transport. No correction for the finite-size density has been applied.

(g) \textit{Diagnostics.} Overlaps do not determine topological order. The orbital entanglement spectrum has been computed for eight electrons at $n=1.37$ only (Appendix~\ref{app:es}); at that size it does not identify the phase, and no quasihole counting or other direct probe has been computed.

(h) \textit{Transport.} The analysis uses data digitized from a published figure, with 11 to 17 points per trace over at most one and a half decades in resistance, and a phenomenological form for the saturation. The sample geometry, needed to convert the measured prefactor into a resistivity, is not known to us.

Further progress requires the following.

\textit{Theory.} (i) A two-subband calculation at sizes for which the system is larger than the separation of the charge layers, by the density-matrix renormalization group on the cylinder or sphere or by variational Monte Carlo with Ho's states, to determine the antisymmetric population at $\nu=1/2$ as a function of density. This is the single most important missing piece. (ii) A direct identification of the phase along the path, from the entanglement spectrum at sizes beyond eight electrons or from the quasihole degeneracy. (iii) Charged-excitation gaps for the two-subband state. (iv) Landau-level mixing and the competition with the anti-Pfaffian. (v) A subband calculation beyond Hartree and beyond zero field. (vi) A theory of $R_{xx}(T)$ at the minimum from the activated regime to saturation. (vii) Two-subband exact diagonalization for 10 and 12 electrons. (viii) A two-subband calculation in a tilted field.

\textit{Experiment.} (i) An analysis of the original $R_{xx}(T)$ data at all densities over the full temperature range, with the saturation included and the prefactors and sample geometry reported, and a statement of the temperature windows behind the gaps of Refs.~\onlinecite{Singh2024} and \onlinecite{Singh2025}. (ii) A measurement of the gap that does not rely on activated transport, for example of the chemical-potential jump at $\nu=1/2$. (iii) The measured $\Delta_{\rm SAS}(n)$ in numerical form, for comparison with the Hartree calculation. (iv) A measurement of the subband populations at $\nu=1/2$, which is the quantity the theory cannot supply; the intersubband excitation, whose energy the calculation predicts to fall with density, is accessible to inelastic light scattering. (v) Tuning of the tunneling at fixed density by an in-plane field, with the gap at $\nu=1/2$ and the strength of the $8/17$ state followed together as functions of tilt, as described above. (vi) Thermal Hall or interferometric measurements across the density range, which alone can distinguish the Abelian from the non-Abelian phase. (vii) The gap of the $\nu=1/3$ state as a function of density in the same sample, which would decide between the treatments of the electrostatic terms (Appendix~\ref{app:third}).

The $\nu=1/2$ state in wide wells has been regarded as a two-component state for most of its history and more recently as a non-Abelian one-component state with a remarkably large gap. Our analysis supports neither description in its simple form. The experimental activation gap, about 4~K (and not 6~K as claimed in the experiment), is comparable to the neutral gap of the plain two-subband diagonalization for eight electrons, about 3~K, and much larger than that of the frozen-subband limit, about 1~K for 16 electrons. The state is not a one-component Pfaffian, which in this well has a neutral gap of 1~K or less for 16 electrons and most likely none in the thermodynamic limit, and it is not the 331 state of two layers, which are too far apart. It is a paired state of electrons in two subbands, on the continuous path between the Pfaffian and 331. Our finite-size results are more consistent with the Pfaffian side of that path than with 331, but they are not definitive; the calculations for larger systems that would decide this remain to be done.

\begin{acknowledgments}
The author thanks Nick Read, Steve Simon, and Zlatko Papi\'c for helpful communications. This work is supported by the Laboratory for Physical Sciences. The author takes full responsibility for the work, which benefited from AI assistance in numerics/figures/text from Anthropic Claude Opus 5.5 with checks done by OpenAI GPT-6 Astra.
\end{acknowledgments}

\appendix

\section{Results in the jellium model}
\label{app:jellium}

We collect here the results obtained with the jellium Hartree profile, for comparison with those of the remote-doping model used in the main text.

The single-particle results differ strongly (Table~\ref{tab:hartree} and Figs.~\ref{fig:profiles}--\ref{fig:hartreemaps}). In the jellium model $\Delta_{\rm SAS}$ is larger by a factor of 1.9 to 3.1 over the density range $n=1.0$--1.8 and nearly independent of density, the antisymmetric fraction at zero field is smaller (18--33\% against 33--44\%), and the charge distribution is much less bilayer-like.

The one-component many-body results differ little (Table~\ref{tab:onecomp} and Fig.~\ref{fig:onecompdens}). At $w=72.5$~nm and $n=1.37$ the gaps are 11--17\% larger with jellium and the overlaps agree to better than 0.001. The gap in kelvin increases somewhat faster with density in the jellium model: between $n=1.0$ and 1.8 by 19\% for $N=8$, 29\% for $N=10$, and 27\% for $N=14$, against 7--10\% with remote doping. In both models the ground state is Pfaffian-like at all densities, the gap is of order 1~K for $N\ge8$, and it has no maximum as a function of density.

The two-subband results differ qualitatively (Table~\ref{tab:twosubjellium}). With the jellium subband structure the bare splitting is $\varepsilon_0=0.39\,E_C$ at $n=1.37$, against 0.17 for remote doping, and the system is predominantly one-component at all densities in all three treatments of the uniform terms. The antisymmetric fraction for $N=8$ is at most 7\% in treatments S and T$'$ and at most 16\% in treatment T, the state is near the Pfaffian end of Ho's path ($t^*\le0.3$), and the lowest excitation is in the even sector, with an energy of 1.6--3.3~K that increases with density. The intersubband excitations are at 6~K or more.

\begin{table*}[t]
\caption{Two-subband exact diagonalization with the jellium Hartree profile, $w=72.5$~nm, $N=8$. Columns as in Table~\ref{tab:twosub}.}
\label{tab:twosubjellium}
\begin{ruledtabular}
\begin{tabular}{ccc|ccccc|ccccc|ccccc}
\hline\hline
 & & $B=0$ & \multicolumn{5}{c|}{Treatment T$'$} & \multicolumn{5}{c|}{Treatment T} & \multicolumn{5}{c}{Treatment S}\\
$n$ & $\alpha$ & $n_A/n$ & $n_A/n$ & $O_{\rm Ho}$ & $t^*$ & $\Delta_e$ & $\Delta_o$ & $n_A/n$ & $O_{\rm Ho}$ & $t^*$ & $\Delta_e$ & $\Delta_o$ & $n_A/n$ & $O_{\rm Ho}$ & $t^*$ & $\Delta_e$ & $\Delta_o$\\
\hline
1.00 & 0.183 & 18\% & 3\% & 0.92 & 0.09 & 1.66 & 21.98 & 5\% & 0.91 & 0.12 & 1.78 & 15.52 & 2\% & 0.93 & 0.06 & 1.57 & 32.15\\
1.20 & 0.166 & 24\% & 4\% & 0.92 & 0.11 & 1.81 & 20.02 & 7\% & 0.91 & 0.15 & 2.03 & 12.70 & 2\% & 0.93 & 0.07 & 1.67 & 33.34\\
1.37 & 0.154 & 27\% & 5\% & 0.92 & 0.12 & 1.94 & 18.53 & 9\% & 0.91 & 0.18 & 2.30 & 10.59 & 2\% & 0.93 & 0.07 & 1.75 & 34.64\\
1.60 & 0.141 & 30\% & 6\% & 0.91 & 0.14 & 2.11 & 16.73 & 12\% & 0.91 & 0.23 & 2.77 & 8.18 & 2\% & 0.93 & 0.07 & 1.83 & 36.74\\
1.80 & 0.132 & 33\% & 7\% & 0.91 & 0.15 & 2.26 & 15.32 & 16\% & 0.91 & 0.29 & 3.30 & 6.11 & 2\% & 0.93 & 0.07 & 1.88 & 38.85\\
\hline\hline
\end{tabular}
\end{ruledtabular}
\end{table*}

The comparison shows where the electrostatic model matters. For a calculation restricted to one component it is nearly irrelevant, and for that reason no refinement of a one-component calculation can tell whether the one-component restriction is justified. The answer to that question depends on the bare subband splitting and on the electrostatic energy of charge transfer between the subbands, and the jellium model, which overestimates the measured $\Delta_{\rm SAS}$ by a factor of 1.7 to 2.4, gives the wrong input for both.

\section{Hartree subband structure of the earlier Princeton wide wells}
\label{app:earlier}

The $\nu=1/2$ FQHE in a wide single GaAs well was first reported by Suen \textit{et al.}~\cite{Suen1992} in a 68-nm well, and its dependence on the subband separation was studied by Suen \textit{et al.}~\cite{Suen1994} in a 77-nm well, the sample with which Ref.~\onlinecite{Singh2025} compares its gaps. Here we apply to these two wells exactly the calculation of Sec.~\ref{sec:hartree}: the remote-doping Hartree model with the same parameters ($m^*=0.067\,m_e$, $\epsilon=12.9$, $V_0=224$~meV, symmetric well, zero-field subband populations with spin degeneracy 2, no exchange-correlation), with the well width fixed at the quoted value and the density varied over the experimental range. The purpose is to see whether the subband structure at which the $\nu=1/2$ FQHE was observed in these samples is the same as in the 72.5-nm well of Ref.~\onlinecite{Singh2025}, namely a second subband within about $0.1\,E_C$ of the first and substantially occupied, which Sec.~\ref{sec:hartree} and conclusion (12) identify as the condition for the state.

\begin{figure*}[t]
\includegraphics[width=0.92\textwidth]{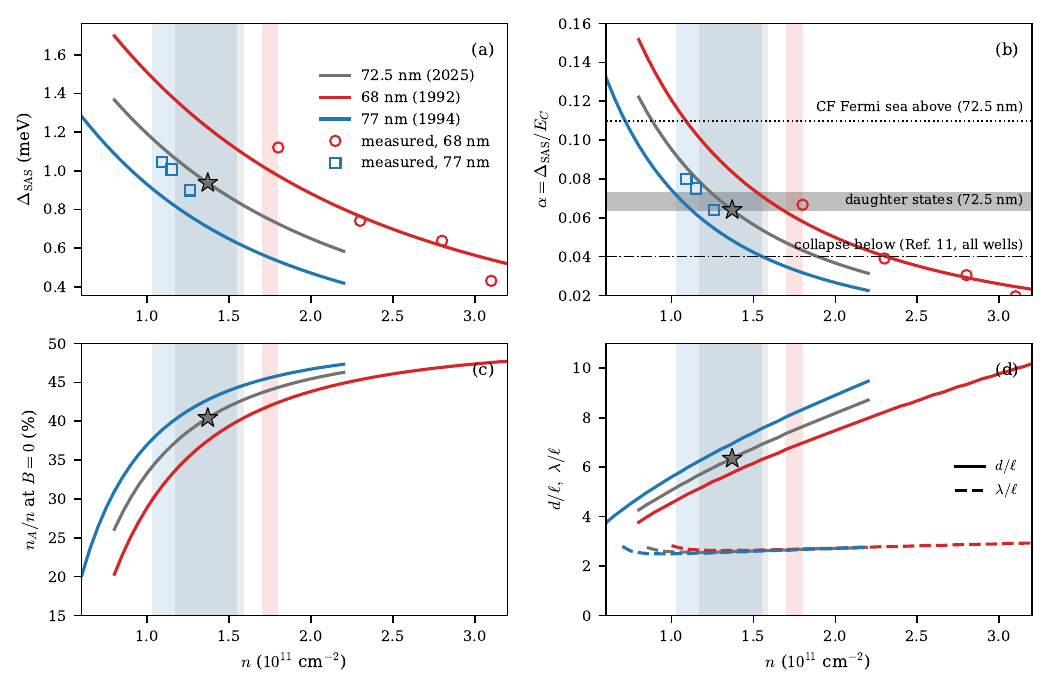}
\caption{Remote-doping Hartree results for the three Princeton wide wells in which the $\nu=1/2$ FQHE has been reported, as functions of the density: 68~nm~\cite{Suen1992} (red), 72.5~nm~\cite{Singh2024,Singh2025} (gray, the well of the main text), and 77~nm~\cite{Suen1994} (blue). (a) Subband splitting $\Delta_{\rm SAS}$; the open symbols are the zero-field values measured from the Shubnikov-de Haas frequencies in Ref.~\onlinecite{Suen1992} (Table~I there; circles) and Ref.~\onlinecite{Suen1994} (squares, converted from the quoted $\alpha$). (b) $\alpha=\Delta_{\rm SAS}/E_C$ with $E_C$ at $\nu=1/2$, with the same measured points; the shaded band and dotted line are the experimental values for the 72.5-nm well as in Fig.~\ref{fig:hartreedens}(b), and the dash-dotted line is $\alpha=0.04$, below which Ref.~\onlinecite{Suen1994} finds the state to collapse in all its wells. (c) Fraction of electrons in the antisymmetric subband at zero field. (d) Separation $d$ of the density maxima (solid) and full width at half maximum $\lambda$ of each layer (dashed) in units of $\ell$ at $\nu=1/2$. The vertical bands mark the density ranges in which the $\nu=1/2$ FQHE was observed in each well: $n=1.7$--1.8 in the 68-nm well~\cite{Suen1992}, 1.03--1.55 in the 77-nm well~\cite{Suen1994}, and 1.17--1.59 in the 72.5-nm well~\cite{Singh2024,Singh2025}; the star marks the gap maximum of the 72.5-nm well.}
\label{fig:earlier}
\end{figure*}

\begin{table}[t]
\caption{Remote-doping Hartree results for the three wells at a density at which the $\nu=1/2$ FQHE was observed: $n=1.8$ in the 68-nm well ($B=15$~T, gap 230~mK~\cite{Suen1992}), 1.37 in the 72.5-nm well (gap maximum~\cite{Singh2025}), and 1.26 in the 77-nm well (the balanced point at which Ref.~\onlinecite{Suen1994} measures $\alpha=0.064$). Measured zero-field values from Refs.~\onlinecite{Suen1992} and \onlinecite{Suen1994} are given in parentheses. Densities in $10^{11}$~cm$^{-2}$, $d$ in nm.}
\label{tab:earlier}
\begin{ruledtabular}
\begin{tabular}{lccc}
\hline\hline
$w$ (nm) & 68 & 72.5 & 77\\
Ref. & \onlinecite{Suen1992} & \onlinecite{Singh2025} & \onlinecite{Suen1994}\\
\hline
$n$ & 1.80 & 1.37 & 1.26\\
$B$ at $\nu=1/2$ (T) & 14.9 & 11.3 & 10.4\\
$E_C$ (meV) & 16.8 & 14.6 & 14.0\\
$\Delta_{\rm SAS}$ (meV) & 0.97 (1.12) & 0.94 & 0.77 (0.90)\\
$\Delta_{\rm SAS}$ (K) & 11.3 (13) & 10.9 & 8.9 (10.4)\\
$\alpha=\Delta_{\rm SAS}/E_C$ & 0.058 (0.067) & 0.064 & 0.055 (0.064)\\
$n_S$ & 1.04 & 0.82 & 0.74\\
$n_A$ & 0.76 & 0.55 & 0.52\\
$n_A/n$ & 42\% & 40\% & 42\%\\
$d$ & 46.4 (44.5) & 48.4 & 52.3\\
$d/\ell$ & 7.0 & 6.3 & 6.6\\
$\lambda/\ell$ & 2.7 & 2.6 & 2.5\\
$n(0)/n_{\rm max}$ & 0.21 & 0.24 & 0.22\\
\hline\hline
\end{tabular}
\end{ruledtabular}
\end{table}

\textit{Comparison with the measured subband splittings.} Both experiments measured $\Delta_{\rm SAS}$ at zero field from the two Shubnikov-de Haas frequencies, exactly the quantity that Eq.~(\ref{eq:sas}) relates to the Hartree populations, so the calculation can be checked directly. For the 68-nm well, Table~I of Ref.~\onlinecite{Suen1992} gives $\Delta_{\rm SAS}=13$, 8.6, 7.4, and about 5~K at $n=1.8$, 2.3, 2.8, and 3.1; the remote-doping model gives 11.3, 8.9, 7.1, and 6.3~K, and for the peak separation 46, 48, 50, and 51~nm against the 44.5, 46, 47.5, and 48.5~nm of the same table. For the 77-nm well, Ref.~\onlinecite{Suen1994} quotes $\alpha=0.080$, 0.075, and 0.064 for the balanced well at $n=1.09$, 1.15, and 1.26; the model gives 0.066, 0.062, and 0.055, i.e., 15--18\% low. The model thus reproduces the measured splittings of both wells to within about 15\% with no adjustable parameter, with the same tendency to lie below the measurement at the lower densities as found for the 72.5-nm well in Sec.~\ref{sec:hartree}, and the measured $\Delta_{\rm SAS}$ decreases with density in both wells as the model predicts. The jellium model would give splittings larger than the measured ones by a factor of about two (Appendix~\ref{app:jellium}). The agreement with every measured zero-field quantity of these two samples, $\Delta_{\rm SAS}$ and its density dependence, the layer separation $d$, and the layer thickness $\lambda$, is thus reasonable, and the calculation of Sec.~\ref{sec:hartree} can be applied to them with the same confidence as to the 2025 sample.

\textit{Where the state is observed.} Figure~\ref{fig:earlier} shows the results for the three wells, those for the 72.5-nm well being repeated from Fig.~\ref{fig:hartreedens}, with the density range in which each experiment observed the $\nu=1/2$ FQHE marked; Table~\ref{tab:earlier} gives the values at a representative density in each well. In the 68-nm well of Ref.~\onlinecite{Suen1992} the state was observed at $n=1.8$ and 1.7, the lowest densities that the back gate allowed, with an activation gap of 230~mK at $n=1.8$. There the model gives $\alpha=0.058$--0.063, 42\% of the electrons in the antisymmetric subband at zero field, $d=6.7$--$7.0\,\ell$, and a central density 21\% of the maximum. These numbers are nearly the same as those at the gap maximum of the 72.5-nm well ($\alpha=0.064$, 40\%, $d=6.3\,\ell$): the two samples, differing in width, density, and magnetic field, and measured 33 years apart, show the $\nu=1/2$ FQHE at nearly the same point in the subband structure. The gaps, 230~mK against $3.9\pm0.5$~K (Sec.~\ref{sec:transport}), differ by more than an order of magnitude at nearly the same $\alpha$; as in the comparison with the 77-nm well in Sec.~\ref{sec:discussion}, this is an effect of sample quality, the mobility of the 1992 sample being $6.5\times10^5$~cm$^2$/V\,s.

In the 77-nm well of Ref.~\onlinecite{Suen1994}, in which the density was tuned with front and back gates and the gap measured as a function of $\alpha$, the state is seen from $n=1.03$ to 1.55, is most stable for $0.045\lesssim\alpha\lesssim0.085$ in the measured $\alpha$, and collapses for $\alpha\lesssim0.04$; over all six wells of that work, of widths 60--96~nm, the state exists only for $0.04\lesssim\alpha\lesssim0.1$. In the remote-doping model the observed density range of the 77-nm well corresponds to $\alpha$ from 0.072 down to 0.040, an antisymmetric-subband fraction from 38\% to 44\%, and $d/\ell$ from 5.7 to 7.6, and the measured $\alpha$ window $0.045$--0.085 corresponds to $n=0.90$--1.44. The same is true of the 72.5-nm well, where the state is observed for $0.05\lesssim\alpha\lesssim0.11$ ($n=1.17$--1.59 in the model, $\alpha=0.079$--0.052, 37--43\% antisymmetric fraction), and of the 68-nm well. In all three samples, over the entire density range in which the $\nu=1/2$ FQHE is observed, the antisymmetric subband is active, i.e., it lies within $0.04$--$0.08\,E_C$ of the symmetric subband and holds 37--44\% of the electrons at zero field; there is no observation of the $\nu=1/2$ FQHE in a wide well at any density at which the second subband is inactive, in these two earlier samples any more than in the 2025 sample. The condition for the state is therefore a window in $\alpha$, not in density or width separately, and it is the same window for all three wells: this is the diagonal strip of Fig.~\ref{fig:hartreemaps}, found empirically in Refs.~\onlinecite{Suen1994} and \onlinecite{Shabani2013}, and the layer thickness $\lambda\approx2.6\,\ell$ that Ref.~\onlinecite{Suen1994} quotes for its wells is the value the model gives for all three [Fig.~\ref{fig:earlier}(d)].

\textit{The two limits.} The experiments also delimit the window from both sides, and the model identifies the two boundaries. On the low-density side $\alpha$ is large, the antisymmetric subband is less occupied and the two layers merge; Refs.~\onlinecite{Suen1994} and \onlinecite{Singh2025} find the compressible composite-fermion state there, in the 72.5-nm well for $\alpha\gtrsim0.11$, which in the model is $n\lesssim0.88$ for that well and $n\lesssim1.09$ and $0.72$ for the 68- and 77-nm wells. On the high-density side $\alpha$ is small, the antisymmetric subband is nearly degenerate with the symmetric one and the layers are far apart; there the state collapses, for $\alpha\lesssim0.04$~\cite{Suen1994}, which in the model is $n\gtrsim2.3$, 1.9, and 1.55 for the 68-, 72.5-, and 77-nm wells, and insulating phases set in~\cite{Suen1992b,Suen1994,Shabani2013,Singh2025}. Reference~\onlinecite{Suen1992} adds a decisive case for the small-$\alpha$ limit: a 120-nm well with $n=1.7$, for which that work quotes $d=96$~nm and $\Delta_{\rm SAS}=0.2$~K, showed no $\nu=1/2$ state at all. For this well the remote-doping model gives $\Delta_{\rm SAS}=0.03$~meV $=0.4$~K, $\alpha=0.002$, $d=98$~nm $=14\,\ell$, and a central density below 1\% of the maximum: two essentially decoupled layers, each at $\nu=1/4$, which is the strict-bilayer limit of conclusion (11) in which no $\nu=1/2$ FQHE exists for these parameters.

\textit{Conclusion.} This Appendix uses only the Hartree calculation and the published experimental data, with no diagonalization, and it establishes the main conclusion of this paper for all three Princeton wide-well samples, not only the 2025 one: the state appears only when the second subband plays an active role. In all three wells, over the density range in which the $\nu=1/2$ FQHE has been observed, the antisymmetric subband lies within $0.04$--$0.08\,E_C$ of the symmetric one and holds 37--44\% of the electrons at zero field, and the charge distribution is bilayer-like with $d\approx5$--$8\,\ell$ and $\lambda\approx2.6\,\ell$. A wide well in which the second subband is inactive, far above the first on the scale of $E_C$, does not show the state: it is a thick single layer with a compressible half-filled Landau level. A wide well in which the second subband is degenerate with the first and the layers are decoupled does not show it either. The state exists only between these limits, where the second subband is active, and in this regime the two-subband diagonalization of Sec.~\ref{sec:twocomp} gives a paired state on Ho's path between the Pfaffian and 331, and not a developed 331 state. The $\nu=1/2$ FQHE of the 1992 and 1994 samples is therefore of the same two-subband character as that of the 2025 sample and, as for that sample [conclusion (10)], more consistent with a Pfaffian-like state than with the Halperin 331 state, contrary to the two-component interpretation given in Ref.~\onlinecite{Suen1994}; the observation in that work that the state is most stable in the balanced well and is destroyed by charge asymmetry does not distinguish the two, since the Pfaffian-like two-subband state also requires the two subbands to be near degeneracy, which asymmetry removes. As in the main text, the zero-field populations are the single-particle input and not the populations at $\nu=1/2$; what the Hartree calculation establishes is that the three experiments probe the same subband structure, so that the many-body analysis of Secs.~\ref{sec:twocomp}--\ref{sec:discussion} applies to all of them.

\section{Numerical details, validation, and the $N=6$ two-subband results}
\label{app:numerics}

\textit{Hartree.} Finite differences on 2001 points over 200~nm, linear mixing with weight 0.1, convergence to $10^{-4}$~meV in the Hartree potential. Doubling the number of points changes $\Delta_{\rm SAS}$ at the reference point by 0.1\% and the peak separation by 0.1~nm.

\textit{Pseudopotentials.} The Legendre coefficients $v_k$ in Eq.~(\ref{eq:sphericalpp}) are obtained by Gauss-Legendre quadrature in the chord distance (300 points), with the transverse integrals of Eqs.~(\ref{eq:vr}) and (\ref{eq:channels}) evaluated on a 321-point grid in $z$; the $3j$ and $6j$ symbols are exact. For the Coulomb interaction the result agrees with the closed form of Ref.~\onlinecite{FanoOrtolaniColombo1986} to $10^{-14}$. The planar pseudopotentials, Eq.~(\ref{eq:vm}), are evaluated by quadrature on 8000 points to $q\ell=40$; for $F=1$ they reproduce $V_1=\sqrt\pi/4$ to ten digits.

\textit{Diagonalization.} Hamiltonians are built in the $L_z=0$ sector in a basis of Slater determinants. Dense diagonalization is used up to dimension 4000 and the implicitly restarted Lanczos method above. For $N=16$ ($D=2\,374\,753$, $3.3\times10^8$ nonzero matrix elements in the upper triangle) the matrix is stored in single precision and the three or four lowest eigenvalues are converged to a relative accuracy of $10^{-7}$, which is $2\times10^{-6}\,E_C$ in the energies and is negligible on the scale of the gaps. The total angular momentum of each eigenstate is obtained from the norm of $L_+|\Psi\rangle$.

\textit{Trial states.} The Pfaffian is the zero mode of the three-body projector; its eigenvalue is zero to $10^{-9}$, it has $L=0$, and the next eigenvalue of the three-body Hamiltonian is 0.85, 0.62, 0.58, 0.50, and 0.48 for $N=6$, 8, 10, 12, and 14. The 331 state is the zero mode of its parent Hamiltonian, with the next eigenvalue at 0.40 ($N=6$) and 0.35 ($N=8$). The component of the 331 state with $N_A=0$ coincides with the Pfaffian to $10^{-15}$.

\textit{Checks.} (i) For $N=4$ at $2Q=9$ with a pure $V_1$ interaction the ground state is the Laughlin state with $E=0$, $L=0$. (ii) The balanced-bilayer gap and 331 overlap at $d=\ell$ are $\Delta=0.05489219$ and $O_{331}=0.98513640$ for $N=6$, and $0.04355004$ and $0.95121298$ for $N=8$; these and the values at $d/\ell=0.5$, 2, and 5 ($N=6$) and 5 ($N=8$) agree to eight digits with an independent calculation. (iii) The two-subband calculation with two thin layers in place of the Hartree wave functions reproduces the bilayer-with-tunneling results (populations, both overlaps, and gap at $d=\ell$, $\Delta_{\rm SAS}=0.03\,E_C$, $N=6$) to four digits. (iv) The operator multiplying $v_0^X$ in Eq.~(\ref{eq:k0}), built from the two-body matrix elements, has the eigenvalues $[(N_L-N_R)^2-N]/2$. (v) The electron-electron Hartree contribution $\delta_H$ of Eq.~(\ref{eq:eps0}) agrees with the same quantity computed directly from the self-consistent potential of the remote-doping model ($-1.517$~meV at $n=1.37$) to four digits. (vi) The Hartree results, the one-component results for $N\le12$ in both conventions, and the bilayer results were recomputed with an independently written program and agree to all digits quoted. The two-subband calculation was checked independently for $N=6$ and 8 with a variant of the uniform terms (the direct terms of T$'$ with $N_aN_b$ in place of $N_a(N_b-\delta_{ab})$, the $X$ term of S); treatments S, T, and T$'$ themselves were not recomputed independently.

\begin{table*}[t]
\caption{Two-subband exact diagonalization for $w=72.5$~nm, remote-doping model, $N=6$. Columns as in Table~\ref{tab:twosub}. This size is aliased in its one-component limit, where the even-sector gap of about 4.7~K has no significance.}
\label{tab:twosub6}
\begin{ruledtabular}
\begin{tabular}{ccc|ccccc|ccccc|ccccc}
\hline\hline
 & & $B=0$ & \multicolumn{5}{c|}{Treatment T$'$} & \multicolumn{5}{c|}{Treatment T} & \multicolumn{5}{c}{Treatment S}\\
$n$ & $\alpha$ & $n_A/n$ & $n_A/n$ & $O_{\rm Ho}$ & $t^*$ & $\Delta_e$ & $\Delta_o$ & $n_A/n$ & $O_{\rm Ho}$ & $t^*$ & $\Delta_e$ & $\Delta_o$ & $n_A/n$ & $O_{\rm Ho}$ & $t^*$ & $\Delta_e$ & $\Delta_o$\\
\hline
0.70 & 0.140 & 21\% & 7\% & 0.92 & 0.19 & 4.69 & 12.32 & 13\% & 0.93 & 0.28 & 4.74 & 7.64 & 4\% & 0.93 & 0.12 & 4.74 & 18.24\\
0.80 & 0.122 & 26\% & 10\% & 0.93 & 0.24 & 4.74 & 9.72 & 20\% & 0.95 & 0.39 & 4.86 & 5.28 & 5\% & 0.93 & 0.14 & 4.72 & 16.67\\
0.90 & 0.107 & 30\% & 14\% & 0.94 & 0.30 & 4.85 & 7.50 & 27\% & 0.95 & 0.51 & 4.42 & 3.72 & 6\% & 0.93 & 0.16 & 4.71 & 15.30\\
1.00 & 0.095 & 33\% & 19\% & 0.95 & 0.38 & 5.00 & 5.74 & 33\% & 0.94 & 0.61 & 2.89 & 2.48 & 7\% & 0.93 & 0.18 & 4.71 & 14.10\\
1.10 & 0.085 & 36\% & 25\% & 0.95 & 0.46 & 4.80 & 4.45 & 38\% & 0.91 & 0.69 & 1.82 & 1.51 & 8\% & 0.93 & 0.20 & 4.73 & 13.04\\
1.20 & 0.076 & 38\% & 29\% & 0.94 & 0.54 & 3.64 & 3.46 & 41\% & 0.87 & 0.76 & 1.14 & 0.81 & 9\% & 0.94 & 0.22 & 4.77 & 12.09\\
1.30 & 0.069 & 39\% & 34\% & 0.92 & 0.61 & 2.64 & 2.57 & 43\% & 0.83 & 0.80 & 0.71 & 0.31 & 10\% & 0.94 & 0.24 & 4.84 & 11.24\\
1.37 & 0.064 & 40\% & 36\% & 0.90 & 0.65 & 2.10 & 2.07 & 44\% & 0.81 & 0.83 & 0.50 & 0.05 & 11\% & 0.94 & 0.25 & 4.89 & 10.70\\
1.45 & 0.059 & 41\% & 38\% & 0.88 & 0.69 & 1.62 & 1.61 & 45\% & \multicolumn{4}{c}{ground state in odd sector} & 12\% & 0.95 & 0.27 & 4.97 & 10.12\\
1.50 & 0.056 & 42\% & 39\% & 0.86 & 0.71 & 1.39 & 1.37 & 46\% & \multicolumn{4}{c}{ground state in odd sector} & 13\% & 0.95 & 0.28 & 5.03 & 9.78\\
1.60 & 0.051 & 43\% & 41\% & 0.83 & 0.75 & 1.02 & 0.97 & 46\% & \multicolumn{4}{c}{ground state in odd sector} & 14\% & 0.95 & 0.30 & 5.15 & 9.15\\
1.70 & 0.047 & 44\% & 43\% & 0.81 & 0.78 & 0.76 & 0.66 & 47\% & \multicolumn{4}{c}{ground state in odd sector} & 15\% & 0.95 & 0.32 & 5.29 & 8.58\\
1.80 & 0.043 & 44\% & 44\% & 0.79 & 0.81 & 0.57 & 0.43 & 48\% & \multicolumn{4}{c}{no $L=0$ ground state} & 16\% & 0.95 & 0.34 & 5.43 & 8.06\\
1.90 & 0.040 & 45\% & 45\% & 0.77 & 0.83 & 0.44 & 0.25 & 49\% & \multicolumn{4}{c}{no $L=0$ ground state} & 18\% & 0.96 & 0.36 & 5.59 & 7.59\\
2.00 & 0.037 & 45\% & 45\% & 0.76 & 0.85 & 0.34 & 0.11 & 49\% & \multicolumn{4}{c}{no $L=0$ ground state} & 19\% & 0.96 & 0.37 & 5.74 & 7.17\\
\hline\hline
\end{tabular}
\end{ruledtabular}
\end{table*}

\textit{$N=6$ two-subband results.} Table~\ref{tab:twosub6} gives the results for $N=6$. The populations and $t^*$ follow those for $N=8$. In treatment T the ground state for $N=6$ ceases to be in the even sector at $n\approx1.4$ and no longer has $L=0$ at $n\ge1.8$; no such breakdown occurs for $N=8$ in the range studied, where the gaps in treatment T fall below 0.1~K at $n=1.9$.

\section{One-component calculations: density, width, and model of the transverse wave function}
\label{app:width}

\begin{figure*}[t]
\includegraphics[width=0.92\textwidth]{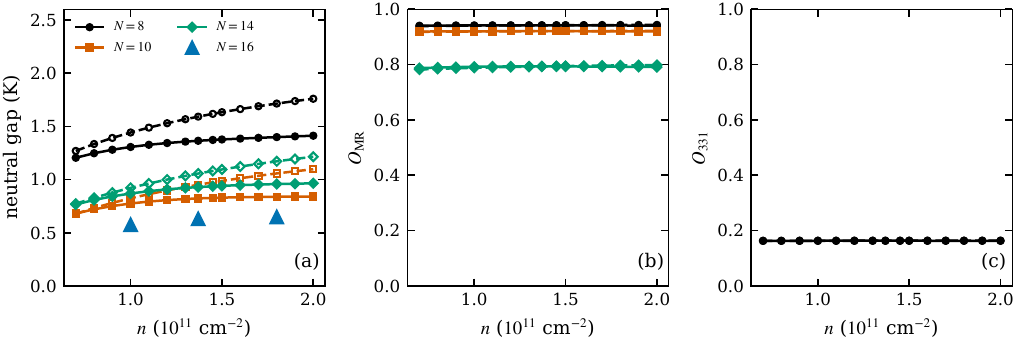}
\caption{Frozen-subband limit at $w=72.5$~nm in the spherical convention with $\rho_S$. (a) Neutral gap, (b) overlap $O_{\rm MR}$ with the Pfaffian, and (c) overlap $O_{331}=|\langle\Psi_{331}|\Psi_{\rm MR}\rangle|\,O_{\rm MR}$ with the 331 state ($N=8$ only; see Sec.~\ref{sec:onecomp}), versus density for $N=8$, 10, and 14 with the remote-doping (solid lines, filled symbols) and jellium (dashed lines, open symbols) profiles. Triangles: $N=16$, remote doping.}
\label{fig:onecompdens}
\end{figure*}

\begin{figure*}[t]
\includegraphics[width=0.92\textwidth]{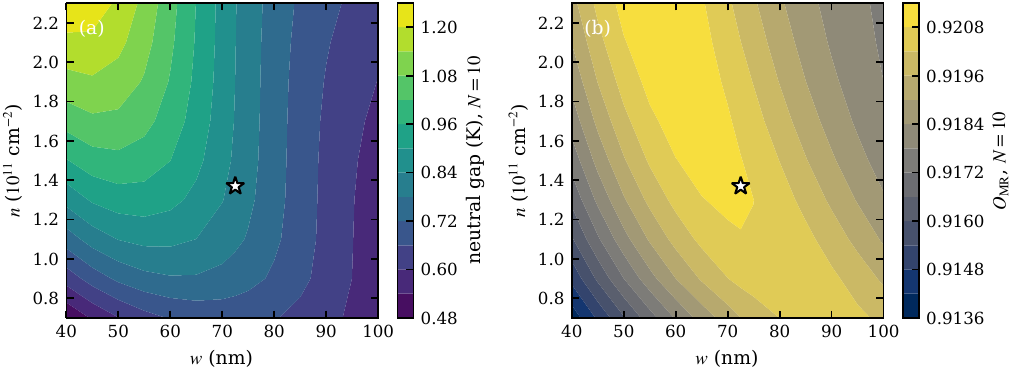}
\caption{Frozen-subband limit, $N=10$, remote-doping profile, spherical convention: (a) neutral gap in kelvin and (b) Pfaffian overlap over the plane of well width and density. The star marks the experimental point.}
\label{fig:onecompmap}
\end{figure*}

Figures~\ref{fig:onecompdens} and \ref{fig:onecompmap} give the dependence of the frozen-subband results of Sec.~\ref{sec:onecomp} on density and width for the Hartree profiles. Over the experimental range $n=1.0$--1.8 the gap in the remote-doping model increases by 7\% for $N=8$ (1.31 to 1.40~K), 9\% for $N=10$ (0.77 to 0.84~K), and 10\% for $N=14$ (0.87 to 0.96~K), the decrease of the gap in units of $E_C$ being compensated by the increase of $E_C$; the overlaps change by less than 0.003. Over $w=40$--100~nm and $n=0.7$--2.3 the $N=10$ gap stays between 0.5 and 1.25~K and the overlap between 0.914 and 0.921.

The rest of this Appendix gives the width dependence in three models of the transverse wave function.

\begin{figure*}[t]
\includegraphics[width=0.92\textwidth]{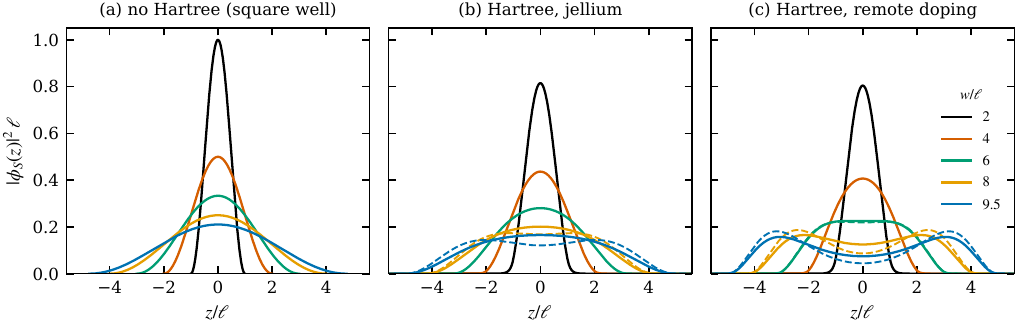}
\caption{Transverse density for well widths $w/\ell=2$, 4, 6, 8, and 9.5 in three models: (a) no Hartree (infinite square well), (b) Hartree with jellium, (c) Hartree with remote doping. Solid lines: lowest subband, $|\phi_S(z)|^2$, which is the density used in the one-component diagonalization. Dashed lines: total zero-field density of the two occupied subbands, shown where the second subband is occupied. Panels (b) and (c) are for $n=1.37$.}
\label{fig:appdens}
\end{figure*}

\begin{figure*}[t]
\includegraphics[width=0.92\textwidth]{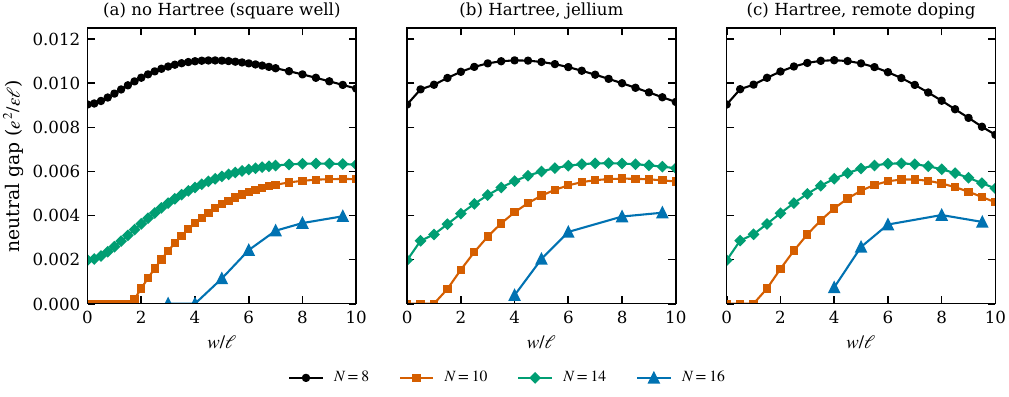}
\caption{Neutral gap of the one-component $\nu=1/2$ state versus well width in three different models of the noninteracting basis: (a) no Hartree, (b) Hartree in the jellium model, (c) Hartree in the remote-doping model. The no-Hartree results agree with Ref.~\onlinecite{StorniMorfDasSarma2010} when the widths are compared as rms widths (see text).}
\label{fig:appgap}
\end{figure*}

The gap is computed from zero width to $w=10\,\ell$ in three models, the first of which is the strict single-layer model with no self-consistency.

\textit{Models.} In all three the electrons occupy a single transverse state $\phi_S(z)$ and interact through Eq.~(\ref{eq:vr}) with $\rho=|\phi_S|^2$. (a) \textit{No Hartree}: the ground state of an infinite square well of width $w$, $\phi_S(z)=\sqrt{2/w}\cos(\pi z/w)$. (b) \textit{Hartree, jellium} and (c) \textit{Hartree, remote doping}: the lowest subband of the self-consistent calculations of Sec.~\ref{sec:hartree}, at fixed density $n=1.37$ ($\ell=7.62$~nm) and varying width, with the finite barrier of 224~meV. The densities are shown in Fig.~\ref{fig:appdens}. For narrow wells the three are similar, the Hartree densities being somewhat broader than that of the square well of the same nominal width because the wave function penetrates the finite barrier (the penetration depth is 1.6~nm, or $0.21\,\ell$, on each side). For wide wells they differ. The square-well density stays singly peaked. In the jellium model the lowest-subband density flattens but also stays singly peaked at all widths, because the Hartree potential there comes only from the difference between the electron density and the uniform background; the total zero-field density (dashed lines) acquires two maxima once the antisymmetric subband is occupied, $4.0\,\ell$ apart at $w=9.5\,\ell$ with a central density that is 84\% of the maximum. In the remote-doping model the electrostatic repulsion of the electrons produces a central barrier, and the lowest-subband density itself has two maxima, $4.3\,\ell$ apart at $w=8\,\ell$ and $6.1\,\ell$ apart at $w=9.5\,\ell$, the width of the well of Ref.~\onlinecite{Singh2025}; the total density there has maxima $6.4\,\ell$ apart and a central density that is 24\% of the maximum. The second subband becomes occupied at zero field between $w=4$ and $6\,\ell$ in the remote-doping model and between $w=6$ and $8\,\ell$ in the jellium model; it is ignored in this Appendix.

\textit{Calculation.} Spherical convention, $2Q=2N-3$, $N=8$, 10, 14, and 16. The gap plotted in Fig.~\ref{fig:appgap} is the difference of the two lowest energies when the ground state has $L=0$, and zero otherwise. For $N=16$ the calculation was done at a few widths only, shown by the symbols.

\textit{Results.} (i) The three models give the same picture. The two Hartree models agree with each other to within 5\% for $w\le7\,\ell$ for $N=8$, 10, and 14. The square-well gaps follow the Hartree gaps with the width shifted by about half a magnetic length, which is the effect of the finite barrier: for $N=10$ the gap opens at $w=1.64\,\ell$ in the square well and at $1.11\,\ell$ in both Hartree models. For $w\ge5\,\ell$ the square-well and jellium gaps agree to within 8\% for these sizes. For $N=16$ the gap opens between $w=4$ and $5\,\ell$ in the square well and just below $4\,\ell$ in the Hartree models, and close to the opening the three models differ by large factors, as they must; at $w=6\,\ell$ the $N=16$ gaps are 0.0024, 0.0033, and 0.0036 for models (a), (b), and (c).

(ii) At the width of the experimental well, $w=9.5\,\ell$, the gaps in models (a), (b), and (c) are 0.0099, 0.0094, and 0.0080 for $N=8$; 0.0057, 0.0056, and 0.0048 for $N=10$; 0.0063, 0.0062, and 0.0055 for $N=14$; and 0.0040, 0.0041, and 0.0037 for $N=16$, in units of $E_C$. The square-well and jellium values agree to within 6\%. The remote-doping value, for which the charge distribution is bilayer-like, is lower than the square-well value by 14--19\% for $N\le14$ and by 7\% for $N=16$.

(iii) The gap is small. Apart from $N=8$ it does not exceed $0.0064\,E_C$ (1.1~K at $n=1.37$) in any model at any width, and for $N=16$ it does not exceed $0.0041\,E_C$ (0.7~K).

(iv) Finite width helps, but the four sizes do not agree on how much is needed. For $N=8$ and 14 the ground state has $L=0$ at all widths including zero, with gaps of 0.0090 and 0.0020 at zero width that increase with width up to $w\approx4$--$8\,\ell$ and then saturate or decrease. For $N=10$ the ground state at small width has $L=4$ and the gap opens at $w=1.1$--$1.6\,\ell$. For $N=16$ the ground state has $L=8$ up to $w\approx4\,\ell$. All four sizes are gapped for $w\ge5\,\ell$, and at no width is the dependence of the gap on $N$ monotonic.

\textit{Comparison with Ref.~\onlinecite{StorniMorfDasSarma2010}.} Figure~2(d) of that paper shows the gap at $\nu=1/2$ as a function of a width parameter $w/\ell$ for $N=8$, 10, 14, 16, and 18; the transverse wave function used is not specified there. If its width parameter is the rms width of the transverse density, as in Ref.~\onlinecite{MorfdAmbrumenilDasSarma2002}, it corresponds to $0.181$ times the width of a square well. With that conversion our square-well results reproduce the published curves: the zero-width gaps for $N=8$ and 14 (0.009 and 0.002), the widths at which the gap opens for $N=10$ and 16 (rms widths of $0.30\,\ell$ and $0.79\,\ell$ here; about 0.3 and 0.8 there), and the ordering and approximate heights of the curves (maxima of 0.011, 0.0064, and 0.0057 for $N=8$, 14, and 10 here; about 0.010, 0.006, and 0.005 there). The comparison is made by reading a published figure and is good to about 15\%.

The conclusion for the wide well is that of the main text: a one-component state in the 72.5-nm well has a gap of about 1~K or less, in any model of the transverse wave function, a factor of four or more below the measured gap at its maximum.

\section{Strict bilayer at the maxima of the Hartree density}
\label{app:modelD}

The calculations of this paper for the well are of three kinds: the strict single-layer model (Appendix~\ref{app:width}), the frozen-subband limit, and the two-subband model. We add here a fourth, which is biased toward the 331 state by construction as the frozen-subband limit is biased toward the Pfaffian. The charge distribution of the remote-doping Hartree calculation has two maxima (Fig.~\ref{fig:profiles}). We replace the well by two layers of zero width at the positions of the maxima, with no tunneling between them and with half of the electrons in each. The model is the zero-thickness bilayer of Sec.~\ref{sec:bilayer} at $\Delta_{\rm SAS}=0$, with interactions $1/r$ within a layer and $1/\sqrt{r^2+d^2}$ between layers, and its only parameter is $d/\ell$, where $d$ is the separation of the maxima. The well width and the density enter through $d/\ell$ and through $E_C$ only.

\begin{table*}[t]
\caption{Strict bilayer at the maxima of the remote-doping Hartree density. $w$ and $d$ in nm, $n$ in $10^{11}$~cm$^{-2}$; $\Delta$ is the neutral gap in units of $E_C$ and $\Delta_K$ the same in kelvin. For comparison, at $d=\ell$: $\Delta=0.0549$, $O_{331}=0.985$ ($N=6$) and $\Delta=0.0436$, $O_{331}=0.951$ ($N=8$).}
\label{tab:modelD}
\begin{ruledtabular}
\begin{tabular}{cccc|ccc|ccc}
 & & & & \multicolumn{3}{c|}{$N=6$} & \multicolumn{3}{c}{$N=8$}\\
$w$ & $n$ & $d$ & $d/\ell$ & $\Delta$ & $\Delta_K$ & $O_{331}$ & $\Delta$ & $\Delta_K$ & $O_{331}$\\
\hline
60 & 1.00 & 31.3 & 3.50 & 0.00434 & 0.631 & 0.838 & 0.00628 & 0.913 & 0.640\\
 & 1.37 & 35.6 & 4.67 & 0.00140 & 0.237 & 0.755 & 0.00201 & 0.342 & 0.590\\
 & 1.60 & 37.1 & 5.26 & 0.00091 & 0.166 & 0.736 & 0.00123 & 0.226 & 0.580\\
70 & 1.00 & 42.7 & 4.79 & 0.00127 & 0.184 & 0.750 & 0.00181 & 0.263 & 0.588\\
 & 1.37 & 45.9 & 6.03 & 0.00058 & 0.098 & 0.721 & 0.00071 & 0.121 & 0.573\\
 & 1.60 & 47.3 & 6.71 & 0.00041 & 0.076 & 0.714 & 0.00046 & 0.085 & 0.570\\
72.5 & 1.00 & 45.3 & 5.07 & 0.00103 & 0.150 & 0.741 & 0.00143 & 0.207 & 0.583\\
 & 1.37 & 48.4 & 6.35 & 0.00049 & 0.083 & 0.718 & 0.00058 & 0.098 & 0.571\\
 & 1.60 & 49.8 & 7.06 & 0.00035 & 0.065 & 0.712 & 0.00038 & 0.070 & 0.569\\
80 & 1.00 & 52.8 & 5.92 & 0.00061 & 0.089 & 0.723 & 0.00076 & 0.110 & 0.574\\
 & 1.37 & 55.9 & 7.34 & 0.00032 & 0.054 & 0.710 & 0.00033 & 0.056 & 0.568\\
 & 1.60 & 57.3 & 8.12 & 0.00023 & 0.043 & 0.707 & 0.00023 & 0.042 & 0.566\\
90 & 1.00 & 63.0 & 7.06 & 0.00035 & 0.051 & 0.712 & 0.00038 & 0.056 & 0.569\\
 & 1.37 & 66.0 & 8.66 & 0.00020 & 0.033 & 0.706 & 0.00018 & 0.031 & 0.566\\
 & 1.60 & 67.4 & 9.55 & 0.00015 & 0.027 & 0.705 & 0.00013 & 0.024 & 0.565\\
100 & 1.00 & 72.9 & 8.17 & 0.00023 & 0.033 & 0.707 & 0.00022 & 0.032 & 0.566\\
 & 1.37 & 75.9 & 9.96 & 0.00013 & 0.022 & 0.704 & 0.00011 & 0.019 & 0.565\\
 & 1.60 & 77.3 & 10.96 & 0.00010 & 0.018 & 0.703 & 0.00008 & 0.015 & 0.564\\

\end{tabular}
\end{ruledtabular}
\end{table*}

\textit{Calculation.} The Hartree calculation of Sec.~\ref{sec:hartree} (remote doping, zero field) was done for $w=60$, 70, 72.5, 80, 90, and 100~nm and $n=1.0$--1.6 in steps of 0.05, and at $n=1.37$. For each width and density the bilayer Hamiltonian was diagonalized on the sphere at $2Q=2N-3$ for $N=6$ (three electrons per layer) and $N=8$ (four per layer). We report the neutral gap, the difference of the two lowest energies, and the overlap $O_{331}$ of the ground state with the 331 state. The ground state has $L=0$ and the first excited state $L=1$ at every point.

\begin{figure*}[t]
\includegraphics[width=0.92\textwidth]{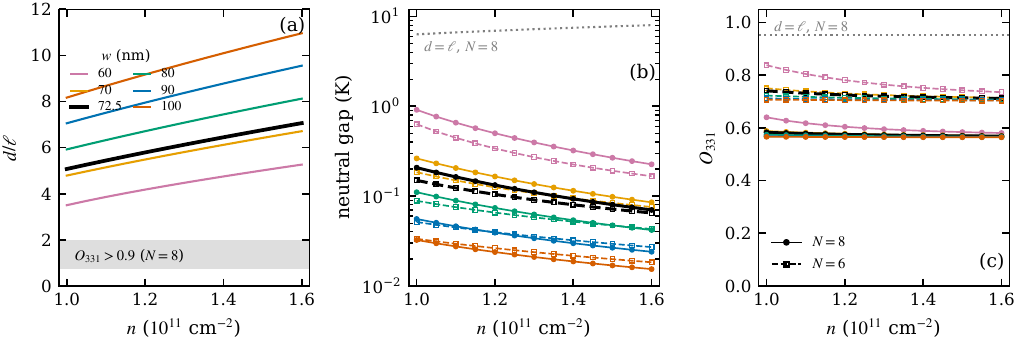}
\caption{Strict zero-width bilayer without tunneling, with the two layers at the maxima of the remote-doping Hartree density, versus density, for well widths of 60--100~nm; the heavy black line is the 72.5-nm well. (a) Separation $d$ of the density maxima in units of the magnetic length at $\nu=1/2$. The shaded band is the range of $d/\ell$ in which the overlap of the bilayer ground state with 331 exceeds 0.9 for $N=8$. (b) Neutral gap. (c) Overlap with the 331 state. In (b) and (c): $N=8$ (solid lines, filled circles) and $N=6$ (dashed lines, open squares); the dotted line is the $N=8$ value at $d=\ell$.}
\label{fig:modelD}
\end{figure*}

\textit{Results.} Figure~\ref{fig:modelD} and Table~\ref{tab:modelD} show the results.

(i) The layers are far apart. In the 72.5-nm well $d$ increases from 45.3~nm at $n=1.0$ to 49.8~nm at $n=1.6$ while $\ell$ decreases from 8.9 to 7.1~nm, so that $d/\ell$ increases from 5.1 to 7.1. Over all widths and densities studied $d/\ell$ lies between 3.5 ($w=60$~nm, $n=1.0$) and 11.0 ($w=100$~nm, $n=1.6$). The bilayer ground state has an overlap above 0.9 with 331 only for $0.75\le d/\ell\le2$ ($N=8$; Sec.~\ref{sec:bilayer}). No point of the range studied is within a factor of 1.7 of that interval, and the 72.5-nm well is a factor of 2.5--3.5 above it.

(ii) The ground state is not the 331 state. In the 72.5-nm well the overlap is 0.58--0.57 for $N=8$ and 0.74--0.71 for $N=6$, against 0.951 and 0.985 at $d=\ell$. These are the values for two decoupled layers: at $d=11\,\ell$ the overlap is 0.564 ($N=8$) and 0.703 ($N=6$), and the 72.5-nm values exceed these by 0.02 and 0.04 at most. The overlap decreases with system size, and it varies little with density and width: over the whole range of Fig.~\ref{fig:modelD} it is 0.56--0.64 for $N=8$ and 0.70--0.84 for $N=6$.

(iii) The gap is very small. In the 72.5-nm well it is 0.0014--0.0004~$E_C$, i.e., 0.21~K at $n=1.0$ falling monotonically to 0.07~K at $n=1.6$ ($N=8$), against $0.044\,E_C$ at $d=\ell$. At $n=1.37$ it is 0.10~K, a factor of 40 below the measured gap. There is no maximum as a function of density. Wider wells give smaller gaps, down to 0.02~K at 100~nm.

(iv) Narrower wells are closer to 331 but are not bilayers. The largest overlap and gap in the range studied, 0.64 and 0.9~K ($N=8$), occur for the narrowest well at the lowest density, $w=60$~nm and $n=1.0$, where $d=3.5\,\ell$. At that point the density at the center of the well is 71\% of its maximum and $\Delta_{\rm SAS}=2.3$~meV, so that a model of two layers without tunneling does not describe the well. Where the charge distribution is bilayer-like the layers are too far apart for 331, and where the maxima are closer there are no separate layers.

The conclusion is that this construction, although built to favor 331, does not produce it: for wells of 60--100~nm at $n=1.0$--1.6 the separation of the two layers is too large throughout, the two layers are nearly independent, the overlap with 331 is that of two uncorrelated quarter-filled layers, and the gap in the 72.5-nm well is a factor of 20--40 below the measured one. The 331-like correlations found in the two-subband calculation (Sec.~\ref{sec:twosub}) therefore do not come from the separation of the density maxima. They require what the strict bilayer leaves out: the width of each layer, $\lambda\approx2.6\,\ell$, which is comparable to the distance between them, and the tunneling.

\section{Benchmark: the Laughlin state at $\nu=1/3$ in the same well}
\label{app:third}

The two-subband calculation of Sec.~\ref{sec:twosub} has no independent check, because the nature of the $\nu=1/2$ state is the question being asked. At $\nu=1/3$ the situation is different: the one-subband ground state is the Laughlin state, its identity is not in doubt, it converges rapidly with system size, and it has no one-component competitor. We have therefore repeated the two-subband calculation at $\nu=1/3$ for the same well, with the same Hartree wave functions and bare splitting and over the same range of density. The purpose is twofold: to see what the second subband does to a state whose character is known, and to test the three treatments of the uniform interaction terms where the answer in the one-subband limit is known. The fate of the $\nu=1/3$ state in thick electron layers is an old question. A collapse of the $\nu=1/3$ FQHE with increasing layer thickness was reported experimentally~\cite{Shayegan1990} and was studied theoretically in one-subband models~\cite{He1990,MorfdAmbrumenilDasSarma2002}; to our knowledge it has not been studied with two subbands.

\textit{Calculation.} The Hamiltonian is Eq.~(\ref{eq:twosub}) on the sphere at the Laughlin flux, $2Q=3N-3$, for $N=6$ and 7 (Hilbert-space dimensions 34\,056 and 370\,992, each in two sectors of the parity of $N_A$). The magnetic field is that of $\nu=1/3$, $B=3nh/e$. At $n=1.37$ this is 17.0~T, so that $\ell=6.22$~nm, $E_C=208$~K, the well is $11.7\,\ell$ wide, the maxima of the charge density are $7.8\,\ell$ apart, $\alpha=0.052$, and the bare splitting, 2.45~meV as at $\nu=1/2$, is $0.137\,E_C$. The uniform terms are treated in the three ways of Sec.~\ref{sec:twosubmethod}: with the sphere's own values (S), with those of a plane of the area of the sphere (T), and with those of a plane at the physical density (T$'$). We also computed the one-subband limit, with all electrons in the symmetric subband, for $N=6$--10.

The ground state $\Psi_0$ is characterized by the fraction $n_A/n$ of electrons in the antisymmetric subband, by its overlap $O_{\rm L}$ with the Laughlin state in the symmetric subband, and by a second overlap that does not depend on the subband populations. The Laughlin state in the symmetric subband, $\Psi_{\rm L}$, is one member of a multiplet of $N+1$ states, $(T^-)^k\Psi_{\rm L}$ with $k=0,\dots,N$ and $T^-=\sum_m c^\dagger_{Am}c^{\phantom\dagger}_{Sm}$, which all have the Laughlin orbital wave function and have $k$ electrons in the antisymmetric subband; in the bilayer language this is the 333 state with an arbitrary pseudospin orientation. We denote by $O_{333}$ the norm of the projection of $\Psi_0$ on this multiplet. It plays the role here that the overlap with Ho's family plays at $\nu=1/2$: it measures the Laughlin correlations whatever the subband populations. The gap is the smaller of $\Delta_e$ and $\Delta_o$ when the ground state has $L=0$ and is set to zero otherwise.

\textit{Checks.} At zero width the overlap of the Coulomb ground state with the Laughlin state is 0.99645 for $N=6$ and 0.99636 for $N=7$, the known values~\cite{FanoOrtolaniColombo1986}. The states of the multiplet are annihilated, to $10^{-16}$, by the subband-independent hard-core interaction of which the Laughlin state is the exact ground state.

\begin{figure}[t]
\includegraphics[width=\columnwidth]{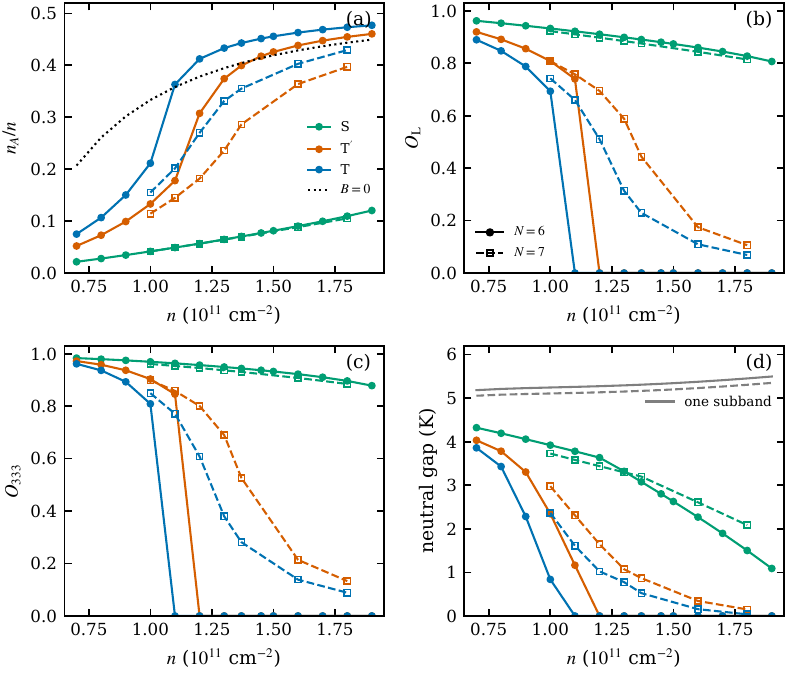}
\caption{Two-subband exact diagonalization at $\nu=1/3$ for the 72.5-nm well (remote-doping model) versus density, for the three treatments of the uniform interaction terms: plain diagonalization S (green), T$'$ (red), and T (blue); $N=6$ (solid lines, filled circles) and $N=7$ (dashed lines, open squares). (a) Fraction of electrons in the antisymmetric subband; the dotted line is the zero-field Hartree value. (b) Overlap $O_{\rm L}$ with the Laughlin state in the symmetric subband. (c) Overlap $O_{333}$ with the multiplet of Laughlin states of arbitrary subband population. (d) Neutral gap; the gray lines are the one-subband limit. Overlaps and gap are set to zero where the ground state does not have $L=0$.}
\label{fig:third}
\end{figure}

\begin{table*}[t]
\caption{Two-subband exact diagonalization at $\nu=1/3$, $w=72.5$~nm, remote-doping model. $n$ in $10^{11}$~cm$^{-2}$; $n_A/n$ in percent; $O_{\rm L}$ and $O_{333}$ as defined in the text; $\Delta$ is the neutral gap in kelvin. An entry 0 means that the ground state does not have $L=0$. The last column is the gap of the one-subband limit.}
\label{tab:third}
\begin{ruledtabular}
\begin{tabular}{cc|cccc|cccc|cccc|c}
 & & \multicolumn{4}{c|}{Treatment S} & \multicolumn{4}{c|}{Treatment T$'$} & \multicolumn{4}{c|}{Treatment T} & One subband\\
$n$ & $N$ & $n_A/n$ & $O_{\rm L}$ & $O_{333}$ & $\Delta$ & $n_A/n$ & $O_{\rm L}$ & $O_{333}$ & $\Delta$ & $n_A/n$ & $O_{\rm L}$ & $O_{333}$ & $\Delta$ & $\Delta$\\
\hline
1.00 & 6 & 4 & 0.93 & 0.97 & 3.9 & 13 & 0.81 & 0.90 & 2.3 & 21 & 0.69 & 0.81 & 0.8 & 5.2\\
 & 7 & 4 & 0.92 & 0.96 & 3.7 & 11 & 0.81 & 0.90 & 3.0 & 16 & 0.74 & 0.85 & 2.4 & 5.1\\
1.20 & 6 & 6 & 0.91 & 0.96 & 3.6 & 31 & 0 & 0 & 0 & 41 & 0 & 0 & 0 & 5.3\\
 & 7 & 6 & 0.90 & 0.95 & 3.4 & 18 & 0.69 & 0.80 & 1.6 & 27 & 0.51 & 0.61 & 1.0 & 5.1\\
1.37 & 6 & 7 & 0.89 & 0.94 & 3.1 & 40 & 0 & 0 & 0 & 44 & 0 & 0 & 0 & 5.3\\
 & 7 & 7 & 0.88 & 0.93 & 3.2 & 29 & 0.44 & 0.53 & 0.9 & 36 & 0.23 & 0.28 & 0.5 & 5.2\\
1.60 & 6 & 9 & 0.86 & 0.92 & 2.3 & 44 & 0 & 0 & 0 & 46 & 0 & 0 & 0 & 5.4\\
 & 7 & 9 & 0.84 & 0.91 & 2.6 & 36 & 0.17 & 0.21 & 0.3 & 40 & 0.11 & 0.14 & 0.2 & 5.2\\
1.80 & 6 & 11 & 0.83 & 0.90 & 1.5 & 45 & 0 & 0 & 0 & 47 & 0 & 0 & 0 & 5.5\\
 & 7 & 11 & 0.81 & 0.89 & 2.1 & 40 & 0.11 & 0.13 & 0.1 & 43 & 0.07 & 0.09 & $<0.1$ & 5.3\\

\end{tabular}
\end{ruledtabular}
\end{table*}

\textit{Results.} Figure~\ref{fig:third} and Table~\ref{tab:third} show the results.

(i) \textit{One-subband limit.} The ground state is the Laughlin state, with an overlap of 0.995 for $N=6$ and 7 and 0.990 for $N=10$ at all densities. The gap at $n=1.37$ is 0.024--0.026~$E_C$ for $N=6$--10, i.e., 4.9--5.3~K, and it is nearly independent of density in kelvin. At zero width the gap of the same systems is 0.075--0.082~$E_C$. A layer of this thickness, 10--13 magnetic lengths, therefore reduces the Laughlin gap by a factor of three and does not destroy the state. This agrees with the conclusion of Ref.~\onlinecite{MorfdAmbrumenilDasSarma2002}, the benchmark for the one-subband problem, where the charged gap of the Laughlin state, extrapolated from systems of up to 12 electrons, was found to decrease smoothly with the rms width of the layer, without a transition, out to widths of tens of magnetic lengths. That work noted that a wide well may split into an effective double layer; the two-subband calculation below is the test of that possibility.

(ii) \textit{Plain diagonalization.} The ground state has $L=0$ at all densities, with 4--11\% of the electrons in the antisymmetric subband for $n=1.0$--1.8. Its component with the antisymmetric subband empty is the Laughlin state to 0.98 or better, the overlap $O_{\rm L}$ falls from 0.93 to 0.83 because the probability of that component falls, and $O_{333}$ is 0.97--0.89. The two sizes agree to within 0.02 in the overlaps and 0.6~K in the gap. The gap decreases with density, from 3.9~K at $n=1.0$ to 3.1~K at $n=1.37$ and 1.5~K at $n=1.8$ for $N=6$ (3.7, 3.2, and 2.1~K for $N=7$): it is a quarter below the one-subband value at $n=1.0$ and 60--70\% below it at $n=1.8$. The second subband lowers the Laughlin gap. This is the opposite of what it does at $\nu=1/2$, where the admixture of the antisymmetric subband raises the even-sector gap threefold (Sec.~\ref{sec:twosub}). The Laughlin state is already the favored state of the symmetric subband and has nothing to gain from the second one; the paired state at $\nu=1/2$ has.

(iii) \textit{Corrected treatments.} With the uniform terms of the infinite system the antisymmetric population rises steeply toward its zero-field Hartree value, and the Laughlin state is lost. For $N=6$ the loss is abrupt: the ground state changes from $L=0$ to $L=2$ between $n=1.1$ and 1.2 in treatment T$'$ and between $n=1.0$ and 1.1 in treatment T, the antisymmetric population jumps to 31\% and 36\%, and the gap just before the change has fallen to 1.2 and 0.8~K. For $N=7$ the ground state keeps $L=0$ and the change is gradual and occurs at higher density: $O_{333}$ is 0.80, 0.69, 0.53, and 0.21 at $n=1.2$, 1.3, 1.37, and 1.6 in treatment T$'$, and 0.61, 0.38, 0.28, and 0.14 in treatment T, and the gap is below 1~K from $n=1.37$ on. In both sizes the state at high density has 36--48\% of the electrons in the antisymmetric subband, no Laughlin correlations, and no gap of significance. It is a state of two charge layers about eight magnetic lengths apart, each at filling $1/6$.

(iv) \textit{Comparison with experiment.} Reference~\onlinecite{Singh2025} does not report on $\nu=1/3$. It reports insulating phases below $\nu=1/2$, at $\nu<8/17$ from $n=1.35$ and on both flanks of $\nu=1/2$ by $n=1.45$, and attributes them to a pinned bilayer Wigner crystal. The loss of the Laughlin state in the corrected treatments is the same decoupling into two layers, and it occurs at $n=1.1$--1.4, at or somewhat below the density at which the experiment finds the insulating phases; it moves to higher density from $N=6$ to $N=7$. The plain diagonalization gives no indication of any such change up to $n=1.9$. To the extent that the insulating phases observed below $\nu=1/2$ extend to $\nu=1/3$, a filling that the experiment did not reach at these densities, the benchmark favors the corrected treatments over the plain diagonalization, which underestimates the antisymmetric population, and it supports the ordering inferred for $\nu=1/2$ in Sec.~\ref{sec:discussion}: the plain diagonalization errs on the one-component side and the corrected treatments, at these sizes, somewhat on the two-component side.

(v) \textit{A test.} The two kinds of treatment make different predictions for a measurable quantity. In the plain diagonalization the $\nu=1/3$ state survives to the highest density with a gap that falls smoothly by about a factor of two between $n=1.0$ and 1.8. In the corrected treatments it is strongly suppressed between $n=1.1$ and 1.4. A measurement of the $\nu=1/3$ gap as a function of density in the sample of Ref.~\onlinecite{Singh2025} would decide between them. It requires fields of 12~T at $n=1.0$ to 22~T at $n=1.8$; in Ref.~\onlinecite{Singh2025} the available field did not reach $\nu=1/2$ at $n=1.78$, which is at 14.7~T, so that $\nu=1/3$ was within its range only for $n\lesssim1.2$. Because the same electrostatic terms control the antisymmetric population at $\nu=1/2$, it would also calibrate the position of the $\nu=1/2$ state on the path between the Pfaffian and 331, which the present calculations cannot fix.

\textit{Limitations.} The calculation is at the Laughlin flux only. A two-component state at total filling $1/3$, such as the Halperin $(5,5,1)$ state, occurs at a different flux on the sphere, and crystalline states are not represented at all; the state that replaces the Laughlin state is therefore not identified, and the abrupt change for $N=6$ and the gradual one for $N=7$ show that its description at these sizes is not converged. The two-subband calculation is limited to six and seven electrons. The gaps are neutral gaps of the ideal system.

\section{Entanglement spectrum of the two-subband ground state}
\label{app:es}

The overlaps of Sec.~\ref{sec:twosub} compare the ground state with particular trial states. The orbital entanglement spectrum~\cite{LiHaldane2008} characterizes its correlations without reference to a trial state, and its low-lying part reflects the edge theory of the phase. We computed it for the two-subband ground state of the sample (remote-doping model, $w=72.5$~nm, $n=1.37$, $N=8$) in the three treatments of the uniform terms, to see whether it locates the state on Ho's path more precisely than the overlaps.

\textit{Calculation.} The 14 orbitals of the sphere ($2Q=13$) are cut into two hemispheres; region ${\cal A}$ contains the seven orbitals of lowest $L_z$ in both subbands. The reduced density matrix of ${\cal A}$ is block-diagonal in the number $N_{\cal A}$ of electrons in ${\cal A}$, their angular momentum $L_z^{\cal A}$ (measured from the lowest orbital), and the parity of the number of them in the antisymmetric subband (even or odd). The entanglement energies $\xi$ are minus the logarithms of its eigenvalues. The same cut is applied to the states of Ho's family, Eq.~(\ref{eq:ho}), for $0\le t\le1.2$; $t=0$ is the Pfaffian in the symmetric subband, whose levels are all in the even sector, and $t=1$ is 331. To locate a ground state on the path we compare its low-lying levels with those of $|\Psi_{\rm Ho}(t)\rangle$ in the sectors $N_{\cal A}=3$, 4, and 5, at the three lowest values of $L_z^{\cal A}$ at which the trial states have levels and in both parities: 30 levels in 17 sectors, the number in each sector being the number of levels of the 331 state below $\xi=6$. The distance $D(t)$ is the rms difference of these levels, with levels above $\xi=12$, or absent, set to 12, and $t_{\rm ES}$ is the value of $t$ that minimizes it. Applied to $|\Psi_{\rm Ho}(t_0)\rangle$ the procedure returns $t_0$.

\begin{figure*}[t]
\includegraphics[width=0.72\textwidth]{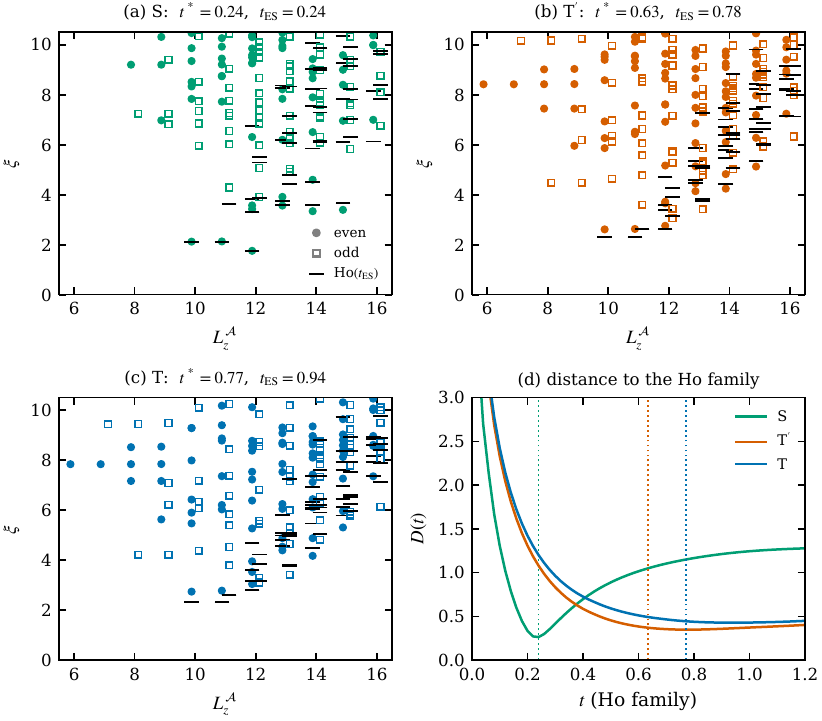}
\caption{Orbital entanglement spectrum of the two-subband ground state at $n=1.37$, $N=8$, remote-doping model, for the hemisphere cut, in the sector $N_{\cal A}=4$. (a)--(c) Treatments S, T$'$, and T: levels in the even (filled circles) and odd (open squares) sector, and the levels of the Ho state at $t=t_{\rm ES}$ (horizontal bars). The lowest $L_z^{\cal A}$ of the trial states is 10. (d) Distance $D(t)$ between the low-lying levels of the ground state and those of the Ho state with parameter $t$; the dotted lines mark the parameter $t^*$ of the best overlap.}
\label{fig:es}
\end{figure*}

\textit{Results.} (i) \textit{Plain diagonalization.} The spectrum has the structure of a Pfaffian-side paired state [Fig.~\ref{fig:es}(a)]. For $N_{\cal A}=4$ the lowest even-sector levels at $L_z^{\cal A}=10$, 11, and 12 number 1, 1, and 3, the counting of the Pfaffian; odd-sector levels, which the Pfaffian does not have and 331 has, appear at $\xi=3.9$--4.3; and the levels not represented in the trial states begin at $\xi\approx6.8$, so that the low-lying part is separated from the rest by an entanglement gap. The minimum of $D(t)$ is sharp, $D=0.26$ at $t_{\rm ES}=0.24$, and $D<0.36$ only for $0.20\le t\le0.28$. This is the value of $t^*$ found from the overlaps (0.24).

(ii) \textit{Corrected treatments.} With the uniform terms of the infinite system the low-lying levels resemble those of 331 more closely: more odd-sector levels, at lower $\xi$. The minimum of $D(t)$ is at $t_{\rm ES}=0.78$ (T$'$) and 0.94 (T), somewhat beyond the overlap values $t^*=0.63$ and 0.77, but it is shallow: $D$ is within 0.1 of its minimum for $t$ from about 0.55 to beyond 1. There is no entanglement gap: levels not represented in the trial states appear at $\xi=4.5$ (T$'$) and 4.2 (T), among the low-lying ones. This agrees with the smaller overlaps with Ho's family in these treatments (0.74 and 0.64): the state is leaving the family of paired states toward two weakly coupled layers.

(iii) \textit{What the spectrum adds.} The entanglement spectrum places each ground state where the overlaps place it, at the same $t$ for the plain diagonalization and somewhat closer to 331 for the corrected treatments. It does not locate the sample more precisely, because the uncertainty in its location is not an uncertainty about the character of a given ground state but about which ground state is the right one, and that is decided by the electrostatic terms of Sec.~\ref{sec:twosubmethod}, not by the diagnostic. At eight electrons the spectrum also cannot distinguish a phase transition along the path from a crossover. It does establish that the ground state of the plain diagonalization has the low-lying entanglement structure of a triplet-paired state on the Pfaffian side of the path, with a clear entanglement gap, whereas the corrected ground states have none.

\end{document}